\documentclass[10pt]{article}
\usepackage[utf8]{inputenc}
\usepackage[T1]{fontenc}
\usepackage{fullpage}
\usepackage{cite}
\usepackage{rotating}
\usepackage{subcaption}
\usepackage{appendix}

\usepackage{algorithm,algpseudocode}%
\algnewcommand{\algorithmicforeach}{\textbf{for each}}
\algdef{SE}[FOR]{ForEach}{EndForEach}[1]
  {\algorithmicforeach\ #1\ \algorithmicdo}
  {\algorithmicend\ \algorithmicforeach}
  
 \algnewcommand{\algorithmicifeach}{\textbf{if}}
\algdef{SE}[IF]{IfEach}{EndIfEach}[1]
  {\algorithmicifeach\ #1\ \algorithmicdo}
  {\algorithmicend\ \algorithmicifeach}

\usepackage{multirow}
\usepackage{colortbl}
\usepackage{times}
\usepackage{graphicx}
\usepackage{bm}
\usepackage{amsmath}
\usepackage{amssymb}
\usepackage[dvipsnames]{xcolor}
\usepackage[normalem]{ulem} 
\usepackage[mathlines]{lineno}
\usepackage{bbm}
\usepackage{cleveref}
\usepackage{physics}

\newlength{\marginadjust}
\definecolor{ao(english)}{rgb}{0.0, 0.5, 0.0}

\newcommand{\Frac}[2]{{{#1}/{#2}}}  
\def\sMid{\,|\,}  
\def\xvec{\mathbf{x}}
\def\lvec{\mathbf{p}_{\rm l}} 
\def\pvec{\mathbf{p}_{\rm s}} 
\def\cvec{\mathbf{p}_{\rm f}} 

\def\svec{\mathbf{s}}
\def\sfg{\svec_{\rm fg}}
\def\sbg{\svec_{\rm oc}}
\def\psivecfg{\boldsymbol{\psi}_{\rm fg}}
\def\psivecbg{\boldsymbol{\psi}_{\rm oc}}
\def\thetavec{\boldsymbol{\theta}}
\def\alphafg{a_{\rm fg}}
\def\alphabg{a_{\rm oc}}
\def\rfg{r_{\rm fg}}
\def\rbg{r_{\rm oc}}
\def\pmin{\theta_{\rm min}}
\def\pmax{\theta_{\rm max}}
\def\pmid{\theta_{\rm mid}}
\def\xvec{\mathbf{x}}
\def\ffg{f_{\rm fg}}
\def\gfg{g_{\rm fg}}
\def\fbg{f_{\rm bg}}
\def\gbg{g_{\rm bg}}
\def\smatfg{\mathbf{S}_{\rm fg}}
\def\smatbg{\mathbf{S}_{\rm bg}}

\def\lnorm{\mathbf{n}_{\lvec}}
\def\cnorm{\mathbf{n}_{\cvec}}
\def\planenorm{\mathbf{n}_{\rm plane}}
\def\pplane{\pvec_{\rm plane}}
\def\pnorm{\mathbf{n}_{\pvec}}
\def\psivecfghat{\widehat{\boldsymbol{\psi}}_{\rm fg}}
\def\psivecbghat{\widehat{\boldsymbol{\psi}}_{\rm bg}}

\def\sfg{\svec_{\rm fg}}
\def\sbg{\svec_{\rm oc}}
\def\thetavec{\boldsymbol{\theta}}
\def\psivecfg{\boldsymbol{\psi}_{\rm fg}}
\def\psivecbg{\boldsymbol{\psi}_{\rm oc}}

\def\psivecfghat{\widehat{\boldsymbol{\psi}}_{\rm fg}}
\def\psivecbghat{\widehat{\boldsymbol{\psi}}_{\rm bg}}

\def\ffg{f_{\rm fg}}
\def\gfg{g_{\rm fg}}
\def\fbg{f_{\rm bg}}
\def\gbg{g_{\rm bg}}
\def\smatfg{\mathbf{S}_{\rm fg}}
\def\smatbg{\mathbf{S}_{\rm bg}}
\def\ael{a_{\rm e}}
\def\bel{b_{\rm e}}
\def\cel{c_{\rm e}}

\def\pvec{\mathbf{p}_{\rm s}}
\def\cvec{\mathbf{p}_{\rm f}}
\def\albedo{a}
\def\deltaArc{\Delta_{\rm arc}^{i,k}}
\def\lvec{\mathbf{p}_{\rm l}}
\def\cvecmid{\bar{\mathbf{p}}_{\rm f}}
\def\pvecmid{\bar{\mathbf{p}}_{\rm s}}
\def\lnorm{\mathbf{n}_{\rm l}}
\def\cnorm{\mathbf{n}_{\rm f}}
\def\pnorm{\mathbf{n}_{\rm s}}

\def\cvecmidi{\bar{\mathbf{p}}_{{\rm f},n}}

\def\mline{m_{\rm line}}
\def\cline{b_{\rm line}}
\def\xvec{\mathbf{x}}
\def\vvec{\mathbf{v}}
\def\yvec{\mathbf{y}}

\def\svec{\mathbf{s}}
\def\qvec{\mathbf{q}}
\def\bvec{\mathbf{b}}
\def\vfg{\vvec_{\rm fg}}
\def\vbg{\vvec_{\rm bg}}

\def\deltaArc{\Delta_{\rm arc}^{n,k}}

\def\planenorm{\mathbf{n}_{\rm plane}}
\def\pplane{\mathbf{p}_{\rm plane}}
\def\camerapix{\mathcal{P}}
\def\hiddensurf{\mathcal{S}}
\def\tmin{\theta_{\rm min}}
\def\tmax{\theta_{\rm max}}
\def\tmid{\theta_{\rm mid}}
\def\dmin{d_{\rm min}}
\def\dmax{d_{\rm max}}
\def\dint{d_{\rm int}}
\def\Astart{A_{\rm start}}
\def\Astop{A_{\rm stop}}
\def\Amid{A_{\rm mid}}
\def\Bstart{B_{\rm start}}
\def\Bstop{B_{\rm stop}}
\def\Bmid{B_{\rm mid}}
\def\alphafg{\albedo_{\rm fg}}
\def\alphabg{\albedo_{\rm oc}}
\def\rfg{r_{\rm fg}}
\def\rbg{r_{\rm oc}}
\def\pmin{\theta_{\rm min}}
\def\pmax{\theta_{\rm max}}
\def\pmid{\theta_{\rm mid}}
\def\stheta{\svec_\theta}

\def\sigmavec{\boldsymbol{\Sigma}}
\def\fmexact{\svec_{\rm exact}}
\def\fmapprox{\svec_{\rm fast}}
\newcommand{\sign}[1]{{\rm sign}( #1)}

\def\Aground{A_{\rm ground}}
\def\Bground{B_{\rm ground}}

\def\T{{\top}}

\title{Non-Line-of-Sight Tracking and Mapping with an Active Corner Camera}
\author{Sheila Seidel$^{1,2,+,}$\thanks{Corresponding author: sseidel@bu.edu. $^+$These authors contributed equally to this work} ,
Hoover Rueda-Chac\'{o}n$^{1,+}$,
Iris Cusini$^3$,
Federica Villa$^3$, \\
Franco Zappa$^3$,
Christopher Yu$^2$, and
Vivek K Goyal$^1$ \\\\
$^1$ Electrical and Computer Engineering, Boston University, 8 St. Mary's Street, Boston, MA, 02215, USA\\
$^2$ Charles Stark Draper Laboratory, 55 Technology Square, Cambridge, MA, 02139, USA \\
$^3$ Dip. Elettronica, Informazione e Bioingegneria, Politecnico di Milano, \\ Piazza Leonardo Da Vinci, 32, Milano, I-20133, Italy
}

\begin{document}
\date{}
\maketitle

\section{The Active Corner Camera}
The ability to form non-line-of-sight (NLOS) images of changing scenes could be transformative in a variety of fields, including search and rescue, autonomous vehicle navigation, and reconnaissance.
Most existing active NLOS methods illuminate the hidden scene using a pulsed laser directed at a relay surface and collect time-resolved measurements of returning light~\cite{Faccio2020}.
The prevailing approaches include raster scanning of a rectangular grid on a vertical wall opposite the volume of interest to generate a collection of confocal measurements~\cite{OToole2018,Lindell2019_Wavebased,Heide2019_nlos,Liu2019_pField}.
These and a recent method that uses a horizontal relay surface~\cite{Rapp2020} are inherently limited by the need for laser scanning.
Methods that avoid laser scanning
track the moving parts of the hidden scene
as one or two point targets~\cite{Chan2017,Gariepy2016}.
In this work,
based on more complete optical response modeling yet still without multiple illumination positions,
we demonstrate accurate reconstructions of objects in motion and a `map’ of the stationary scenery behind them.
The ability to count, localize, and characterize the sizes of hidden objects in motion, combined with mapping of the stationary hidden scene, could greatly improve indoor situational awareness in a variety of applications.

The challenge of both active and passive NLOS imaging techniques is that measured light returns to the sensor after multiple diffuse bounces. With each bounce, light is scattered in all directions, eliminating directional information, and attenuating light by a factor proportional to the inverse-square of the path length. 
Particularly in the passive setting, where no illumination is introduced, occluding structures that limit possible light paths have been used to help separate light originating from different directions in the hidden scene \cite{Torralba2014, Bouman2017,Thrampoulidis2018_exploiting, Xu2018,Baradad2018, Saunders2019computational,Yedidia2019_using,Tanaka2020,Lin2020}. 
Useful structures include the aperture formed by an open window~\cite{Torralba2014} or the inverse pinhole~\cite{Cohen1982} created when a once-present object moves between measurement frames.
Unlike other occluding structures,
whose shapes must be estimated or somehow known~\cite{Yedidia2019_using,Aittala2019ComputationalMB, Saunders2019computational,Xu2018,Thrampoulidis2018_exploiting},
vertical wall edges have a known shape and are often present when NLOS vision is desired.
An edge occluder blocks light as a function of its azimuthal incident angle around the corner and, as a result, enables computational recovery of azimuthal information about the hidden scene.
This was first demonstrated in the passive setting~\cite{Bouman2017,Seidel2019_corner}, where 1D (in azimuthal angle) reconstructions of the hidden scene were formed from photographs of the floor adjacent to the occluding edge; 2D reconstruction was demonstrated in a controlled static environment, although the longitudinal information present in the passive measurement was found to be weak~\cite{seidel2020}.

In the active setting, most of the approaches proposed to date scan a pulsed laser over a set of points on a planar Lambertian relay wall and perform time-resolved sensing with a single-photon detector to collect transient information~\cite{Velten2012, Buttafava2015, Pediredla2017, Arellano2017, Heide2019_nlos, ahn_2019_convolutionalNLOS, OToole2018, Lindell2019_Wavebased, Xin2019_Fermat, Liu2019_pField}.
To reconstruct large-scale scenes, these approaches generally require scanning a large area of the relay wall and thus a large opening into the hidden volume.
To partially alleviate these weaknesses,
edge-resolved transient imaging (ERTI)~\cite{Rapp2020}
combines the use of an edge occluder
 from passive NLOS imaging with the
transient measurement
abilities of active systems.
ERTI
scans a laser on the floor along an arc around a vertical edge, incrementally illuminating more of the hidden scene with each scan position.
Differences between measurements at consecutive scan positions are processed together to reconstruct a large-scale stationary hidden scene.
However, the laser scanning requirement is still a limiting constraint.
An earlier work using the floor as a relay surface shortens acquisition time by using a 32\,$\times$\,32 pixel SPAD array in conjunction with a stationary laser~\cite{Gariepy2016}.
Simultaneous measurements from the 1024 pixels
and background subtraction are used to
track the horizontal position of a hidden object in motion, modeled as a point reflector.

\begin{figure}
    \centering
    \includegraphics[width=1\linewidth]{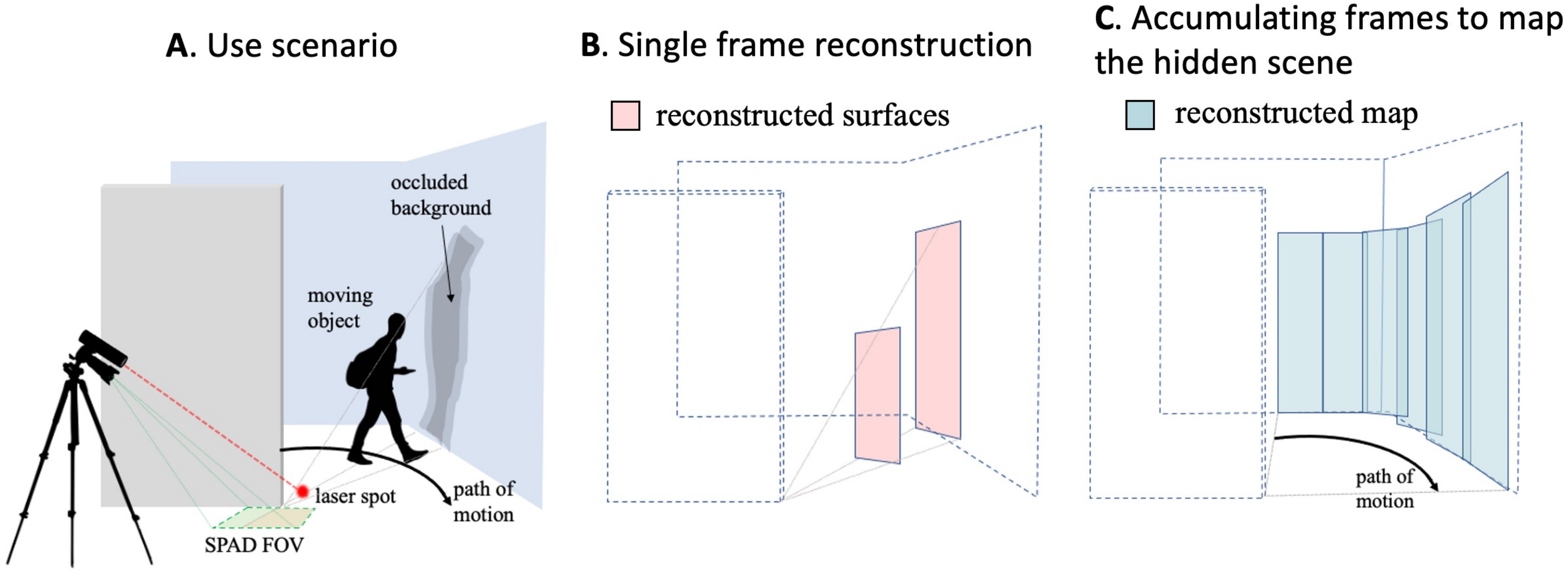}
    \caption{An active corner camera use scenario is shown in (A).
    A pulsed laser pointed at the floor illuminates the hidden scene while a SPAD camera adjacent to the occluding wall measures the temporal response of returning light. An initial reference measurement is acquired to characterize the response of the  stationary scene. When the moving object enters, the new measurement includes added photon counts due to the object and reduced photon counts at more distant ranges due to the occluded background region behind it. Using these changes, we reconstruct objects in motion as well as the occluded background regions behind them (B). By accumulating frames as an object moves through the hidden scene, we form a map of the stationary hidden scene.
    }
    \label{fig:acquisition_strategy}
\end{figure}

In this work, we use similar hardware  as in~\cite{Gariepy2016} and also use a floor as a relay surface. 
As illustrated in \Cref{fig:acquisition_strategy}A, our desire for NLOS vision is caused by an occluding wall;
unlike in~\cite{Gariepy2016}, the edge of the wall is explicitly modeled and exploited to enable reconstruction of moving objects in the hidden scene. Like the passive corner-camera systems in~\cite{Bouman2017,Seidel2019_corner,seidel2020,Krska:ICCP2022}, we position the SPAD field of view (FOV) adjacent to the wall edge, as shown in \Cref{fig:acquisition_strategy}A, to derive azimuthal resolution from the occluding edge. As in~\cite{Rapp2020}, we derive longitudinal resolution from the temporal response to the pulsed laser. However, unlike~\cite{Rapp2020}, our proposed system acquires data for each frame in a single snapshot without scanning, allowing us to track hidden objects in motion.
Consider \Cref{fig:acquisition_strategy}A and note that a moving target not only adds reflected photons to the measurement, but also reduces photons due to the shadow it casts on the stationary scene behind it. Through additional modeling of occlusion within the hidden scene itself, we use these changes to reconstruct occluded background regions for each frame (\Cref{fig:acquisition_strategy}B).  As an object moves through the hidden scene, reconstructions of occluded background regions may be accumulated to form a \textit{map} of the hidden scene (\Cref{fig:acquisition_strategy}C). In contrast to~\cite{Gariepy2016}, where $x$ and $y$ coordinates are estimated for a hidden target in motion at an assumed height, our algorithm counts hidden objects in motion and reconstructs their shape (i.e., height and width), location, and reflectivity while simultaneously mapping the stationary hidden scenery occluded by them. 

In our setup, the measurement rate at the $n$th spatial pixel in the $k$th time bin is Poisson distributed 
\begin{equation}
    \xvec^{n,k}\sim {\rm Poisson}\!\left(\bvec^{n,k} + \sfg^{n,k}(\psivecfg)-\sbg^{n,k}(\psivecfg,\psivecbg)\right),
    \label{eq:model}
\end{equation}
where $\bvec\in\mathbb{R}^{N\times K}$ is the rates due to stationary scenery, $\sfg\in\mathbb{R}^{N\times K}$ is the response of the foreground object, 
and $\sbg\in\mathbb{R}^{N\times K}$ is the response of the occluded background region, before the object enters.
We assume that $\bvec$ is approximately known through a reference measurement acquired before moving objects enter the hidden scene
or through other means.
Vectors $\psivecfg$ and $\psivecbg$ contain parameters that describe the foreground objects and corresponding occluded background regions. We seek to recover the parameters $\psivecfg$ and $\psivecbg$ from the measurement $\xvec$, for each measurement frame.

\begin{figure}
    \centering
    \includegraphics[width=1\linewidth]{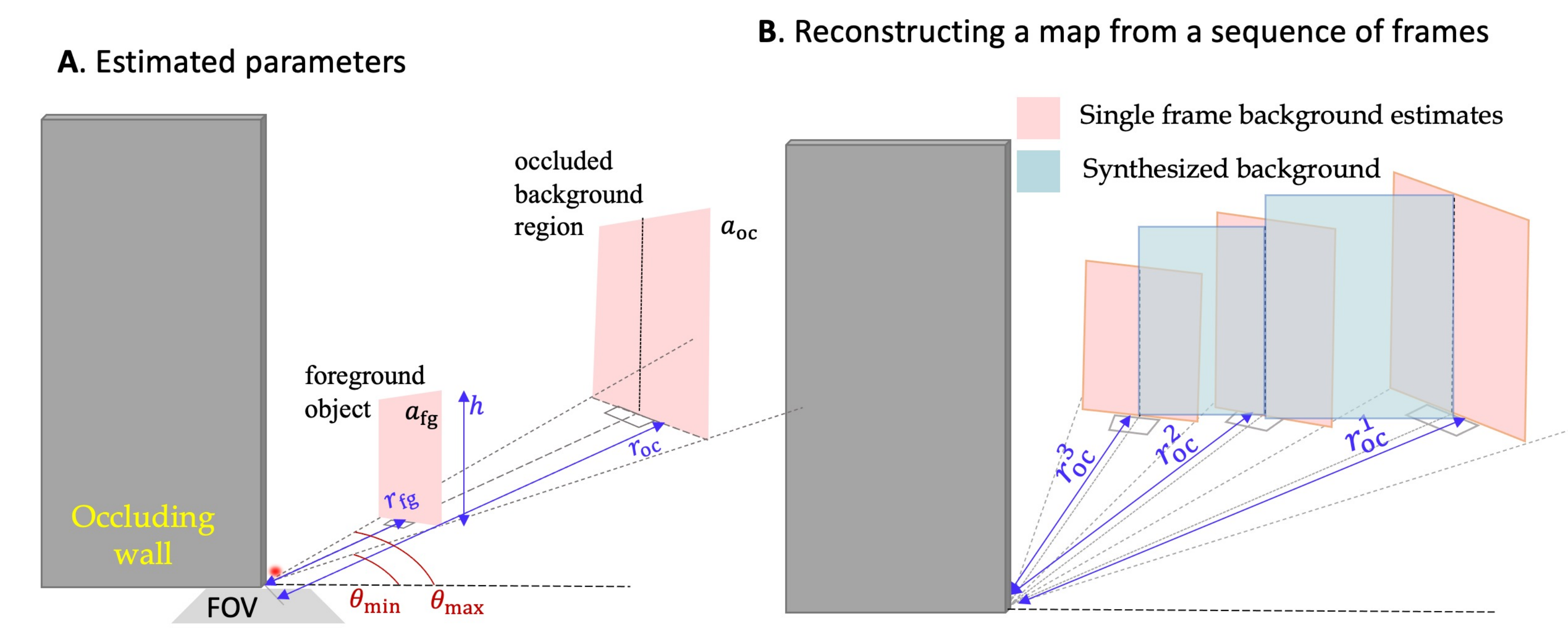}
    \caption{Our facet-based model describes moving objects and the occluded background regions behind them as edge-facing rectangular, planar facets characterized by the parameters shown in (A). When parameters have been estimated for a sequence of frames, estimates are post-processed together to form a map reconstruction (B). The vertical lines going through the center of estimated occluded regions (light red) are connected to form the map reconstruction (blue).
    }
    \label{fig:algorithm_concept}
\end{figure}

As shown in \Cref{fig:algorithm_concept}A for a single moving object, we model moving objects and their occluded background regions each as a single vertical, planar, rectangular facet resting on the ground. 
We assume there are $M$ moving objects with parameters $\psivecfg = \{(\thetavec^m,\alphafg^m,\,\rfg^m,\,h^m), \,\,\,\, m = 1,\ldots,M \}$.
Marked in \Cref{fig:algorithm_concept}A, $\alphafg^m$ is the albedo, $\rfg^m$ is range, and $h^m$ is height of the $m$th object.
Angles $\thetavec^m = (\pmin^m, \pmax^m)$ are the minimum and maximum polar angles of the foreground facet, measured around the occluding edge in the plane of the floor.
The $m$th occluded region is described by range $\rbg^m$ and albedo $\alphabg^m$ parameters $\psivecbg= \{(\alphabg^m,\,\rbg^m), \,\,\,\, m = 1,\ldots,M\}$.
The height of the occluded region is not a separate parameter; it depends upon its range $\rbg$ and the corresponding moving object's range $\rfg$ and height $h$.
When parameters $\psivecfg$ and $\psivecbg$ have been estimated for a sequence of measurement frames, the vertical lines running through the centers of estimated occluded background regions (light red) are joined by planar facets (blue) to form a contiguous map of the background, as shown in \Cref{fig:algorithm_concept}B. 

\begin{figure}
    \centering
    \includegraphics[width=.4\linewidth]{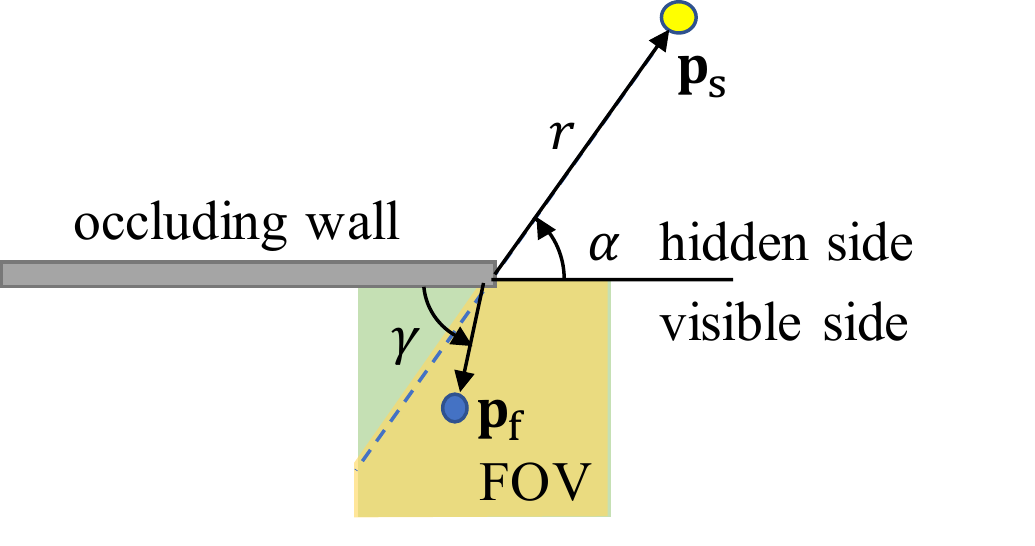}
    \caption{A bird's eye view of the vertical edge occluder.
    The edge blocks light from scene point $\pvec$ as a function of its azimuthal angle $\alpha$, measured around the corner.
    A point $\cvec$ in the SPAD FOV at angle $\gamma$ is illuminated by $\pvec$ if $\gamma\geq\alpha$.
    }
    \label{fig:forward_model}
\end{figure}

A method to quickly compute the rates due to a planar rectangular facet resting on the ground (i.e., $\sfg(\psivecfg)$ and $\sbg(\psivecfg,\psivecbg)$) is a key part of our inversion algorithm.
Take $\lvec$ to be the position of the laser illumination and $\cvec$ to be a point on the floor in the area of the $n$th camera pixel $\camerapix_n$.
The flux
during the $k$th time bin
at the $n$th camera pixel
due to hidden surface $\hiddensurf$ is
\begin{align}
    \svec^{n,k}
    =
    \int_{(k-1)\Delta_t}^{k\Delta_t}
    \int_{\camerapix_n}
    &\int_{\hiddensurf} v(\pvec,\cvec)a(\pvec)\frac{G(\pvec,\,\lvec,\,\cvec)}{\norm{\lvec-\pvec}^2\norm{\cvec-\pvec}^2}\nonumber\\
    &w\Big(t-t_0 - \frac{\norm{\lvec-\pvec}+\norm{\cvec-\pvec}}{c}\Big) d\pvec
    \,d\cvec
    \,dt
    ,
    \label{eq:exactFM}
\end{align}
where $a(\pvec)$ is the surface albedo at point $\pvec$, $w(\cdot)$ is the pulsed illumination waveform, $\Delta_t$ is the duration of a time bin,
$t_0$ is the time the pulse hits the laser spot,
and $c$ is the speed of light.
The factor $G(\cdot,\cdot,\cdot)$ is the Lambertian bidirectional reflectance distribution function (BRDF) and is the product of foreshortening terms (i.e., the cosine of the angle between the direction of incident light and the surface normal), as described in Appendix~\ref{app:fast_approx}.
The factor $v(\pvec,\cvec)$ is the `visibility function' that describes the occlusion provided by the occluding edge between hidden scene point $\pvec$ and SPAD FOV point $\cvec$.
As shown in the bird's eye view of \Cref{fig:forward_model},
point $\cvec$ is located at angle $\gamma$ measured from the occluding wall in the plane of the floor. Point $\pvec$ is at azimuthal angle $\alpha$, in the plane of the floor, measured around the corner from the the boundary between hidden and visible sides of the wall.
Thus, point $\cvec$ is only visible to $\pvec$ if $\gamma\geq\alpha$:
\begin{equation}
    v(\pvec,\cvec) = 
        \begin{cases}
    1, & \mbox{if $\gamma\geq\alpha$} \\
    0, & \mbox{otherwise}.
    \end{cases}
    \label{eq:visibilityFunction}
\end{equation}
The yellow region in the SPAD FOV is the collection of all points $\cvec$ not occluded from point $\pvec$ by the wall, where $v(\pvec,\cvec) = 1$.
In the green region, light from $\pvec$ is blocked by the wall and $v(\pvec,\cvec) = 0$.
This fan-like pattern is the `penumbra' exploited by the passive corner camera in~\cite{Bouman2017,Seidel2019_corner,seidel2020,Krska:ICCP2022}. 

In some previous works, computation time is reduced by making a confocal approximation~\cite{OToole2018,Rapp2020} (i.e., assuming the laser and detector are co-located).
Under this assumption, the set of points $\pvec$ in the scene that correspond to equal round-trip travel time from $\lvec$, to $\pvec$, and back to $\cvec$, lie on a sphere. 
In contrast, as in~\cite{Gariepy2016},
we seek to exploit the spatial diversity of our sensor array and thus require a more general ellipsoidal model that arises when $\lvec$ and $\cvec$ are not co-located.
When $\hiddensurf$ is a vertical, rectangular, planar facet, the intersection of a given round trip travel time (the ellipsoid) and the plane containing our facet is an ellipse. Using \cite{Klein2012}, we write that ellipse in translational form, enabling us to perform the integration in \eqref{eq:exactFM} in polar coordinates.
This method, described further in Appendix~\ref{app:fast_approx}, allows us to compute $\sfg(\psivecfg)$ and $\sbg(\psivecfg,\psivecbg)$ quickly enough to implement our inversion algorithm.

\begin{figure}
    \centering
    \includegraphics[width=1\linewidth]{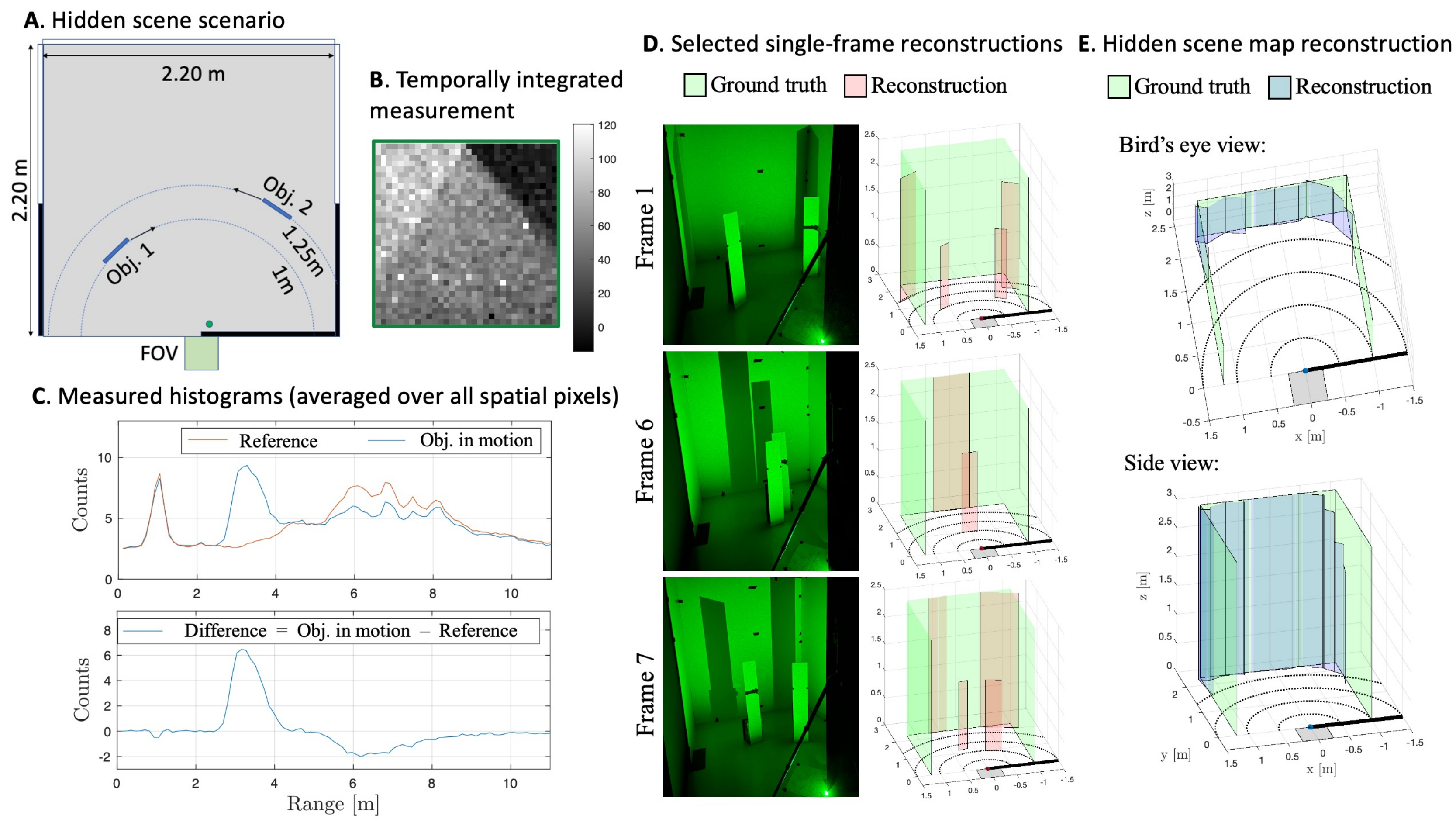}
    \caption{Sample measurements and reconstruction results for a scenario where two objects move through the hidden scene (A).
    Temporally integrated measurements in (B) show the penumbra pattern, with a distinct shadow due to each of the two hidden objects.
    Spatially averaged measurements for the stationary scene (red) and one motion frames (blue) are shown on the top axis of (C), with their difference shown below. The peak near 3\,m is due to the moving objects; the dip near 6\,m is due to the background occlusion.
    Selected single-frame reconstructions are shown in (D).
    Two views of the map reconstructions, accumulated over 8 frames, are shown in (E).
   }
    \label{fig:reconstruction_results}
\end{figure}

Before estimating parameters $\psivecfg$ and $\psivecbg$ for a given frame, we estimate the number of moving objects $M$.
The passive corner camera processing of~\cite{Seidel2019_corner}
is applied to the temporally integrated difference measurement (e.g., \Cref{fig:reconstruction_results}B) to produce a 1D reconstruction of change in the hidden scene as a function of azimuthal angle $\alpha$.
The intervals where this 1D reconstruction is above some threshold are counted to determine $M$.
Parameters $\psivecfg$ and $\psivecbg$ are then estimated from time-resolved measurement $\xvec$ using a
maximum likelihood estimate (MLE)
constrained over broad, realistic ranges of $\psivecfg$ and $\psivecbg$.
To approximate the constrained MLE, the Metropolis-Hastings algorithm is applied in two stages: first to estimate foreground parameters $\psivecfg$, assuming no occlusion of the background, and second to estimate the parameters of the occluded background region $\psivecbg$, assuming $\psivecfg = \psivecfghat$. Further details about our procedure for estimating $N$, $\psivecfg$, and $\psivecbg$ are included in Appendix~\ref{app:inversion_algorithm}.

In \Cref{fig:reconstruction_results}, we show reconstruction results for eight measurement frames acquired as two hidden objects move along arcs toward and then past each other as shown in \Cref{fig:reconstruction_results}A. 
In this demonstration, the \textit{integration time} (i.e., the total time over which the camera collects meaningful data) used for each new frame was 0.4\,s.
Integration time for the reference measurement was 30\,s.
The \textit{acquisition time} (i.e., the total time required to collect, accumulate, and transfer data) was longer; see Appendix~\ref{app:acq_time_analysis}.
Measurements averaged spatially over all pixels are shown in \Cref{fig:reconstruction_results}C\@.
The top plot shows the stationary scene measurement (red) with the measurement acquired after objects have moved into the scene in Frame~1 (blue); their difference is shown on the axis below.
A peak in the difference around 3 meters is due to the additional photon counts introduced by the two moving objects; a dip at 6 meters is due to their occluded background regions.
Although it is impossible to separate the contributions from each target in this
spatially integrated
view of the data,
the vertical edge occluder casts two distinct shadows in the temporally integrated measurement shown in \Cref{fig:reconstruction_results}B\@.
Our processing exploits spatiotemporal structure of the data that is not apparent from the projections in \Cref{fig:reconstruction_results}B and C\@.

Single-frame reconstruction results are shown for three different frames in \Cref{fig:reconstruction_results}D\@.
In Frames 1 and 7, two targets are resolved with accurate heights, widths, and ranges.
The reconstructed occluded background regions are placed accurately in range.
In Frame 6, the closer target passes in front of the more distant one, and the single reconstructed target is placed at the range of the front-most object.
Two views of the reconstructed maps (blue), accumulated over all eight measurement frames, are shown in \Cref{fig:reconstruction_results}E to closely match the true wall locations (green).

\begin{figure}
    \centering
    \includegraphics[width=1\linewidth]{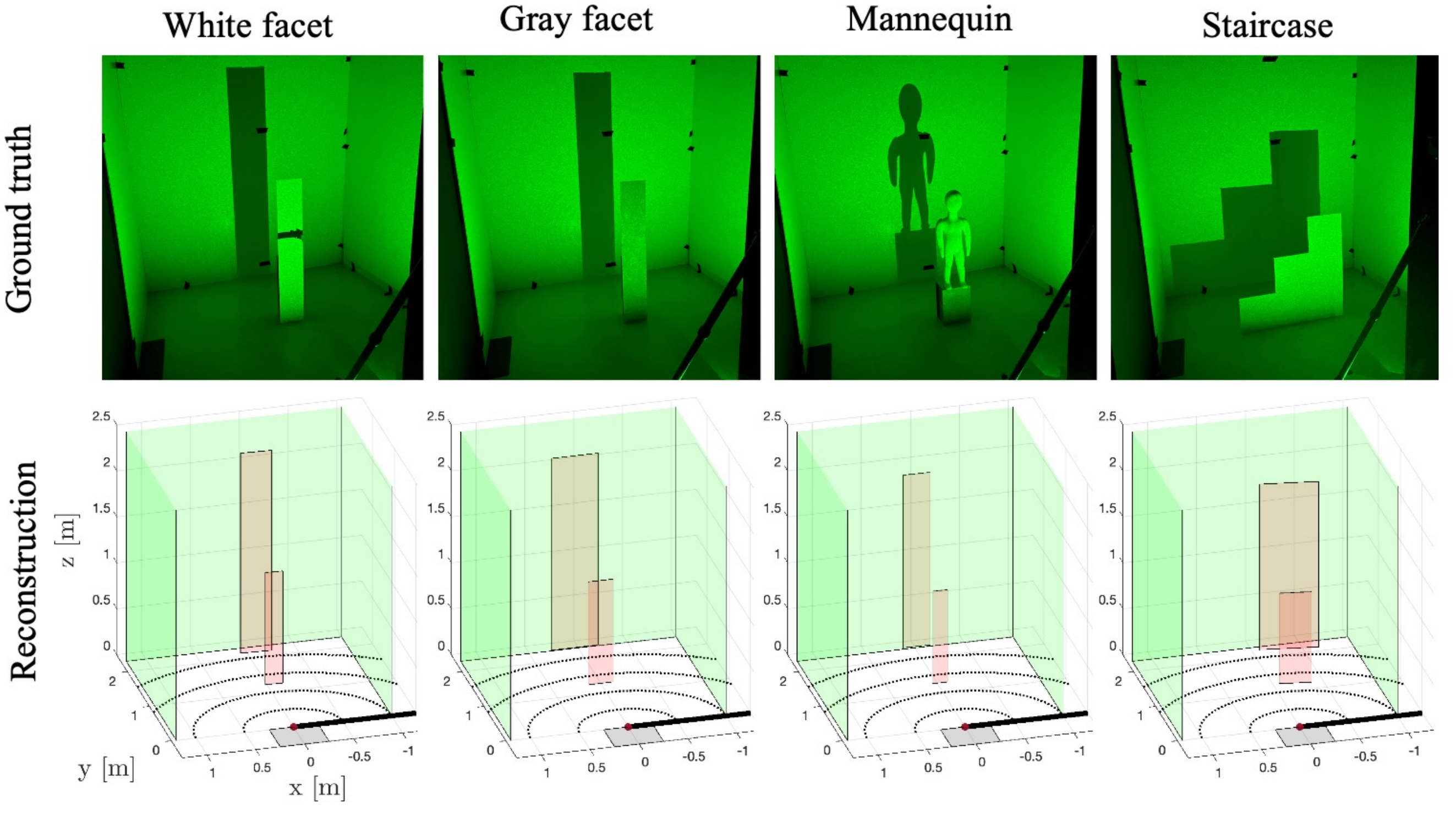}
    \caption{Single-frame reconstruction results for four different hidden objects, including a less reflective gray target, a non-planar mannequin, and a non-rectangular staircase. In all cases, our model allows us to accurately locate both the object and the stationary scene in the background.}
    \label{fig:robustness_demo}
\end{figure}

In \Cref{fig:robustness_demo}, we demonstrate that our reconstruction algorithm works with dimmer moving objects as well as with objects that do not match our rectangular, planar facet model.
Single-frame reconstruction results are shown for the white facet, a darker gray facet, a mannequin, and a staircase shaped object.
In all four cases, the reconstructed foreground object is correctly placed in range.
Although our model does not allow us to reconstruct the varying height profile of the stairs, we correctly reconstruct it to be wider and more to the right than the other hidden objects.
In Appendix~\ref{app:results}, we demonstrate that our algorithm works under a wide range of conditions, including different hidden object locations, frame lengths, and lighting conditions.

In this work, we present an active NLOS method to accurately reconstruct both objects in motion and
a map of stationary hidden scenery behind them. This innovation is made possible through careful modeling of occlusion due to the vertical edge and within the hidden scene itself.
The algorithm presented in \cite{Gariepy2016} attempts only to identify a single occupied point in the hidden scene, making detailed modeling of the scene response unnecessary.
In this work, we also make no assumptions about light returning from the visible scene, allowing arbitrary visible scenery to be placed at the same ranges as the hidden objects of interest. This is true in \cite{Rapp2020} as well, however in their setup, with the single-element SPAD fixed in position and a very small laser scan radius, the contribution to the measurement from the visible side may be assumed constant across all measurements. In our configuration, the SPAD array has a non-negligible spatial extent resulting in a visible-side contribution that varies across the measurements. Our use of a stationary scene measurement allows us to effectively remove the contribution due to unknown visible-side scenery; our modeling of occlusion within the hidden scene itself allows us to perform this `background subtraction' without losing all information about the stationary hidden scenery.

Although we have successfully demonstrated our acquisition method, various aspects of our system and algorithm could be improved upon.
Our current algorithm processes each frame independently, using only broad constraints on the unknown parameters.
An improved system could jointly process frames and benefit from inter-frame priors.
Such priors could incorporate continuity of motion, the fact that object height, width, and albedo are unlikely to change between frames, and the fact that that walls in the hidden scene are typically smooth and continuous. 
In our demonstration, we use a thin occluding wall and do not model wall thickness.
The thin-wall assumption is illustrated in \Cref{fig:forward_model}, where the angle $\alpha$ is measured around the same point regardless of the location of $\pvec$. When the the wall has appreciable thickness, cases $\alpha\in [0,\,\pi/2)$ and $\alpha\in[\pi/2,\,\pi]$ require different modeling. 
One could incorporate wall thickness into the model or estimate wall thickness as an additional unknown parameter.
A method might also be designed to produce higher resolution reconstructions of each moving target.
Each target could be divided horizontally into several vertical segments, each with an unknown albedo and height to be estimated.
This type of algorithm might better resolve the staircase object in \Cref{fig:robustness_demo}.
Through further analysis, it might also be possible to optimize certain parameters in our setup. For example, certain FOV sizes and positions or laser locations might produce a better balance between the different sources of information in the data.

The demonstrations in this work employed a sensor with 32\,$\times$\,32 SPAD pixels, 390 ps timing resolution, 3.14\% fill factor, and $\sim$17\,kHz frame rate, limited by the USB 2.0 link~\cite{villa_2014_32x32SPAD}.
A frame length of 10\,$\mu$s and a gate-on period of 800\,ns yielded a duty cycle of 8\%. 
Particularly, the spatial and temporal resolution limit the precision of the estimated facet parameters, whereas the fill factor and frame rate limit the signal-to-noise ratio for a given acquisition time and, thus, the ability to track faster or farther objects.
We expect the results reported in this manuscript will improve by orders of magnitude with new SPAD technology,
as reviewed in~\cite{Bruschini_2019_review,villa_zappa_2021_review},
where some works have demonstrated up to 1 megapixel SPAD arrays~\cite{Morimoto_2020_MegapixelSPAD},
greater than 100 kHz frame rates~\cite{tosi_2021},
fill factors greater than 50\%~\cite{tosi_2021,Hutchings_2019},
and time resolution finer than 100\,ps~\cite{tosi_2021,velten_tosi_2020}.

\section{Experimental Details}\label{sec11}

\subsection{Setup} 
Illumination is provided using a 120\,mW master oscillator fiber amplifier picosecond laser (PicoQuant VisUV-532) at 532\,nm operating wavelength.
The laser has an $\sim$80\,ps FWHM pulse width and is triggered by the SPAD with a repetition frequency of 50 MHz.
The SPAD array consists of 32\,$\times$\,32 pixels with a fill factor of 3.14\%, with fully independent electronic circuitry, including a time-to-digital converter per pixel~\cite{villa_2014_32x32SPAD}.
At the 532 nm laser wavelength and room temperature, the average photon detection probability is $\sim$30\% and the average dark count rate is 100 Hz.
The array has a 390\,ps time resolution set by its internal clock rate of 160.3\,MHz.
Attached to the SPAD is a lens with focal length of 50\,mm, which yields a 25\,$\times$\,25\,cm field of view when placed at around 1.20\,m above the floor.
We set each acquisition frame length to 10\,$\mu$s, with a gate-on time of 800\,ns, thus yielding an 8\% duty cycle.
During the 800\,ns gate-on time of each frame, 40 pulses (800\,ns * 50\,MHz) illuminate the scene.
The SPAD array has a theoretical frame rate of 100\,kHz, set by the 10\,$\mu$s readout per frame, but experimentally we observed just $\sim$17\,kHz, which was mainly limited by the USB 2.0 connection to the computer.

\subsection{Data acquisition} 
For our demonstrations, we set up a hidden room 2.2\,m wide, 2.2\,m deep and 3\,m high, as shown in \Cref{fig:reconstruction_results}A\@.
Assuming the coordinate system origin is at the bottom of the occluding edge,
the left wall is at $x = -1.20$\,m,
the right wall is at $x=1$\,m,
the back wall is at $y=2.2$\,m,
and the ceiling is at $z=3$\,m.
The walls are made of white foam board and the ceiling is black cloth.
The SPAD array is positioned on the side of the wall, looking down at the occluding edge origin, allowing half of the array to be occluded.
The laser is positioned so that it shines close to the origin. 
To reject the strong ballistic contribution (first bounce) of light reflected from the origin, we punched a hole in the occluding wall and shined the laser through the hole.
The true location of the laser spot on the floor is slightly off the origin, by 6\,cm to the right side.
The latter was found by cross-checking and minimizing the number of ballistic photons measured by the SPAD array.
More recent SPAD arrays incorporate a fast hard gate
to rapidly enable and disable the detector with few hundreds picoseconds width,
which can be tuned to censor the ballistic photons~\cite{velten_tosi_2020, tosi_2021}.

Two test scenarios were analyzed.
For the first, we used two rectangular white foam board facets of size 20\,$\times$\,110\,cm as our moving objects.
For the second, we used four different targets:
a white foam board facet (of size 20\,$\times$\,110\,cm),
a gray foam board facet (white foam board painted with a gray diffuse spray paint),
a fabric mannequin of size 30\,$\times$\,80\,cm,
and a stair-like facet of size 75\,$\times$\,75\,cm.
These objects were used to test our method on targets of different shape, height and albedo.
All tests were conducted with the objects facing the occluding edge.
Before moving objects enter the hidden room, a 30\,s acquisition was collected to form an estimate of $\bvec$, the response of the stationary scene.
Then, new measurement frames were collected with moving objects fixed at discrete points along their trajectories during 0.4\,s.
In Appendix~\ref{app:results}, we demonstrate that these measurements can be acquired over a much shorter period of time with little effect on the reconstruction quality.

\subsection*{Acknowledgments}
This work was supported in part by the US National Science Foundation under grant number 1955219
and in part by the Draper Scholars Program.

\newpage
\appendices
\section{Fast Forward Model Approximation}
\label{app:fast_approx}

\subsection{Transient light transport modeling}
\label{appforward_model_computation}
In our system, a pulsed laser and SPAD array are both pointed at the floor adjacent to an occluding edge.
Take $\lvec$ to be the position of the laser spot, $\cvec$ to be a point in the area of the $n$th camera pixel $\camerapix_n$, and $\pvec$ to be a point on the hidden surface $\hiddensurf$.
The camera measurement rate
integrated over the $k$th time bin
at the $n$th spatial pixel
is
\begin{equation}
    \svec^{n,k}
    =
    \int_{(k-1)\Delta_t}^{k\Delta_t}
    \int_{\camerapix_n}
    \int_{\hiddensurf}
    v(\pvec,\cvec)\albedo(\pvec)
    \frac{G(\pvec,\,\lvec,\,\cvec)}{\norm{\lvec-\pvec}^2\norm{\cvec-\pvec}^2}\,
    w\!\left(t-t_0 - \frac{\norm{\lvec-\pvec}+\norm{\cvec-\pvec}}{c}\right)
    d\pvec\,
    d\cvec\,
    dt
    ,
    \label{eq:exactFM2}
\end{equation}
where $v(\pvec,\cvec)$ is the visibility function defined in the main document,
$\albedo(\pvec)$ is the surface albedo at point $\pvec$,
$w(\cdot)$ is the pulsed waveform,
$\Delta_t$ is the duration of a time bin,
$t_0$ is the time the pulse hits the laser spot,
and $c$ is the speed of light.
The Lambertian bidirectional reflectance distribution function (BRDF) factor $G(\cdot,\cdot,\cdot)$ is given by
\begin{equation}
    G(\pvec,\,\lvec,\,\cvec) = \cos{ (\measuredangle(\pvec-\lvec,\lnorm))} \cos{ (\measuredangle(\lvec-\pvec,\pnorm))} \cos{ (\measuredangle(\cvec-\pvec,\pnorm))} \cos{ (\measuredangle(\pvec-\cvec,\cnorm))}, 
\end{equation}
where $\lnorm$, $\pnorm$, and $\cnorm$ are the surface normal vectors at points $\lvec$, $\pvec$, and $\cvec$, respectively.
When the camera pixel size is sufficiently small,
\eqref{eq:exactFM2} may be approximated by
\begin{equation}
    \svec^{n,k}
    \approx
    \Delta_{\camerapix}
    \int_{(k-1)\Delta_t}^{k\Delta_t}
    \int_{\hiddensurf}
    v(\pvec,\cvecmidi)
    \albedo(\pvec)
    \frac{G(\pvec,\,\lvec,\,\cvecmidi)}{\norm{\lvec-\pvec}^2\norm{\cvecmidi-\pvec}^2}\,
    w\!\left(t-t_0 - \frac{\norm{\lvec-\pvec}+\norm{\cvecmidi-\pvec}}{c}\right)
    d\pvec\,
    dt
    ,
    \label{eq:approxFM-inserted}
\end{equation}
where $\Delta_{\camerapix}$ is the area of the camera pixel and
$\cvecmidi$ is the center of pixel $n$.
When the pulse duration is short relative to the time bin length,
we may replace $w(\cdot)$ by a Dirac impulse function $\delta(\cdot)$ scaled by a pulse intensity $I$ to write
\begin{equation}
    \svec^{n,k}
    \approx
    \Delta_{\camerapix} I \int_{(k-1)\Delta_t}^{k\Delta_t}
    \int_{\hiddensurf}
    v(\pvec,\cvecmidi)
    \albedo(\pvec)
    \frac{G(\pvec,\,\lvec,\,\cvecmidi)}{\norm{\lvec-\pvec}^2\norm{\cvecmid-\pvec}^2}\,
    \delta\!\left(t-t_0 - \frac{\norm{\lvec-\pvec}+\norm{\cvecmidi-\pvec}}{c}\right)
    d\pvec\,
    dt.
    \label{eq:pixelApproxFM}
\end{equation}

\subsection{Fast computation of the response from a vertical planar facet}
\label{app:fast_comp_facet}

\begin{figure}
    \centering
    \begin{subfigure}{0.3\textwidth}
    \includegraphics[width=1\linewidth]{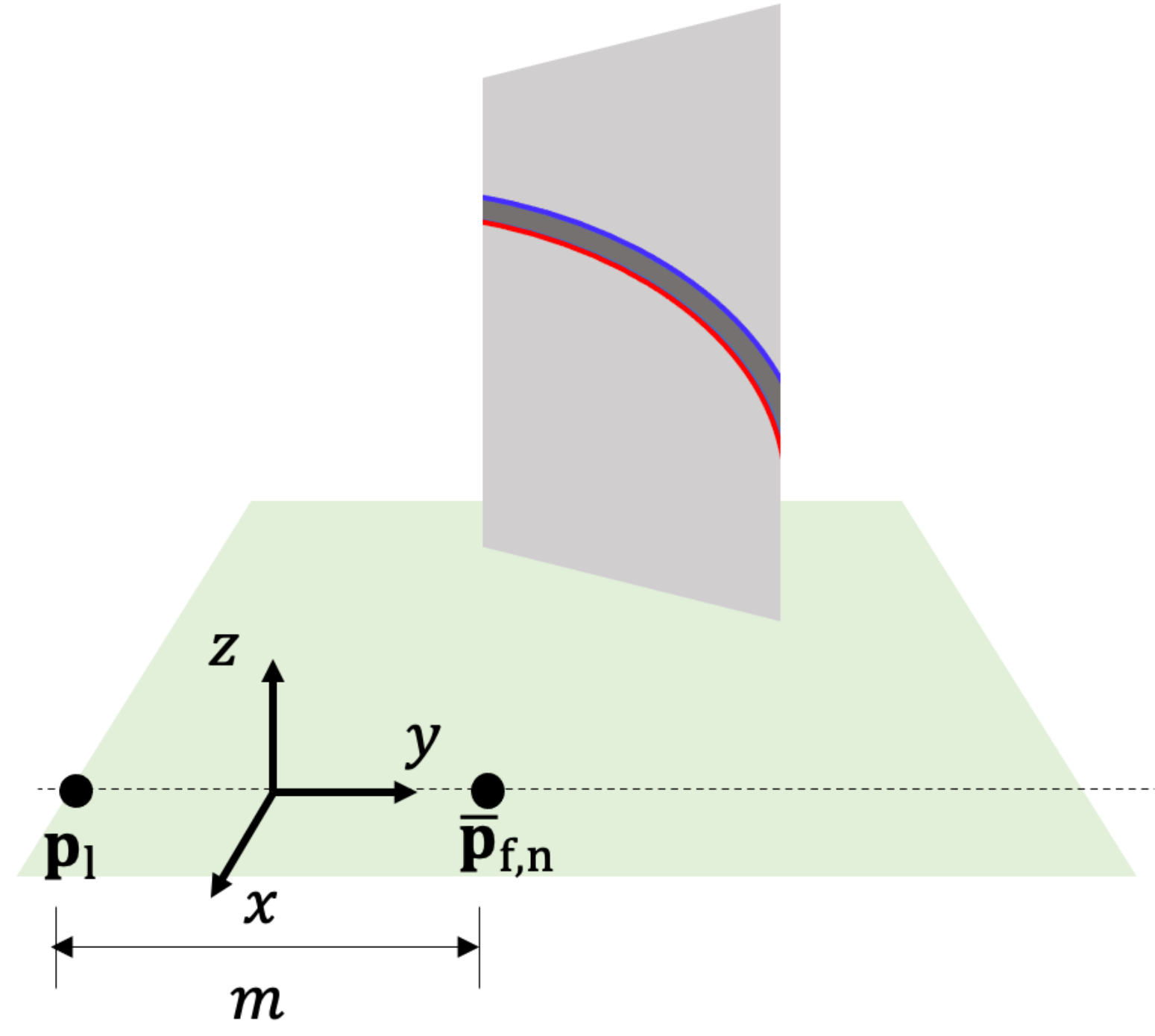}
    \caption{Laser spot $\lvec$ and pixel center $\cvecmidi$ form the focii of an ellipsoid that contains points with equal round-trip travel time. The intersection of an ellipsoid and a planar facet (light gray) is an ellipse segment (ex. red and blue lines). The occluding edge is not pictured.
    }
    \label{fig:coord_system}
    \end{subfigure}
    \hspace{2mm}
    \begin{subfigure}{0.33\textwidth}
    \includegraphics[width=1\linewidth]{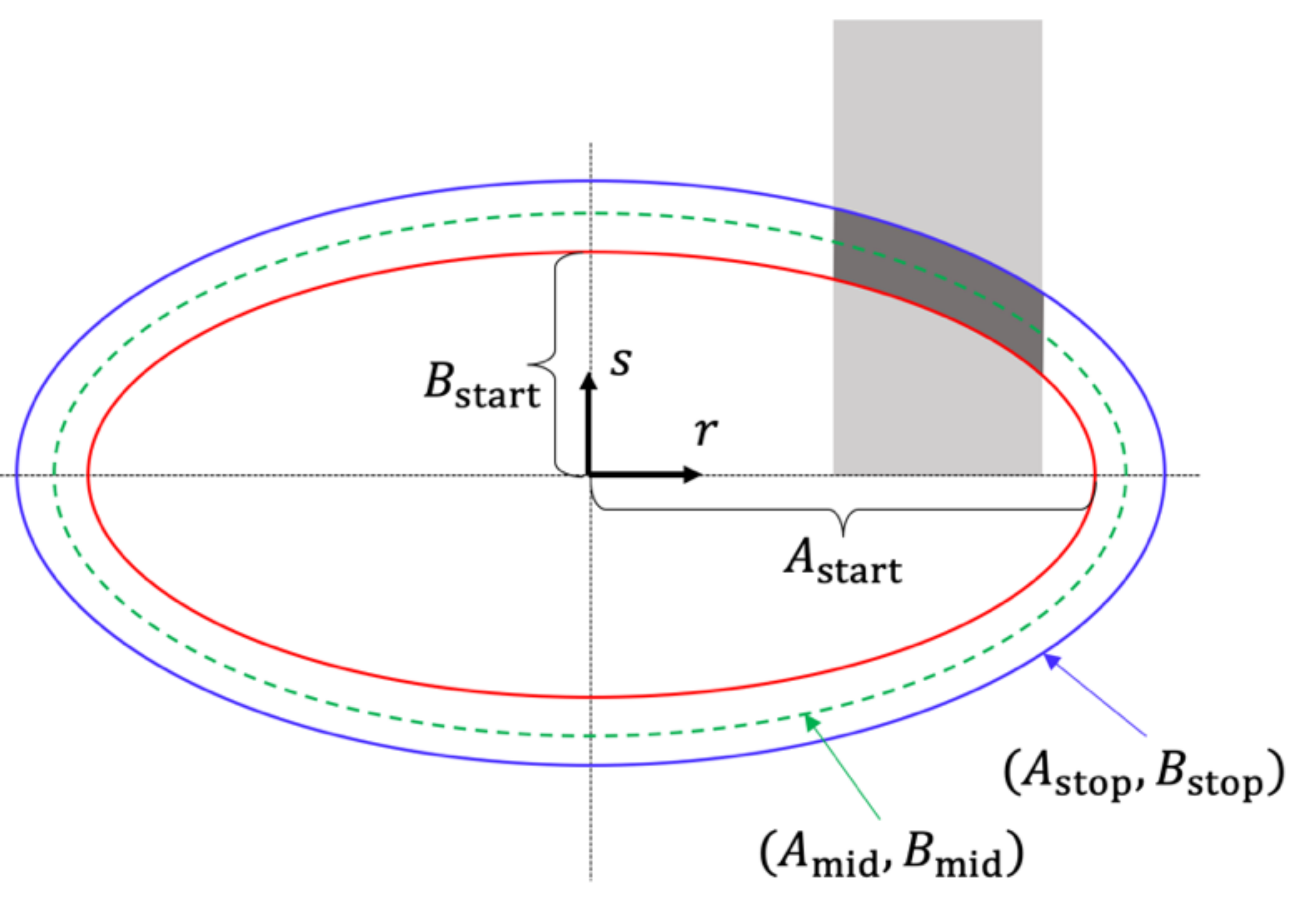}
    \caption{In the $rsu$ coordinate system, the ellipse intersection may be written in translational form with major and minor axis lengths $A$ and $B$. Axis lengths $(\Astart, \Bstart)$, $(\Amid, \Bmid)$, and $(\Astop, \Bstop)$ correspond to ellipses formed by the start, middle, and stop times of a time bin. The dark gray annulus is the region of the facet illuminated over the course of a time bin.}
    \label{fig:ellipse_intersection}
    \end{subfigure}
    \hspace{2mm}
    \begin{subfigure}{0.3\textwidth}
    \includegraphics[width=1\linewidth]{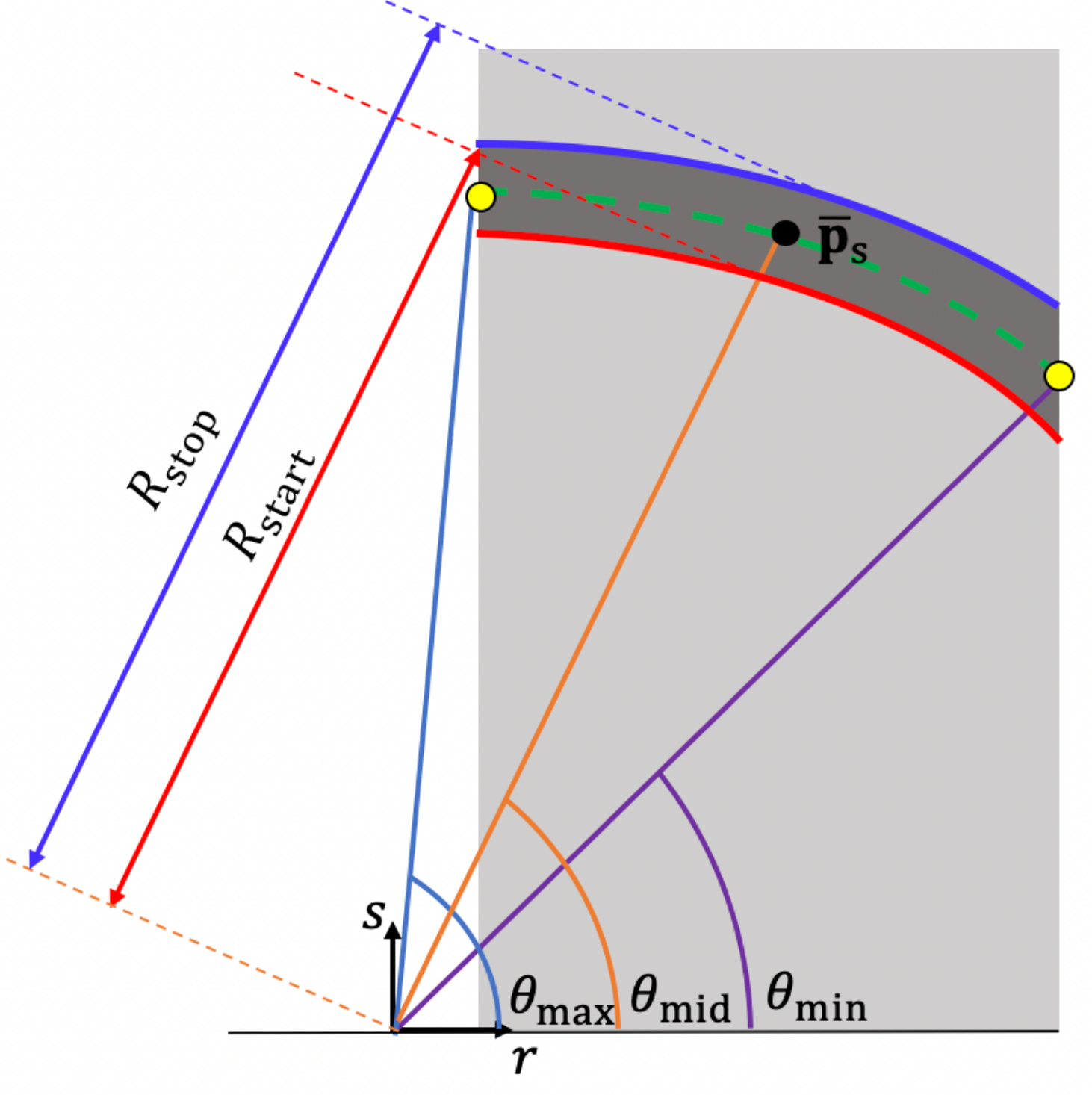}
    \caption{The facet response at a given time bin is the sum of the integrand over the dark gray annulus. The angle $\tmid$ bisects the minimum and maximum angles $\tmid$ and $\tmax$. Ranges $R_{\rm start}$ and $R_{\rm stop}$ are the ranges of ellipses $(\Astart, \Bstart)$ and $(\Astop, \Bstop)$ at angle $\tmid$.}
    \label{fig:ellipse_integration}
    \end{subfigure}
    \caption{The ellipsoid-plane intersection shown in (a) is shown as an ellipse written in translational form in (b). The response of a facet in a given time bin can be thought of as a polar integral over the dark grey region in (c).}
\end{figure}

Our inversion algorithm is based on a planar facet model of the hidden scene. Thus, we are interested in quickly computing the response of a vertical planar facet resting on the floor, shown in grey in \Cref{fig:coord_system}.
The method we develop here achieves a speed gain factor of about 150 over a direct numerical integration with similar accuracy.

We define a coordinate system (different than in the main document) so that the laser spot $\lvec$ and pixel center $\cvecmidi$
are placed a distance $m$ apart along the $y$-axis equidistant from the origin,
as marked in \Cref{fig:coord_system}.
For any $d > m$, points $\lvec$ and $\cvecmidi$ form the foci of an ellipsoid that contains all points $\pvec$ in the scene with round-trip travel distance
$d = \norm{\lvec-\pvec}+\norm{\cvecmidi -\pvec}$.
The intersection of that ellipsoid with the planar facet is an ellipse segment, as shown in red in \Cref{fig:coord_system}.
Now consider the integration in \eqref{eq:pixelApproxFM}.
As $t$ increases,
the $d$ value matching the shift of the Dirac grows and the corresponding ellipse intersection expands. Over the duration of the time bin, the integral represents the light returning from the dark grey annulus, between the red and blue segments, shown in \Cref{fig:coord_system}.
In \cite{Rapp2020},
$\lvec$ and $\cvec$ are treated as being at the origin,
so that the ellipsoid reduces to a sphere and the planar intersection reduces from an ellipse to a circle; the double integral in \eqref{eq:pixelApproxFM} is converted to polar coordinates, resulting in an approximate, closed-form solution to \eqref{eq:pixelApproxFM}.
In this work, since the spatial diversity of our sensor array
is essential, we cannot use this approximation.  Instead, we
develop a quick computation of the rates due to a vertical facet under the more general elliptical model. 

Consider the coordinate system in \Cref{fig:coord_system} with laser position $\lvec$ and pixel center $\cvecmidi$. The ellipsoid corresponding to round-trip travel distance $d$ may be written as
\begin{equation}
    1 = \frac{x^2}{\ael^2}+\frac{y^2}{\bel^2}+\frac{z^2}{\cel^2},
    \label{eq:ellipsoid}
\end{equation}
where $\ael = \cel = \sqrt{\left(\frac{d}{2}\right)^2-\left(\frac{m}{2}\right)^2}$ and $\bel=\frac{d}{2}$.
The planar facet is contained in a plane that may be described by normal vector $\pnorm$ and point $\qvec$.
The intersection of this plane and the ellipsoid in \eqref{eq:ellipsoid} is an ellipse that may be written in translational form~\cite{Klein2012} in a new $[r, \,s,\, u]^\T$ coordinate system:
\begin{equation}
    1 = \frac{r^2}{A^2} + \frac{s^2}{B^2}.
    \label{eq:ellipseTranslationalForm}
\end{equation}
If $\qvec$ is chosen to be interior to the ellipsoid, the procedure in \cite{Klein2012} may be used to find the new $[r,s,u]$ coordinate system and provides formulae for $A$ and $B$ given $\pnorm$, $\qvec$, and the ellipsoid parameters $(\ael, \bel, \cel)$.
\Cref{fig:ellipse_intersection} shows ellipses corresponding to the start (parameters: $\Astart$ and $\Bstart$), middle (parameters: $\Amid$ and $\Bmid$), and stop (parameters: $\Astop$ and $\Bstop$) times of a time bin in the $rs$-plane with the facet in grey.
The double integral in \eqref{eq:pixelApproxFM} sums the integrand over the dark grey annulus.
This double integral can be formulated as integration in polar coordinates over a region in the $rs$-plane as shown in \Cref{fig:ellipse_integration}.
For computational speed, we approximate this integral as
\begin{equation}
     \svec^{n,k} \approx \Delta_{\camerapix} I \deltaArc \albedo \frac{G(\pvecmid,\,\lvec,\,\cvecmidi)}{\norm{\lvec-\pvecmid}^2\norm{\cvecmidi-\pvecmid}^2},
    \label{eq:ellipseApproxFM}
\end{equation}
where $\deltaArc$ is the annulus area. The point $\pvecmid$ is on the ellipse corresponding to the middle of the time bin (i.e., with parameters $\Amid$ and $\Bmid$) at polar angle $\tmid=(\tmin+\tmax)/2$.
Note that we have replaced $\albedo(\pvec)$ with $\albedo$ under the assumption that the facet has uniform albedo.
The area $\deltaArc$ is approximated as a fraction of a circular annulus:
\begin{align}
    \deltaArc
    &\approx \frac{\tmax-\tmin}{2\pi}(\pi R_{\rm stop}^2 -\pi R_{\rm start}^2)
    = \frac{1}{2}(\tmax-\tmin)(R_{\rm stop}^2 - R_{\rm start}^2).
    \label{eq:circleApprox}
\end{align}
Ellipse radii at polar angle $\theta = \tmid$: $R_{\rm start}$ and $R_{\rm stop}$, marked in \Cref{fig:ellipse_integration}, may be computed using
\begin{equation}
    R(\theta,A,B) = \frac{AB}{\sqrt{B^2\cos^2\theta + A^2\sin^2\theta}}.
    \label{eq:ellipseRadius}
\end{equation}
In practice, when a facet is wide, the extent of the annulus may not be well described by a single central point $\pvecmid$.
When the distance $d_{\rm annulus}$ between the edges of the annulus, marked with yellow dots in \Cref{fig:ellipse_integration},
is greater than parameter $d_{\rm max}$, the annulus is divided into smaller annulus segments that each satisfy $d_{\rm annulus}<d_{\rm max}$.
The full procedure for computing the facet response at each of the $N$ camera pixels and $K$ time bins is outlined in \Cref{alg:forwardModel_computation}.

\begin{figure}
\centering
    \begin{subfigure}{0.24\textwidth}
    \includegraphics[width=1\linewidth]{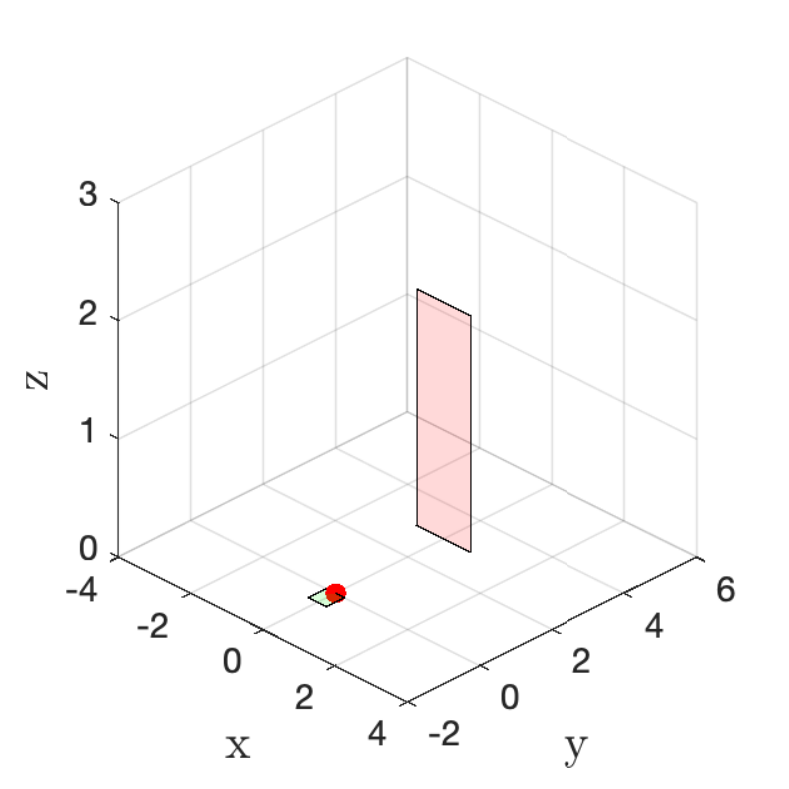}
    \label{fig:person1}
    \end{subfigure}
        \begin{subfigure}{0.24\textwidth}
    \includegraphics[width=1\linewidth]{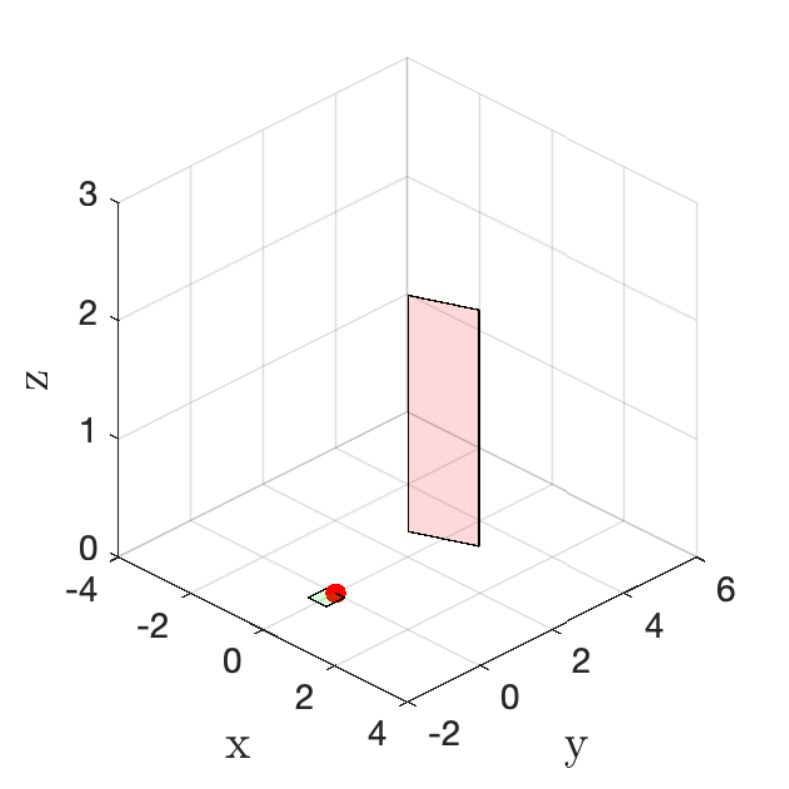}
    \label{fig:person2}
    \end{subfigure}
        \begin{subfigure}{0.24\textwidth}
    \includegraphics[width=1\linewidth]{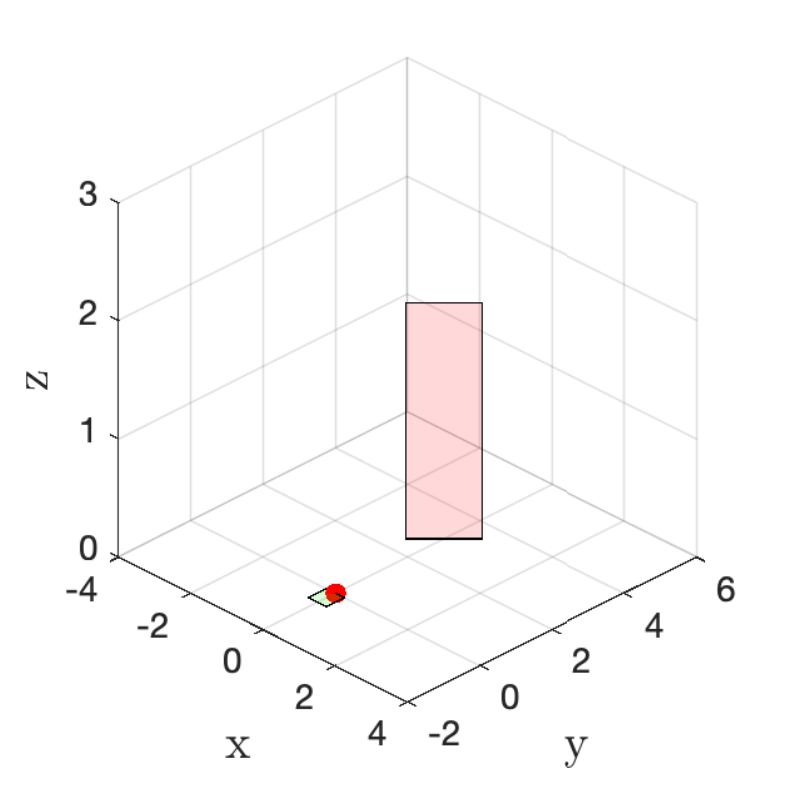}
    \label{fig:person3}
    \end{subfigure}
        \begin{subfigure}{0.24\textwidth}
    \includegraphics[width=1\linewidth]{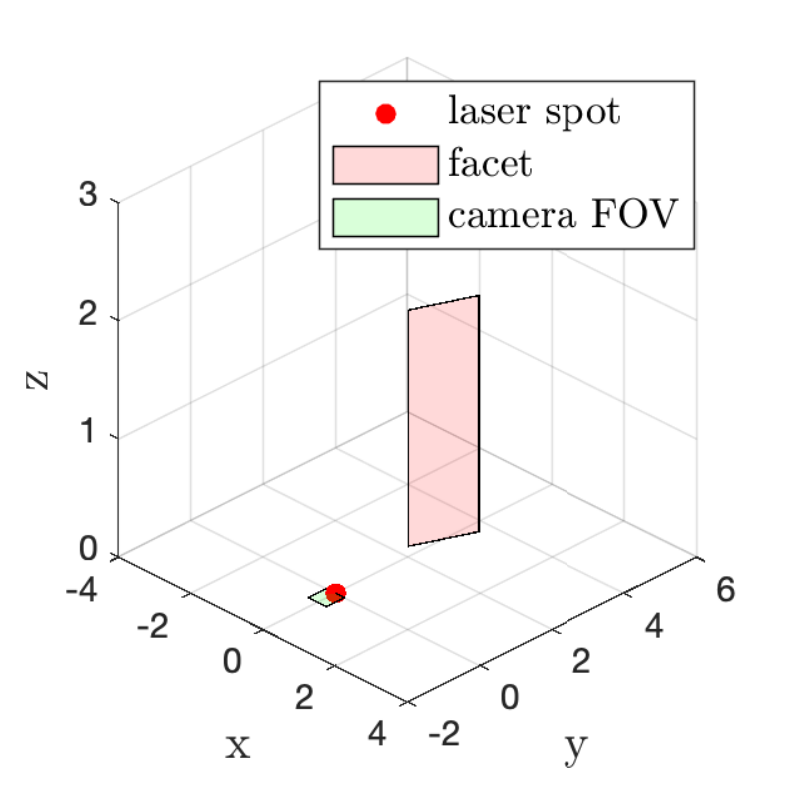}
    \label{fig:person4}
    \end{subfigure}\\
        \begin{subfigure}{0.24\textwidth}
    \includegraphics[width=1\linewidth]{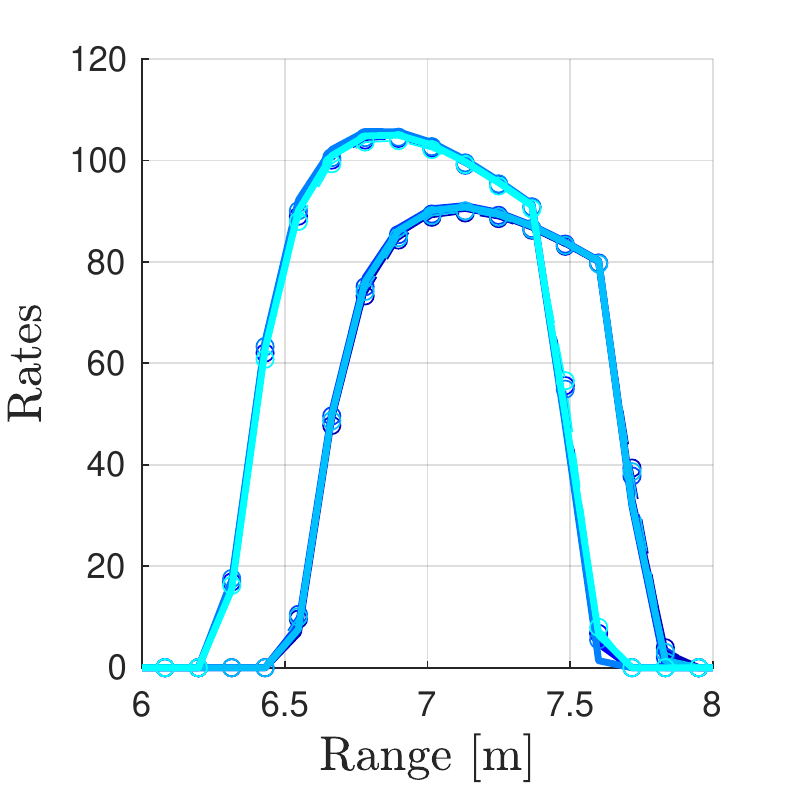}
    \caption{rotation $=0$,\\ $\frac{\norm{\fmexact-\fmapprox}_1}{\norm{\fmexact}_1} = 0.0248$ }
    \label{fig:person1_sim}
    \end{subfigure}
        \begin{subfigure}{0.24\textwidth}
    \includegraphics[width=1\linewidth]{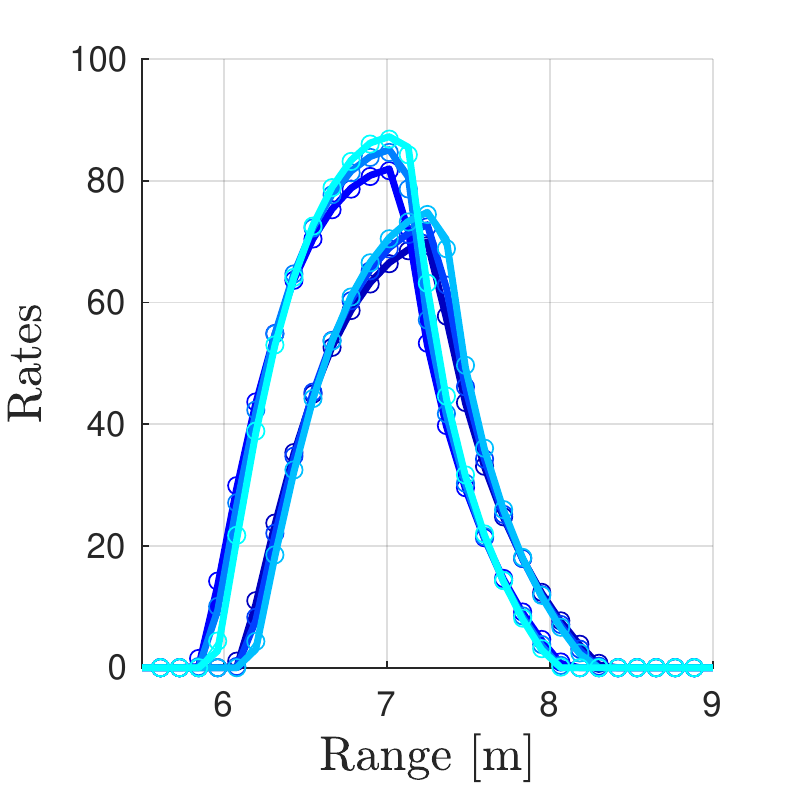}
    \caption{rotation $=\pi/8$,\\ $\frac{\norm{\fmexact-\fmapprox}_1}{\norm{\fmexact}_1} = 0.0074$ }
    \label{fig:person2_sim}
    \end{subfigure}
        \begin{subfigure}{0.24\textwidth}
    \includegraphics[width=1\linewidth]{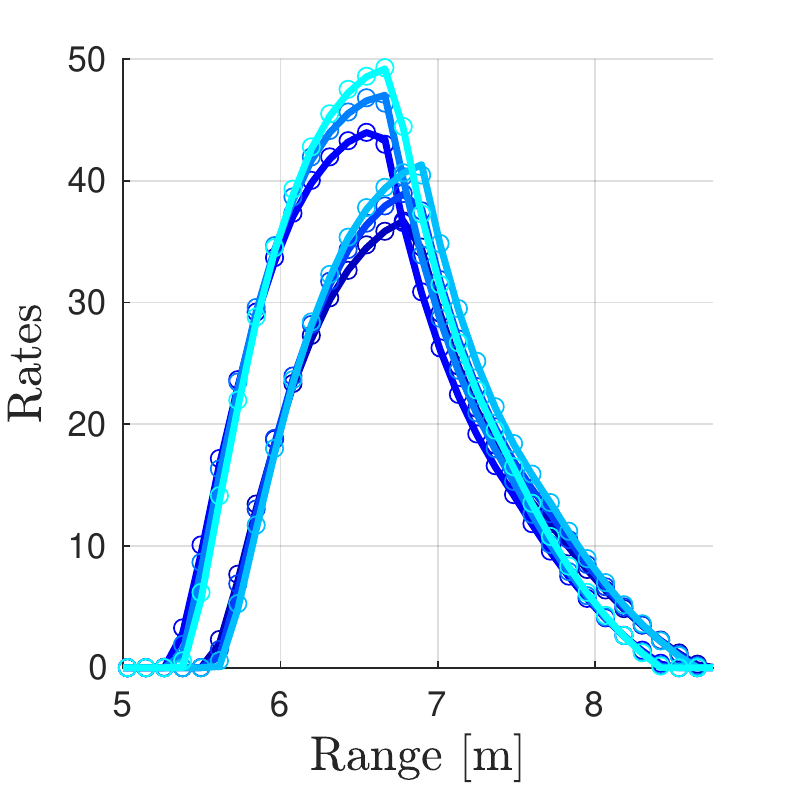}
    \caption{rotation $=2\pi/8$,\\ $\frac{\norm{\fmexact-\fmapprox}_1}{\norm{\fmexact}_1} = 0.0092 $ }
    \label{fig:person3_sim}
    \end{subfigure}
        \begin{subfigure}{0.24\textwidth}
    \includegraphics[width=1\linewidth]{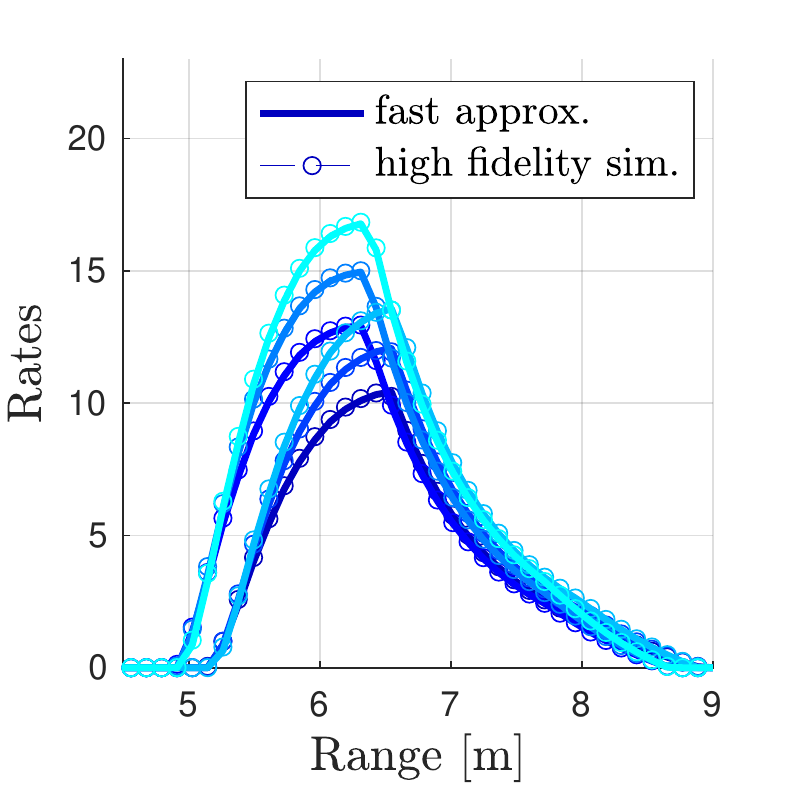}
    \caption{rotation $=3\pi/8$,\\ $\frac{\norm{\fmexact-\fmapprox}_1}{\norm{\fmexact}_1} = 0.0118 $ }
    \label{fig:person4_sim}
    \end{subfigure}
    \caption{\label{fig:rotate_person} A comparison of \Cref{alg:forwardModel_computation} to conventional numerical integration for a person-sized facet at different tilt angles with a width of 0.75\,m and height of 2\,m. The first row of plots shows the position of the SPAD FOV, laser spot, and the facet. The bottom row shows computed rates at select pixels for each method. The SPAD FOV is 0.5 $\times$ 0.5\,m. }
\end{figure}

\begin{figure}
\centering
    \begin{subfigure}{0.24\textwidth}
    \includegraphics[width=1\linewidth]{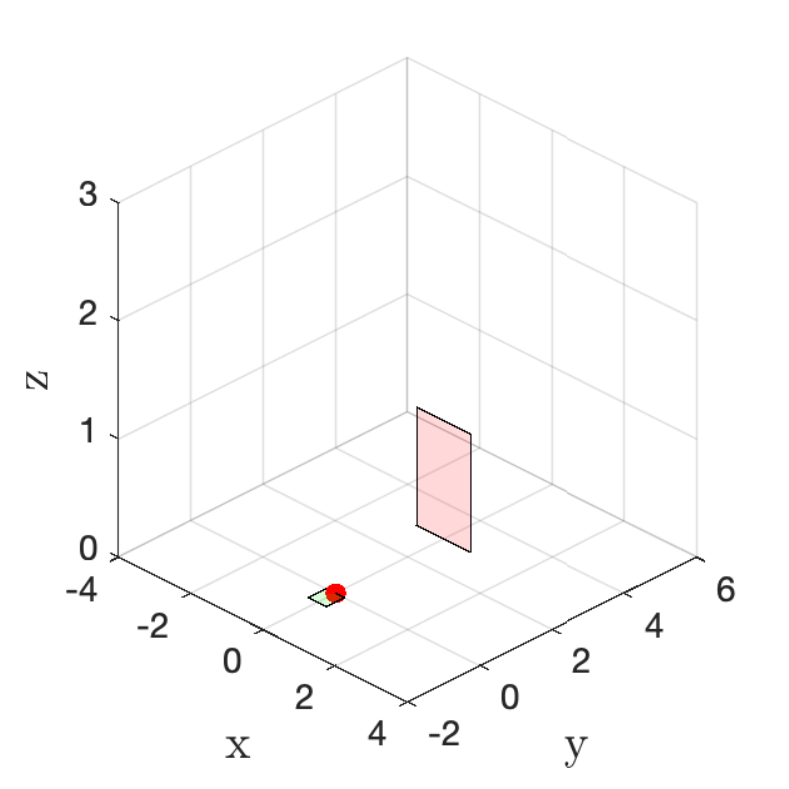}
    \label{fig:shortFacet1}
    \end{subfigure}
        \begin{subfigure}{0.24\textwidth}
    \includegraphics[width=1\linewidth]{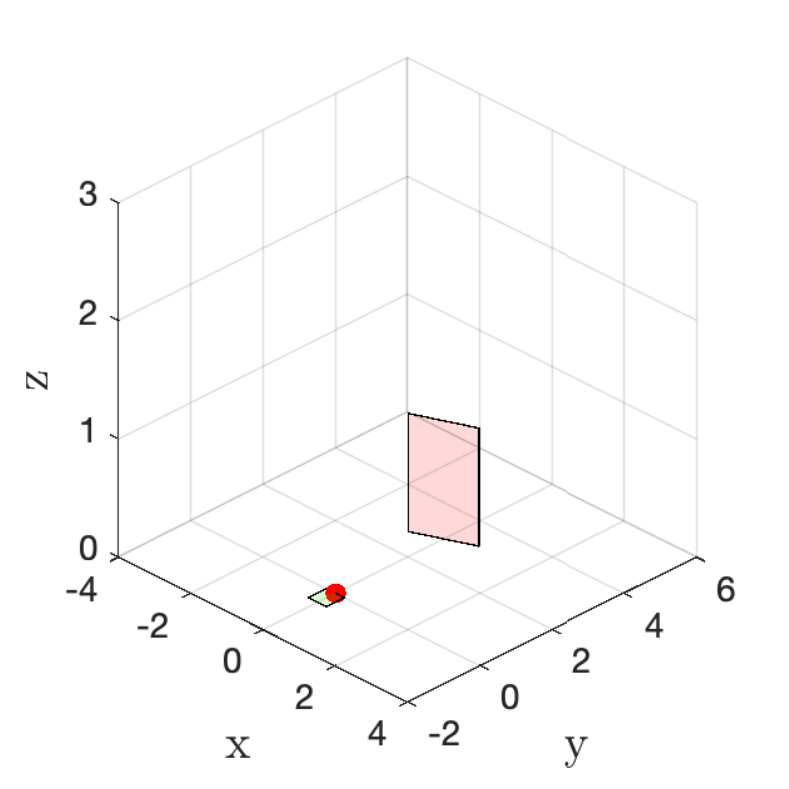}
    \label{fig:shortFacet2}
    \end{subfigure}
        \begin{subfigure}{0.24\textwidth}
    \includegraphics[width=1\linewidth]{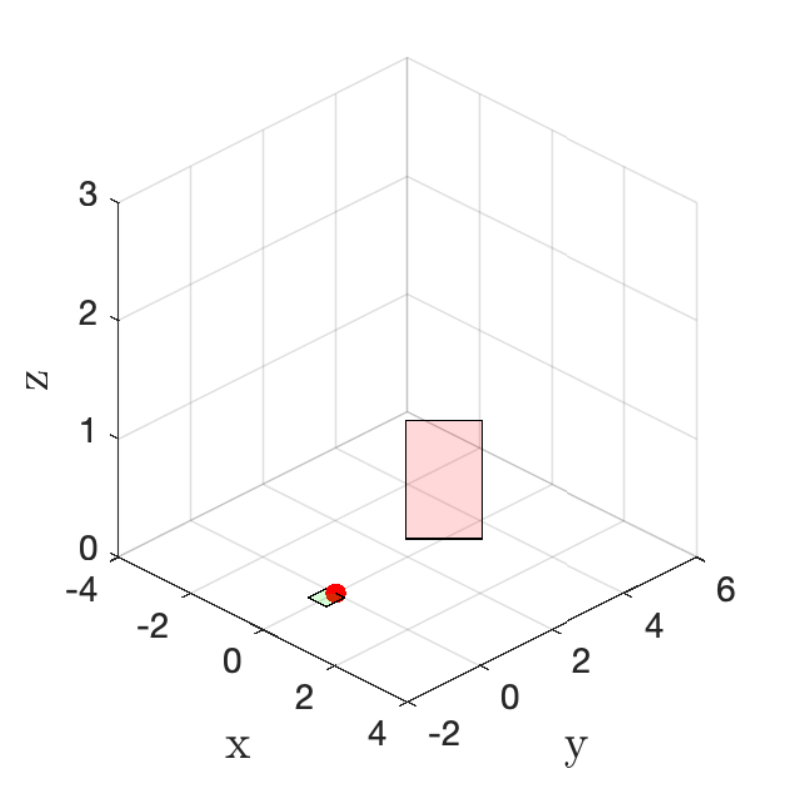}
    \label{fig:shortFacet3}
    \end{subfigure}
        \begin{subfigure}{0.24\textwidth}
    \includegraphics[width=1\linewidth]{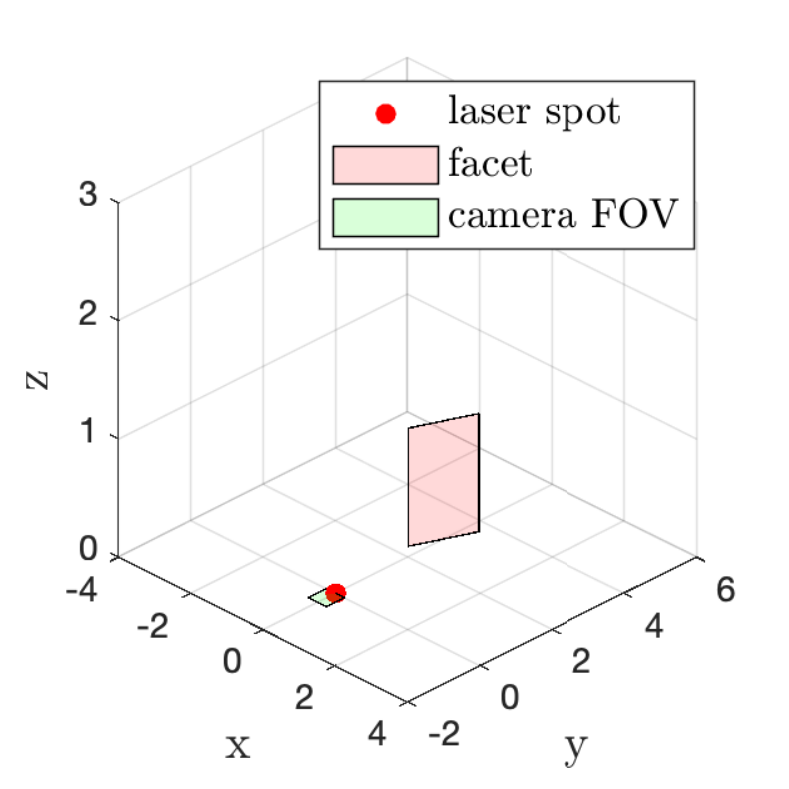}
    \label{fig:shortFacet4}
    \end{subfigure}\\
        \begin{subfigure}{0.24\textwidth}
    \includegraphics[width=1\linewidth]{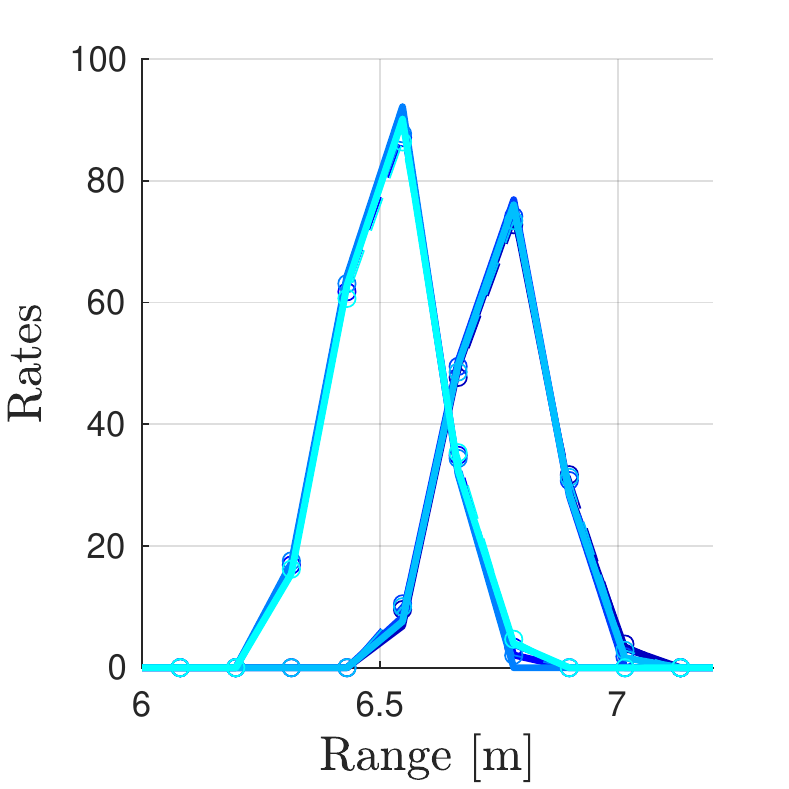}
    \caption{rotation $=0$,\\ $\frac{\norm{\fmexact-\fmapprox}_1}{\norm{\fmexact}_1} = 0.0775$ }
    \label{fig:shortFacet1_sim}
    \end{subfigure}
        \begin{subfigure}{0.24\textwidth}
    \includegraphics[width=1\linewidth]{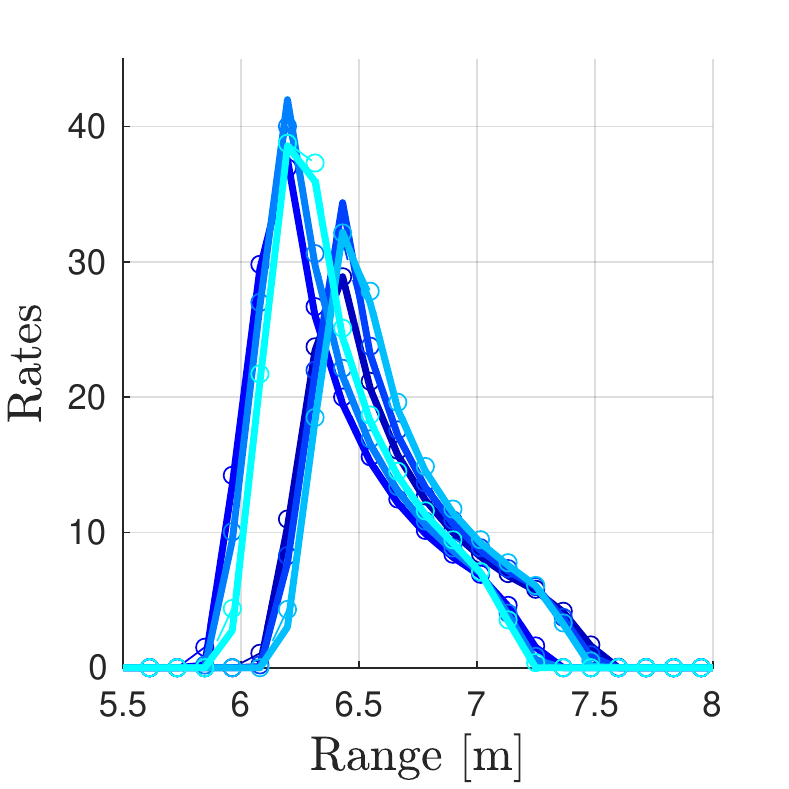}
    \caption{rotation $=\pi/8$,\\ $\frac{\norm{\fmexact-\fmapprox}_1}{\norm{\fmexact}_1} = 0.0300$ }
    \label{fig:shortFacet2_sim}
    \end{subfigure}
        \begin{subfigure}{0.24\textwidth}
    \includegraphics[width=1\linewidth]{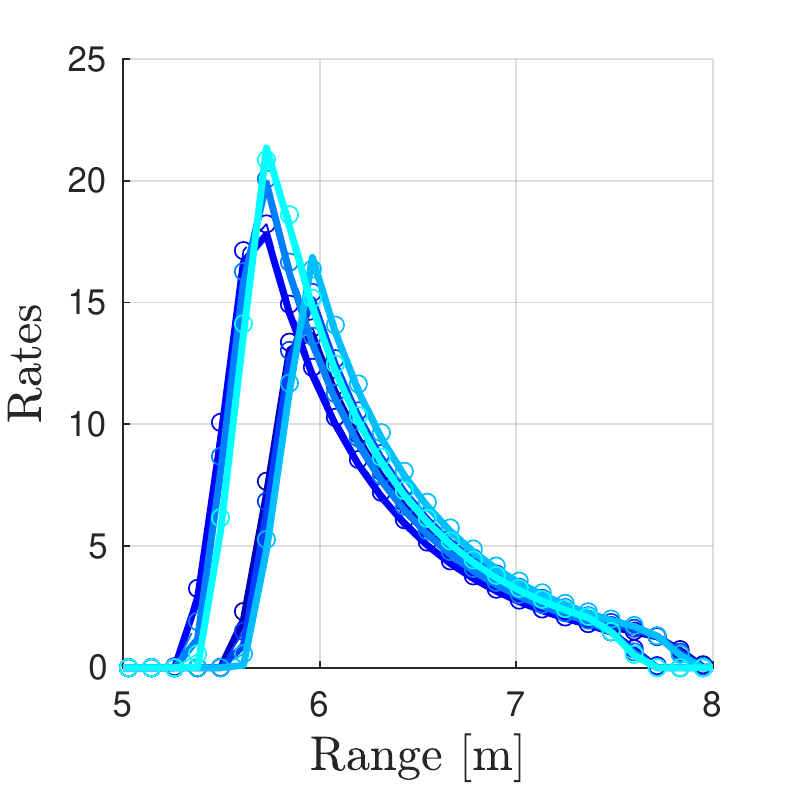}
    \caption{rotation $2\pi/8$,\\ $\frac{\norm{\fmexact-\fmapprox}_1}{\norm{\fmexact}_1} = 0.0316$ }
    \label{fig:shortFacet3_sim}
    \end{subfigure}
        \begin{subfigure}{0.24\textwidth}
    \includegraphics[width=1\linewidth]{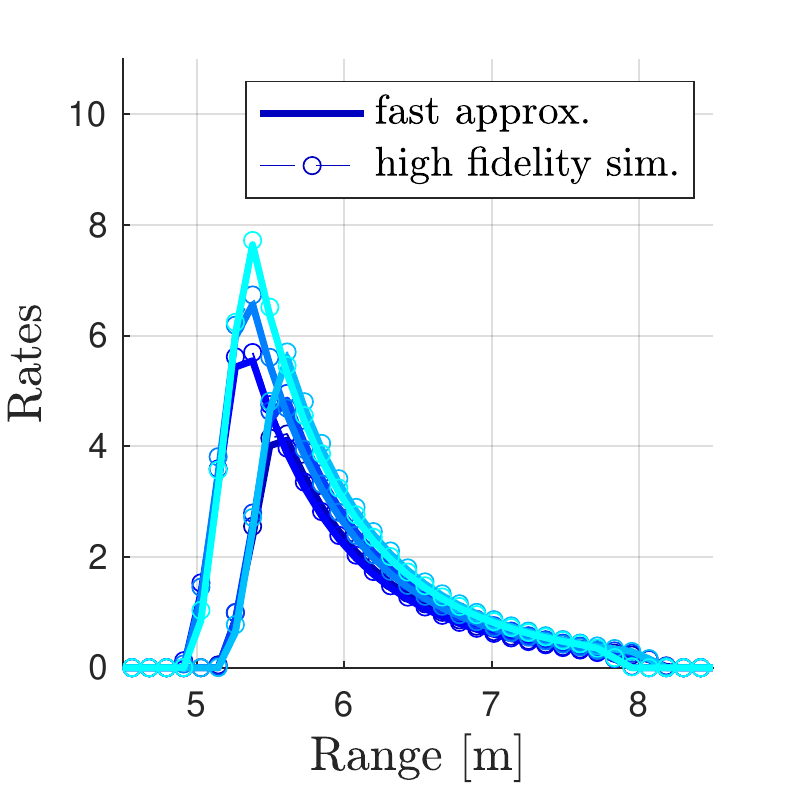}
    \caption{rotation $=3\pi/8$,\\ $\frac{\norm{\fmexact-\fmapprox}_1}{\norm{\fmexact}_1} = 0.0338$ }
    \label{fig:shortFacet4_sim}
    \end{subfigure}
    \caption{\label{fig:rotate_baby} A comparison of \Cref{alg:forwardModel_computation} to conventional numerical integration for a child-sized facet at different tilt angles with a width of 0.75\,m and a height of 1\,m. The first row of plots shows the position of the SPAD FOV, laser spot, and the facet. The bottom row shows computed rates at select pixels for each method. The SPAD FOV is 0.5 x 0.5\,m.}
\end{figure}

\begin{figure}
    \centering
    \begin{subfigure}{0.333\textwidth}
    \includegraphics[width=1\linewidth]{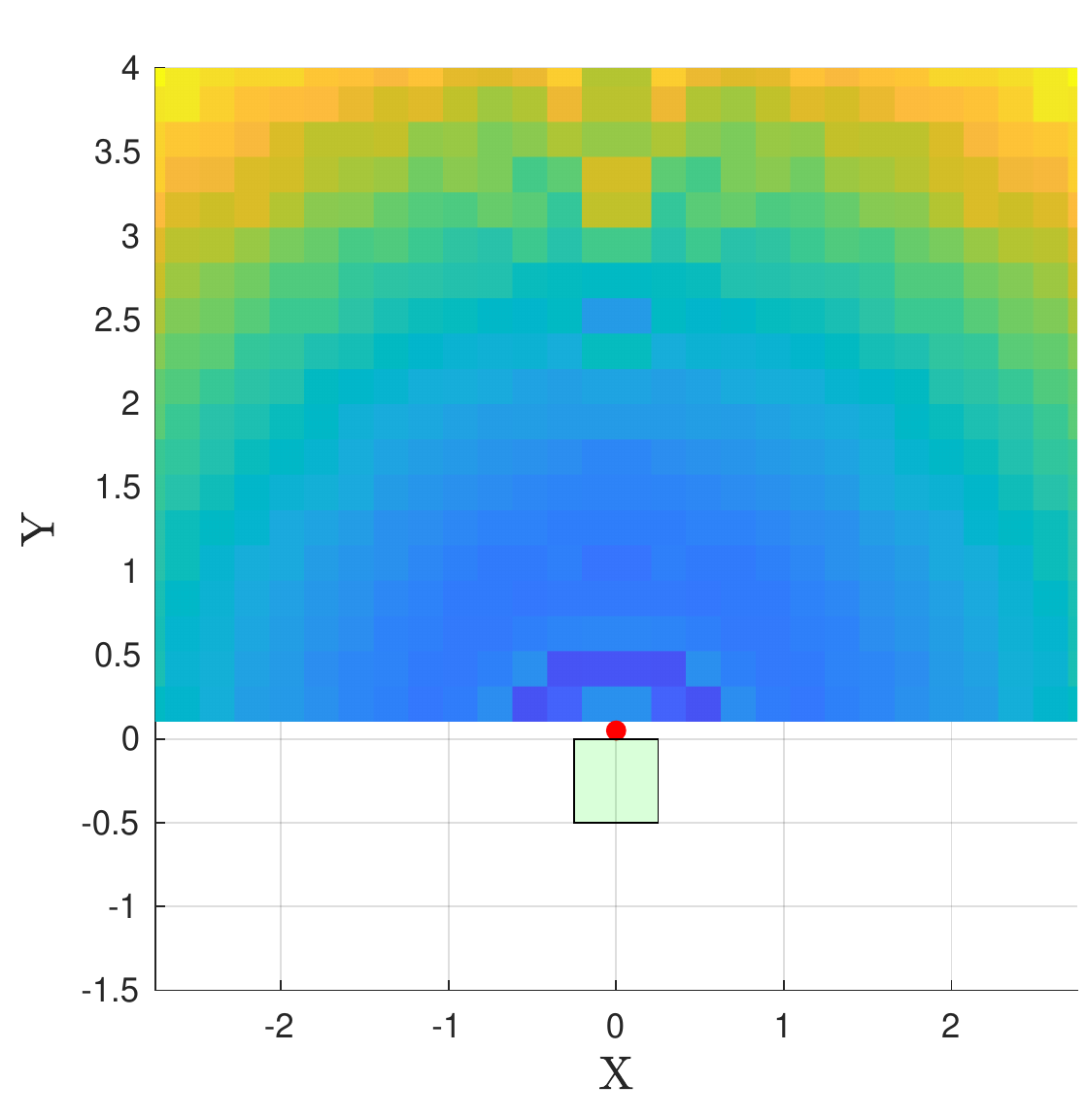}
    \caption{Facet is 1m tall and 1m wide }
    \label{fig:facetGrid_1m}
    \end{subfigure}
     \begin{subfigure}{0.4\textwidth}
    \includegraphics[width=1\linewidth]{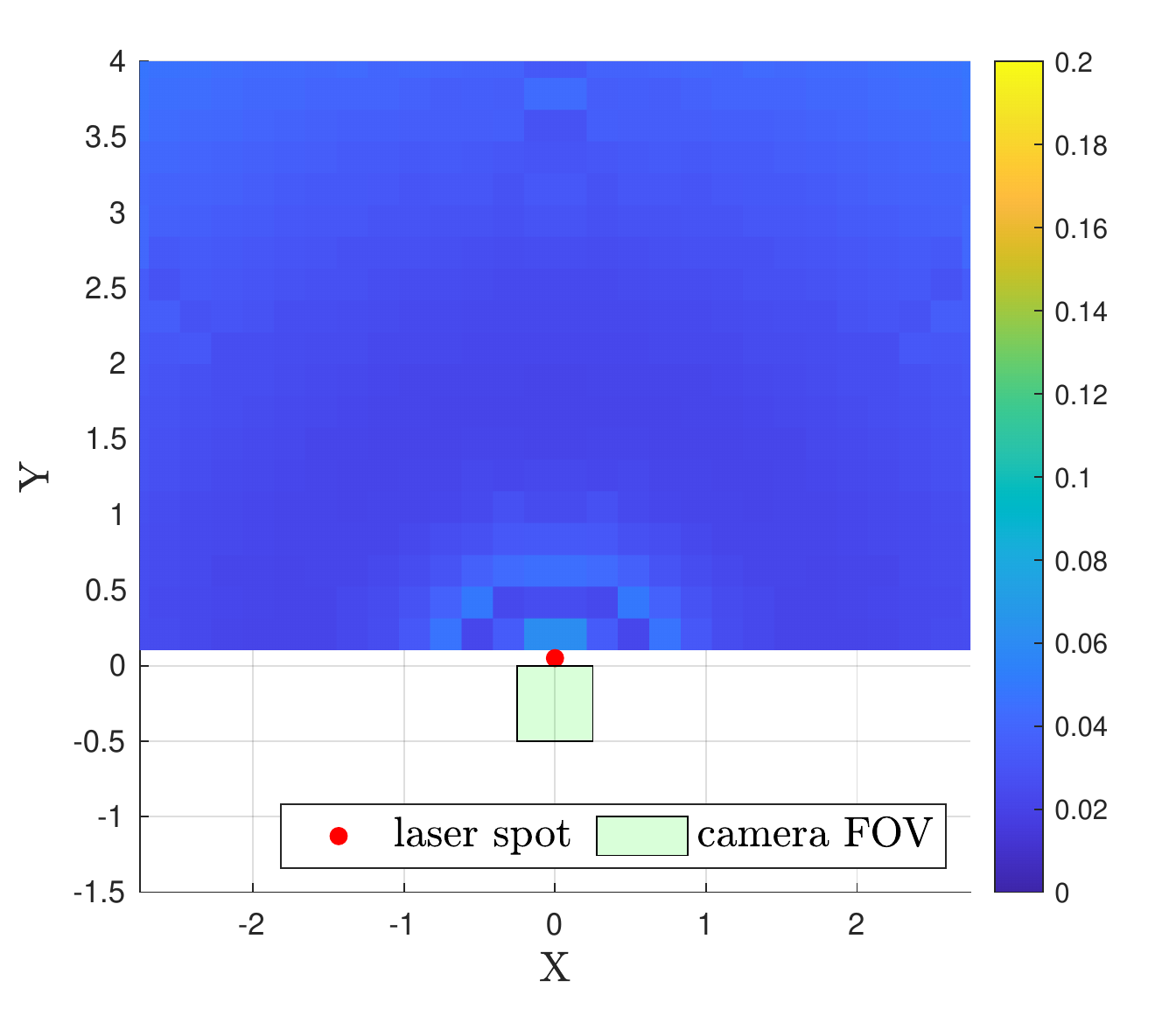}
    \caption{ Facet is 2m tall and 1m wide}
    \label{fig:facetGrid_2m}
    \end{subfigure}
    \caption{Relative error metric $\Frac{\norm{\fmexact-\fmapprox}_1}{\norm{\fmexact}_1}$ plotted for different facet positions in space for a 1\,m tall facet (a) and a 2\,m tall facet (b). All facets face the occluding edge. 
    }
    \label{fig:edgeFacingFacetTest}
\end{figure}

\Cref{fig:rotate_person} compares facet responses computed using \Cref{alg:forwardModel_computation} (solid line) with those computed using a slower
high-fidelity simulation tool based on numerical integration (dashed line with markers) for a person-sized facet at four different azimuthal rotation angles.
Different colored curves denote rates computed at different array pixels.
Note that in all four cases, our fast forward model computation closely matches the higher-fidelity simulator.
In \Cref{fig:rotate_baby}, we show results for a shorter facet, similar in height to the child-sized mannequin we use in our experiments.
The shorter facet also exhibits close match between \Cref{alg:forwardModel_computation} and the higher-fidelity simulation. 
\Cref{fig:edgeFacingFacetTest} shows $\norm{\fmexact-\fmapprox}_1/\norm{\fmexact}_1$ at different points in space for a vertical facet facing the origin.
Here, $\fmapprox$ is the vector of rates computed using \Cref{alg:forwardModel_computation} and $\fmexact$ is generated using the higher-fidelity simulation.
Although the shorter facet (a) has more error (i.e., larger $\norm{\fmexact-\fmapprox}_1/\norm{\fmexact}_1$) than the taller facet (b), both have relatively small error.
In our experimental demonstrations, we use a small scaled-down room with objects similar in size to the one tested in \Cref{fig:facetGrid_1m}.
Although our experimental results in \Cref{app:results} are achieved using a scaled down setup, \Cref{fig:facetGrid_2m} suggests that the forward model computation in \Cref{alg:forwardModel_computation} is even more accurate in a larger, more life-like setting.

\begin{algorithm}
\caption{Fast computation of rates at the SPAD array due to a vertical planar facet resting on the floor}
\label{alg:forwardModel_computation}
\begin{algorithmic}[1]
    \ForEach{$n = [1, 2, \ldots, N]$}    \Comment{loop through camera pixels}
    \State find indices of first and last affected time bins $k_{\rm min}$ and $k_{\rm max}$ according to \Cref{appfacet_time_bounds}
    \ForEach{$k = [k_{\rm min},k_{\rm min}+1, \dots, k_{\rm max}]$} \Comment{loop through affected time bins}
    \State find $rsu$ coordinate system and compute ellipse parameters $(\Amid, \Bmid)$ and $(\Astop, \Bstop)$ using \cite{Klein2012}
    \IfEach{$k == k_{\rm min}$}
    \State compute ellipse parameters $(\Astart, \Bstart)$
    \EndIfEach
    \State find $\tmin$ and $\tmax$ by finding intersections of ellipse $(\Amid, \Bmid)$ with facet edges
    \State compute $\tmid = \frac{1}{2}(\tmin+\tmax)$
    \State compute $R_{\rm start}$ and $R_{\rm stop}$ using \eqref{eq:ellipseRadius} with arguments $(\Astart,\Bstart)$, $(\Astop,\Bstop)$, and $\tmid$
    \State compute approximate facet response in $k$th time bin at $i$th pixel using \eqref{eq:ellipseApproxFM} and \eqref{eq:circleApprox}
    \State $(\Astart,\Bstart) \leftarrow (\Astop,\Bstop)$ \Comment{next time bin starts with end of this one}
    \EndForEach
    \EndForEach
\end{algorithmic}
\end{algorithm}

\subsection{Facet time bounds}
\label{appfacet_time_bounds}

\begin{figure}
\centering
\includegraphics[width=.3\linewidth]{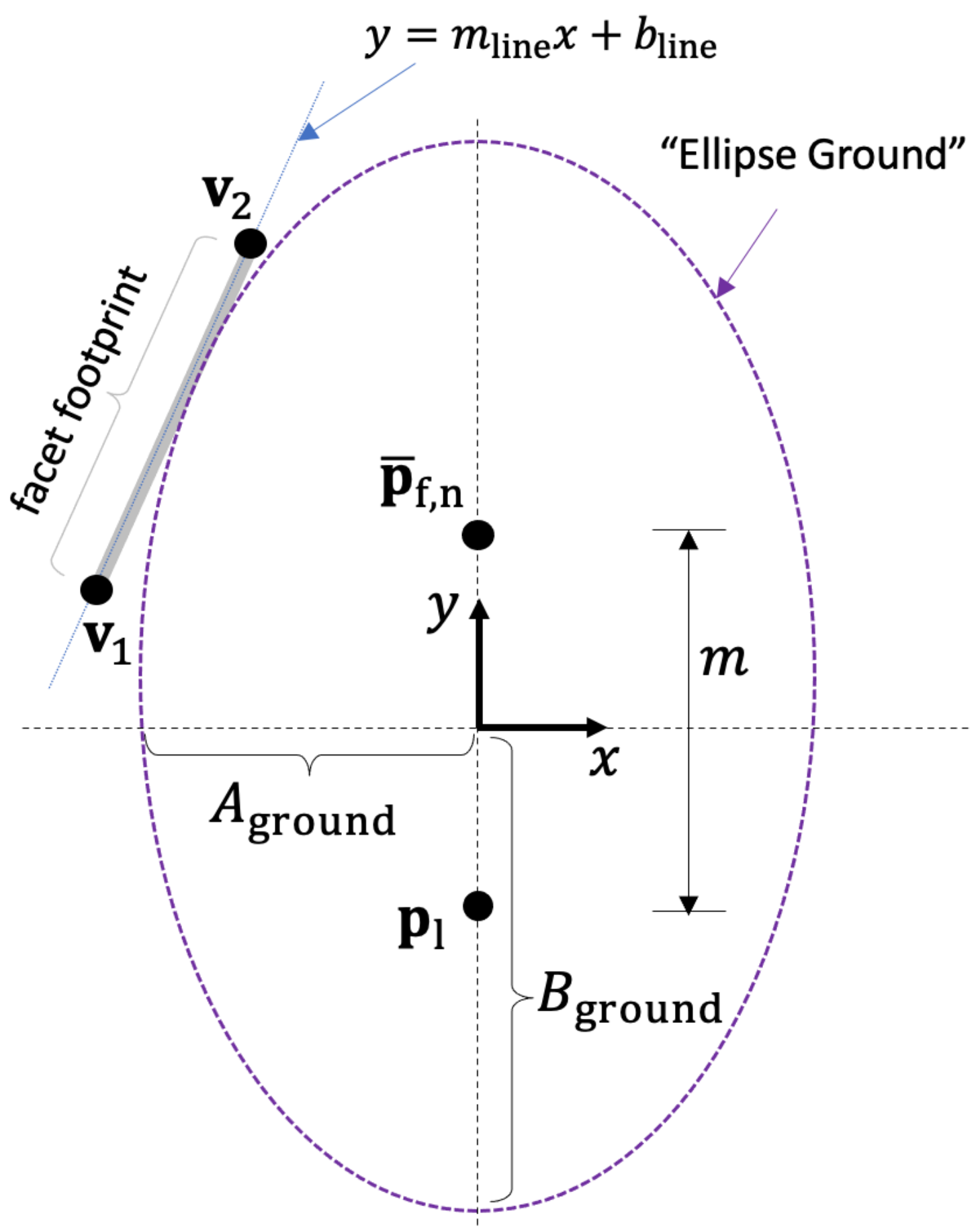}
\caption{With both $\lvec$ and $\cvecmidi$ in the ground plane, and the vertical hidden facet resting on the ground, the part of the facet with the shortest round-trip travel time is also on the ground.
Regions with equal round-trip travel time from $\lvec$, to the hidden scene, and back to $\cvecmidi$ form ellipsoids.
The intersection of this ellipsoid and the ground plane is an ellipse (purple).
The shortest round-trip travel time that causes this ellipse to encounter the bottom of the facet (i.e., the line segment between $\vvec_1$ and $\vvec_2$) is the earliest time of arrival due to the facet.
}
\label{fig:ellipse_ground}
\end{figure}

Take $t_{\rm min}$ and $t_{\rm max}$ to be the smallest and largest arrival times due to the facet, corresponding to round-trip travel distances $\dmin$ and $\dmax$.
We seek $\dmin$ and $\dmax$ in order to compute rates due to the facet at all affected time bins.
To simplify the geometry, we assume we are only interested in vertical rectangular facets resting on the floor.
Under this assumption, the part of the facet with the shortest round-trip travel distance is somewhere along its bottom edge.
The two bottom vertices of the planar facet $\vvec_1$ and $\vvec_2$ are contained in the $xy$-plane (i.e., the ground plane), as shown in \Cref{fig:ellipse_ground}.
When $\vvec_2(1)\neq\vvec_1(1)$, the line that runs along the bottom of the facet, connecting the two lowest vertices ($\vvec_1$ and $\vvec_2$), is given by
\begin{equation}
    y =\mline x + \cline,\label{eq:lineEq}
\end{equation}
where
\begin{subequations}
\begin{align}
    &\mline = \frac{\vvec_2(2)-\vvec_1(2)}{\vvec_2(1)-\vvec_1(1)}, \\
    &\cline =-\vvec_1(1)\mline+\vvec_1(2).
\end{align}
\end{subequations}
The intersection of the ground plane with the ellipsoid in \eqref{eq:ellipsoid} is an ellipse (``Ellipse Ground'') (not to be confused with the ellipse in the $rs$-plane of \eqref{eq:ellipseTranslationalForm}), with foci at $[0,\, \Frac{m}{2}]$ and $[0,\, -\Frac{m}{2}]$.
The equation for this ellipse in the $xy$-plane is
\begin{equation}
    \frac{x^2}{\Aground^2} + \frac{y^2}{\Bground^2} = 1,
    \label{eq:ellipseGround}
\end{equation}
where 
\begin{subequations}
\begin{align}
    \Aground(\dint) &= \sqrt{\big(\frac{\dint}{2}\big)^2-\big(\frac{m}{2}\big)^2},\label{eq:aground}\\
    \Bground(\dint) &= \frac{\dint}{2}.\label{eq:bground}
\end{align}
\end{subequations}
When the condition
\begin{equation}
    \cline^2 = \Aground^2\mline^2+\Bground^2 \label{eq:condition}
\end{equation}
is met, Ellipse Ground \eqref{eq:ellipseGround} and the line in \eqref{eq:lineEq} intersect at a single point, rather than two, corresponding to round-trip travel distance $\dint$.
Substituting \eqref{eq:aground} and \eqref{eq:bground} into \eqref{eq:condition} and solving for $\dint$ yields
\begin{equation}
    \dint = 2\sqrt{\frac{\cline^2+\Big(\frac{\mline m}{2}\Big)^2}{\mline^2+1}}.
\end{equation}
We use $\dint$ in \eqref{eq:aground} and \eqref{eq:bground} to solve for $\Aground$ and $\Bground$ and find the corresponding single point of intersection by solving \eqref{eq:lineEq} and \eqref{eq:ellipseGround} for $(x_{\rm int},y_{\rm int})$:
\begin{subequations}
\begin{align}
    x_{\rm int}&= \frac{-\Aground^2\mline\cline}{\Bground^2+\Aground^2\mline^2}, \\
    y_{\rm int}&= \mline x_{\rm int} + \cline.
\end{align}
\end{subequations}
If the point $(x_{\rm int},y_{\rm int})$ is on the line segment between $\vvec_1$ and $\vvec_2$, then the shortest round-trip travel time from $\lvec$ to the planar facet and back to $\cvec$ is $\dmin = \dint$.
If the point $(x_{\rm int},y_{\rm int})$ is \textit{not} on the line segment between $\vvec_1$ and $\vvec_2$, then the point on the line segment that is closest to the point $(x_{\rm int},y_{\rm int})$ is the point with the shortest round-trip travel time from $\lvec$ to the facet and back to $\cvec$.
This is one of the bottom facet vertices $\vvec_1 $ or $\vvec_2$:
\begin{equation}
    \dmin = \min \bigg\{\norm{\lvec-\vvec_1} + \norm{\cvec-\vvec_1},\,
                            \norm{\lvec-\vvec_2} + \norm{\cvec-\vvec_2}\bigg\}.
\end{equation}
The longest round-trip travel time $\dmax$ corresponds to the furthest of the top two vertices ($\vvec_3$ and $\vvec_4$):
\begin{equation}
    \dmax = \max \bigg\{\norm{\lvec-\vvec_3} + \norm{\cvec-\vvec_3},\,
                            \norm{\lvec-\vvec_4} + \norm{\cvec-\vvec_4}\bigg\}.
\end{equation}
In the special case where $\vvec_2(1) = \vvec_1(1)$, we recall that $\lvec$ and $\cvec$ are equally spaced from the origin along the $y$-axis.
In this case, if $\sign{\vvec_1(2)}\neq \sign{\vvec_2(2)}$, we know that the point on the line segment from $\vvec_1$ to $\vvec_2$ with the shortest round-trip travel time is at $[\vvec_2(1),\, 0,\, \vvec_2(3)]^\T$ and
\begin{equation}
    \dmin = \norm{\lvec-[\vvec_2(1),\, 0,\, \vvec_2(3)]^\T} + \norm{\cvec-[\vvec_2(1),\, 0,\, \vvec_2(3)]^\T}.
\end{equation}
When $\sign{\vvec_1(2)} = \sign{\vvec_2(2)}$, the shortest round-trip travel time is to the vertex $\vvec_1$ or $\vvec_2$ with the smallest magnitude $y$-coordinate:
\begin{equation}
    \dmin = \min \bigg\{\norm{\lvec-\vvec_1} + \norm{\cvec-\vvec_1},\,
                            \norm{\lvec-\vvec_2} + \norm{\cvec-\vvec_2}\bigg\}.
\end{equation}
The indices of the first and last affected time bins are
$k_{\rm min}= {\rm floor}(\Frac{\dmin}{(c\Delta_t)})$ and
$k_{\rm max}= {\rm floor}(\Frac{\dmax}{(c\Delta_t)})$ respectively.

\section{Inversion Algorithm}
\label{app:inversion_algorithm}
In this work, we seek a reconstruction of \textit{change} in the hidden scene from frame to frame.
Until motion occurs in the hidden scene, measurements include light returning from hidden- and visible-side stationary scene content.
When an object enters the foreground of the hidden scene, the new measurement changes to include added rates due to the foreground object as well as a rate reduction due to the object's occlusion of the background.
At each new frame, we use these changes to reconstruct the moving object as well as the stationary background behind it.
As an object traverses the hidden scene, these background segments accumulate to form a reconstruction of the stationary hidden scene.

\subsection{Model and likelihood}
\label{app:model_and_likelihood}

When objects move into the hidden scene, the camera measurement at the $n$th spatial pixel and the $k$th time bin is Poisson distributed,
\begin{equation}
    \xvec^{n,k}\sim {\rm Poisson}(\bvec^{n,k} + \sfg^{n,k}(\psivecfg)-\sbg^{n,k}(\psivecfg,\psivecbg)),
    \label{eq:model2}
\end{equation}
where
$\bvec\in\mathbb{R}^{N\times K}$ is the rates due to stationary scenery,
$\sfg\in\mathbb{R}^{N\times K}$ is the response of the objects, and
$\sbg\in\mathbb{R}^{N\times K}$ is the response of the occluded background region,
before the objects enter.
We assume that we have observed the floor for long enough that
$\bvec$ is approximately known.%
\footnote{It has also been shown that taking the median at each spatial pixel and time bin over a sequence of measurement frames produces a useful proxy for $\bvec$, even when objects are moving in the hidden scene~\cite{Gariepy2016,Bouman2017}.}
Vectors $\psivecfg$ and $\psivecbg$ contain parameters that describe the foreground objects and corresponding occluded background regions, all of which are modeled as vertical, planar, rectangular facets that face the occluding edge.
The foreground facets have parameters 
\begin{equation}
    \psivecfg = \{(\thetavec^m,\alphafg^m,\,\rfg^m,\,h^m), \,\,\,\, m = 1,\ldots,M \},
\end{equation}
where
facet $m$ has albedo $\alphafg^m$, range $\rfg^m$, and height $h^m$;
the number of foreground facets is $M$.
Angles $\thetavec^m = (\pmin^m, \pmax^m)$ are the minimum and maximum polar angles of the foreground facet $m$, measured around the occluding edge in the plane of the floor.
The occluded regions have parameters
\begin{equation}
    \psivecbg= \{(\alphabg^m,\,\rbg^m), \,\,\,\, m = 1,\ldots,M\},
\end{equation}
where occluded background region $m$ has albedo $\alphabg^m$ and range $\rbg^m$.
As described in Appendix~\ref{appendix:occlusion_geometry},
the height of the occluded region depends upon its range $\rbg$ and on the range $\rfg$ and height $h$ of the foreground facet.  Given parameters $\psivecfg$ and $\psivecbg$, the procedure outlined in \Cref{alg:forwardModel_computation} may be used to quickly compute $\sfg$ and $\sbg$.

Because all the Poisson random variables in \eqref{eq:model2} are independent,
the likelihood of the measurement vector $\xvec$ given parameters $\psivecfg$ and $\psivecbg$ is the product
\begin{equation}
    f(\xvec \sMid \psivecfg ,\psivecbg)
    = \prod_{n=1}^N\prod_{k=1}^K
        \frac{\left(\bvec^{n,k} + \sfg^{n,k}(\psivecfg)-\sbg^{n,k}(\psivecfg,\psivecbg)\right)^{\xvec^{n,j}}
        \exp\!\left(-\left(\bvec^{n,k} + \sfg^{n,k}(\psivecfg)-\sbg^{n,k}(\psivecfg,\psivecbg)\right)\right)}
        {\xvec^{n,k}!}.
    \label{eq:likelihood}
\end{equation}

\subsection{Estimation}
\label{app:parameter_estimation}

\begin{figure}
    \includegraphics[width=1\linewidth]{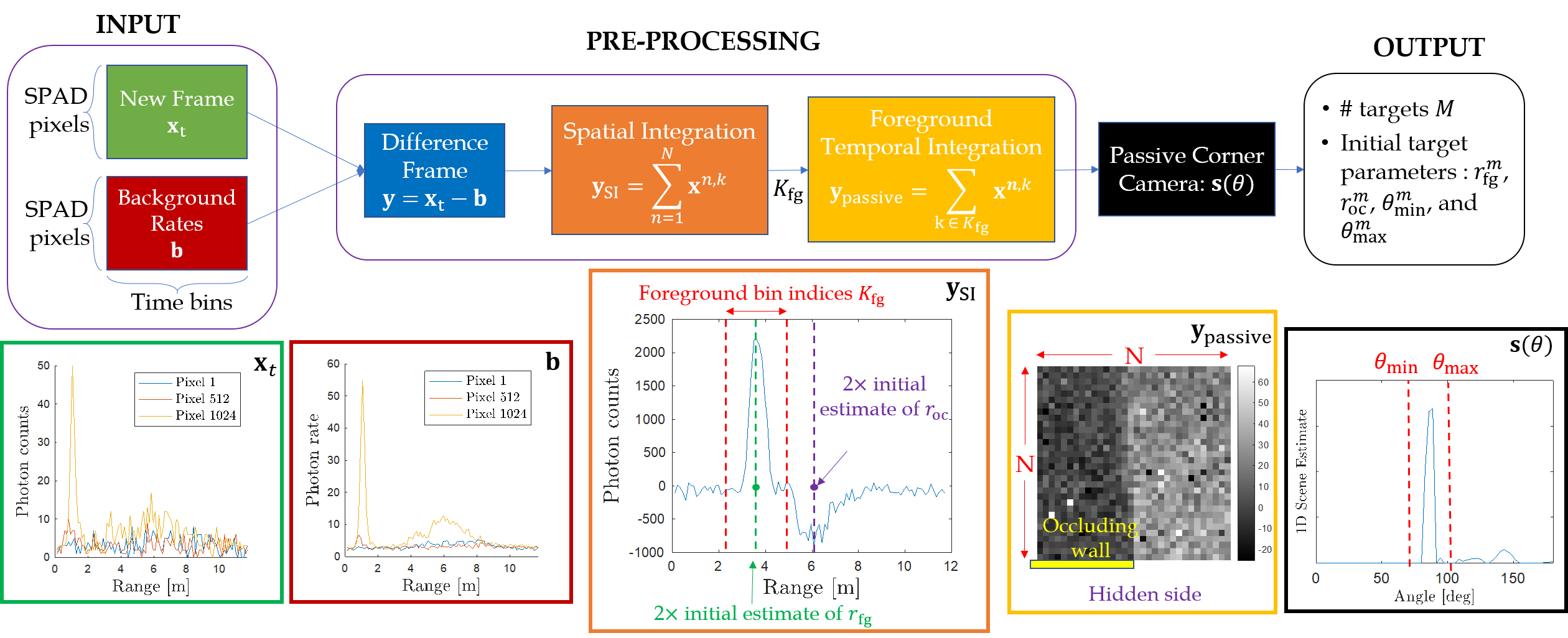}
    \caption{\label{fig:algorithmPipelineInitialization} Outline of the parameter initialization procedure.
    }
\end{figure}

\begin{figure}
    \includegraphics[width=1\linewidth]{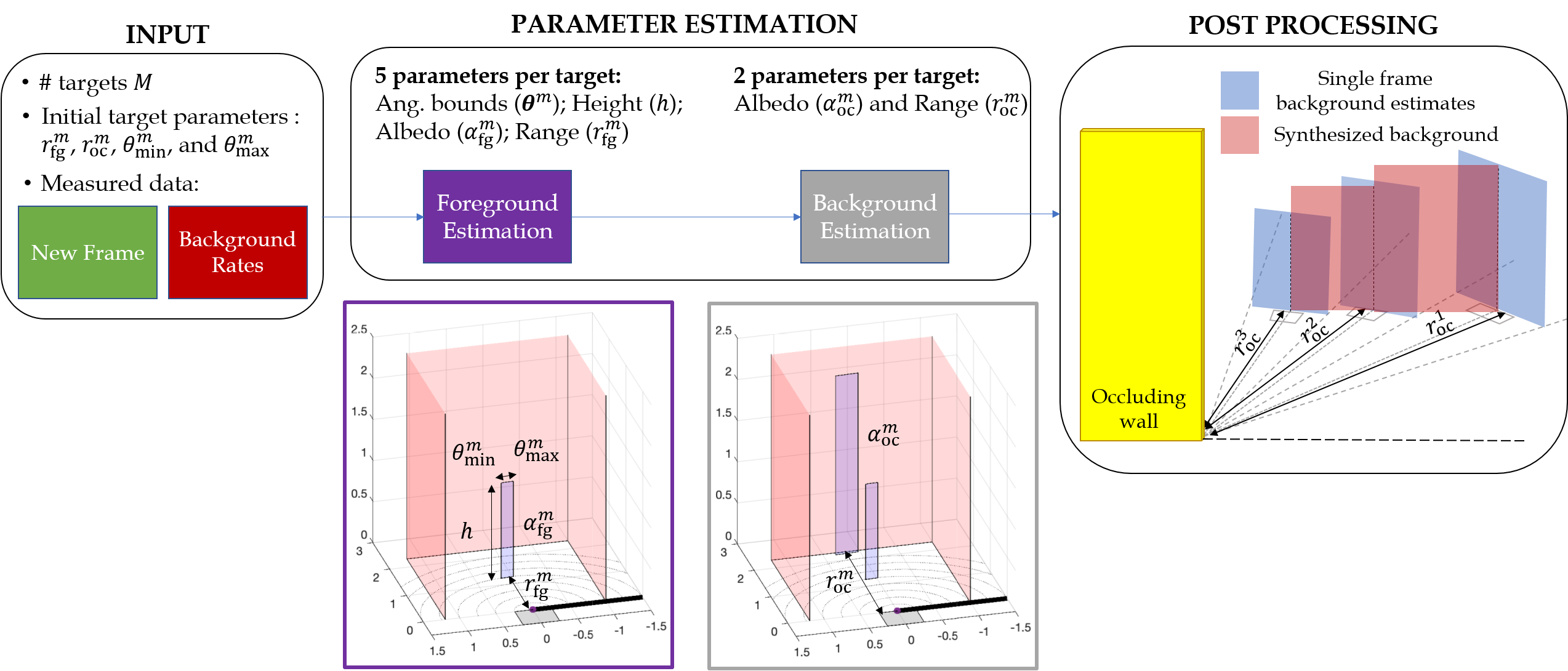}
    \caption{\label{fig:algorithmPipelineEstimation} Outline of the parameter estimation and post processing algorithm steps.
    }
\end{figure}

Our inversion algorithm, outlined in \Cref{fig:algorithmPipelineInitialization} and \Cref{fig:algorithmPipelineEstimation}, estimates the parameters $\psivecfg$ and $\psivecbg$ for each measurement frame where motion has occurred.


\paragraph{Parameter initialization.}
Figure \ref{fig:algorithmPipelineInitialization} covers the parameter initialization procedure.
Rates 
due to stationary scenery on visible and hidden sides of the occluding wall (red) 
are estimated from an initial reference measurement
In the pre-processing stage, a difference frame (blue) $\yvec$
is computed by subtracting the estimated background rates (red) $\bvec$
from the new measurement frame (green) $\xvec_{\rm t}$:  $\yvec = \xvec_{\rm t} - \bvec$.
For each time index $k$, this difference frame is spatially integrated (across all spatial pixels
in the SPAD array)
to create the collapsed profile of photon counts versus range, $\yvec_{\rm SI}$
(orange):
\begin{equation}
    \yvec_{\rm SI}^k = \sum_{n=1}^{N} \yvec^{n,k} .
\end{equation}
The light travel distances corresponding to the maximum and minimum values of $\yvec_{\rm SI}$ are halved and used to initialize parameters $\rfg$ and $\rbg$.
The time bin indices $K_{\rm fg}$ of the foreground object are selected according to
\begin{equation}
    K_{\rm fg} = \left\{k \,\,\,|\,\,\, \, k<p, \,\,\, \yvec_{\rm SI}^{k} \geq \beta_{\rm time} \max(\yvec_{\rm SI})\right\},
\end{equation}
where $p$ is the index of the minimum entry of $\yvec_{\rm SI}$
and $\beta_{\rm time}<1$ is a tuning parameter used to set the threshold.
For each pixel $n$, temporal integration (yellow)
of $\yvec_t$ is performed over the foreground time bin indices to form the equivalent of a passive measurement:
\begin{equation}
    \yvec_{\rm passive}^n = \sum_{k\in K_{\rm fg}}\yvec^{n,k}.
\end{equation}
An example of $\yvec_{\rm passive}$ for a single-object scenario,
with a clear penumbra pattern,
is pictured in the yellow box in \Cref{fig:algorithmPipelineInitialization}.
This `passive' measurement $\yvec_{\rm passive}$ is fed into the one-dimensional passive corner camera algorithm of \cite{Seidel2019_corner}\footnote{Specifically, we apply the algorithm on pg.\ 6, which was developed for use on a uniform floor.}
to produce a one-dimensional profile of the hidden scene $\stheta\in\mathbb{R}^Q$ as a function of azimuthal angle $\theta$ around the corner, where $Q$ is the number of angular bins in our discrete representation of the 1D hidden scene.
In all results presented here, $Q = 64$.
An example of $\stheta$ is shown in the black box in \Cref{fig:algorithmPipelineInitialization}, where a sharp peak is visible at the object location around $\theta = \pi/2$.
The azimuthal resolving power of the vertical edge makes this representation of our data well suited for counting the number of objects $M$ moving in our hidden scene~\cite{seidel2020}.
We compare $\stheta$ to a threshold $(\Frac{\beta_{\rm \theta}}{Q})\sum_{q=1}^Q\stheta^q$, where $\beta_{\rm \theta}$ is a tuning parameter.
Each interval where $\stheta$ is above the  threshold is considered to be a single object.
The angles of the first and last threshold crossing for each object are used to initialize parameters $\pmin$ and $\pmax$.

\paragraph{Parameter estimation.} 
Figure \ref{fig:algorithmPipelineEstimation} outlines the parameter estimation and post-processing steps.
Parameter estimation is performed using the Metropolis--Hastings (MH) algorithm, a type of Markov chain Monte Carlo (MCMC) method, and the likelihood in \eqref{eq:likelihood}.
Note that although all parameters in $\psivecfg$ and $\psivecbg$ could be estimated simultaneously in a single run of the MH algorithm, we choose to separate the parameter estimation into two stages for speed.
Foreground parameters $\psivecfg$ are estimated first, assuming there is no background occlusion.
Next, background parameters $\psivecbg$ are estimated, keeping the foreground parameters $\psivecfg$ fixed.
In this way, each stage only requires us to evaluate the forward model for a single facet per moving object, per iteration of the MH algorithm.
When estimating foreground parameters, we only compute the response $\sfg^{n,k}(\psivecfg)$ for each proposal;
when estimating background parameters, $\sfg^{n,k}({\psivecfghat})$ is fixed using already estimated foreground parameters $\psivecfghat$ and we only evaluate $\sbg^{n,k}(\psivecfghat,\psivecbg)$ at each algorithm iteration.
Separating parameter estimation into two problems is justified by the fact that foreground objects and occluded background regions generally affect very different swaths of time bins.
Additionally, because foreground objects are generally closer, their measured responses are much larger.
Thus, foreground parameters $\psivecfg$ may be accurately estimated without incorporating background occlusion into the model.
Although inter-frame parameter priors could be incorporated into our algorithmic framework, we demonstrate good performance with simple uniform priors on the parameters in $\psivecfg$ and $\psivecbg$, choosing wide bounds to exclude extreme and unrealistic scenarios.

In the foreground parameter estimation stage, we use the MH algorithm to draw samples from the posterior distribution:
\begin{equation}
    \ffg(\psivecfg \sMid \xvec) \propto \ffg(\xvec \sMid \psivecfg) \, \gfg(\psivecfg),
\end{equation}
where $\gfg(\psivecfg)$ is the prior on $\psivecfg$,
\begin{equation}
\gfg(\psivecbg) \propto 
    \begin{cases}
    1, &\text{if\,\,\,\,} \psivecfg\in\smatfg;\\
    0, &\text{otherwise},
    \end{cases}
\end{equation}
with $\smatfg$ defined to be the set of all possible values of $\psivecfg$ that arises from uniform priors on each of the parameters in $\psivecfg$.
The likelihood $\ffg(\xvec \sMid \psivecfg)$ of measurement $\xvec$ given parameter $\psivecfg$ may be approximated as
\begin{equation}
    \ffg(\xvec \sMid \psivecfg)
    \approx \prod_{n=1}^N \prod_{k=1}^K
    \frac{\left(\bvec^{n,k} + \sfg^{n,k}(\psivecfg)\right)^{\xvec^{n,k}}
    \exp\!\left(-\left(\bvec^{n,k} + \sfg^{n,k}(\psivecfg)\right)\right)}
    {\xvec^{n,k}!},
    \label{eq:likelihoodFG}
\end{equation}
where the effects of background occlusion have been excluded from the model for speed.
Candidate parameter values $\psivecfg'$ are drawn according to
$\psivecfg' \sim\mathcal{N}(\psivecfg^t,\,\sigmavec_{\rm fg})$,
where $\psivecfg^t$ is a vector containing the current state of each unknown parameter and $\sigmavec_{\rm fg}$ is a diagonal matrix with an entry corresponding to each parameter's proposal variance.
This matrix is scaled every 100 iterations to achieve an acceptance rate near $23\%$~\cite{Gelman1997}. 

Following the MH algorithm, proposal $\psivecfg'$ is accepted with probability $\omega$:
\begin{align}
    \omega &= \min \bigg[1,\,\,\frac{\ffg(\xvec \sMid \psivecfg') \gfg(\psivecfg')} {\ffg(\xvec \sMid \psivecfg^t)\gfg(\psivecfg^t)}\bigg]\\
    &\overset{\text{(a)}}{=}
    \begin{cases}
    \min \bigg[1,\,\,\frac{\ffg(\xvec\,|\,\psivecfg')}{\ffg(\xvec\,|\,\psivecfg^t)}\bigg],
        &\text{if\,\,\,\,} \psivecfg\in\smatfg;\\
     0, &\text{otherwise},
    \end{cases}
    \label{eq:fg_accept_prob}
\end{align}
where (a) arises from the fact that for $\psivecfg^t$ to have been accepted, $\gfg(\psivecbg^t)>0$, so that
\begin{equation}
    \frac{\gfg(\psivecbg')}{\gfg(\psivecbg^t)} =
    \begin{cases}
    1, &\text{if\,\,\,\,} \psivecfg\in\smatfg\\
    0, &\text{otherwise}.
    \end{cases}
\end{equation}
For each proposal $\psivecbg'$, we evaluate \eqref{eq:likelihoodFG} with
\begin{align}
    \frac{\ffg(\xvec \sMid \psivecbg')}{\ffg(\xvec \sMid \psivecbg^t)}
    &= \exp\bigg[\log\bigg( \frac{\ffg(\xvec \sMid \psivecbg')} {\ffg(\xvec \sMid \psivecbg^t)}\bigg)\bigg]\nonumber\\
    &=\exp\bigg[\log(\ffg(\xvec \sMid \psivecfg')) -\log(\ffg(\xvec \sMid \psivecfg^t))\bigg]\nonumber\\
    &=\exp\bigg[\sum_{n,k}\log(\ffg(\xvec^{n,k} \sMid \psivecfg'))-\sum_{n,k}\log(\ffg(\xvec^{n,k} \sMid \psivecfg^t))\bigg],
\end{align}
where to prevent a computing overflow, we compute $\log(\ffg(\xvec^{n,k} \sMid \psivecfg))$ as
\begin{equation}
    \log(\ffg(\xvec^{n,k} \sMid \psivecfg)) = \xvec^{n,k}\log(\sfg^{n,k})-\sfg^{n,k}-\log(\Gamma(\xvec^{n,k}+1)),
\end{equation}
where $\Gamma(\cdot)$ is the Gamma function.
An approximate constrained maximum likelihood (ML)
estimate $\psivecfghat$ is formed by drawing samples from
$\ffg(\psivecfg \sMid \xvec)$
using the procedure outlined above, binning samples of each parameter in $\psivecfg$ into histograms, and taking the center of the most commonly occurring bin as the estimate of that parameter.

In the background estimation step, we fix the foreground parameter estimate $\psivecfghat$ and estimate the parameters $\psivecbg$ that describe the occluded regions behind them. The posterior distribution of $\psivecbg$ given measurement $\xvec$ is
\begin{equation}
    \fbg(\psivecbg\,|\,\xvec) \propto \fbg(\xvec\,|\,\psivecbg)\gbg(\psivecbg),
    \label{eq:posteriorBG}
\end{equation}
where $g_{\rm bg}(\psivecfg)$ is the prior on $\psivecbg$:
\begin{equation}
\gfg(\psivecbg) \propto 
    \begin{cases}
    1, &\text{if\,\,\,\,} \psivecbg\in\smatbg;\\
    0, &\text{otherwise},
    \end{cases}
\end{equation}
and $\smatbg$ is the set of possible parameter values that arises from uniform priors on the parameters in $\psivecbg$ and from a constraint that that any parameter $\psivecbg$ must yield positive rates:
$\bvec^{n,k} + \sfg^{n,k}(\psivecfghat)-\sbg^{n,k}(\psivecfghat,\psivecbg)>0$.
With the estimated foreground rates $\psivecfghat$ fixed, the likelihood $\fbg(\xvec \sMid \psivecbg)$, including the effects of occlusion within the hidden scene, is approximately
\begin{equation}
    \fbg(\xvec\,|\,\psivecbg)
    \approx \prod_{n=1}^N\prod_{k=1}^K
    \frac{\left(\bvec^{n,k} + \sfg^{n,k}(\psivecfghat)-\sbg^{n,k}(\psivecfghat,\psivecbg)\right)^{\xvec^{n,k}}
    \exp\!\left(-\left(\bvec^{n,k} + \sfg^{n,k}(\psivecfghat)-\sbg^{n,k}(\psivecfghat,\psivecbg)\right)\right)}
    {\xvec^{n,k}!}.
    \label{eq:likelihoodBG}
\end{equation}
Using a procedure similar to the foreground parameter estimation step, we draw samples from the posterior distribution in \eqref{eq:posteriorBG} using the MH algorithm and form an approximate
constrained ML 
estimate $\psivecbghat$. 

\paragraph{Laser power correction.}
In our experiments, we observed the laser power to fluctuate over the duration of our acquisitions.
Before processing the data, we estimate a multiplicative scaling factor to adjust for any laser power fluctuation that occurred between the reference measurement and the motion frame.
Take $\bvec'$ to be the estimated background rates, before correcting for variable laser power.
The motion frame measurement at time $t$ is given by $\xvec_t$.
Assuming that reference and motion frames should have the same rates at close range
(because only more distant parts of the scene are changing),
we compute the scale factor $\kappa$ using all $N$ camera pixels summed over the first 10 time bins:
\begin{equation}
    \kappa = \frac{\displaystyle \sum_{k=1}^{10}\sum_{n=1}^{N}\xvec_t^{n,k}}
                  {\displaystyle \sum_{k=1}^{10}\sum_{n=1}^{N}[\bvec']^{n,k}}.
\end{equation}
The corrected background measurement, used in all previous parts of this section, becomes $\bvec = \kappa\bvec'$.

\subsection{Combining estimates from subsequent frames}
\label{app:post-processing}
As frames accumulate, the sequence of estimates $\psivecfghat$ and $\psivecbghat$ may be processed together to form a reconstruction of the hidden scenery behind the moving objects.
This idea is illustrated in the {\sc post processing} box in \Cref{fig:algorithmPipelineEstimation}.
Here, the estimated occluded regions for three subsequent frames are shown in blue. In our post-processing step, we connect the vertical lines running through the center of these facets to form the combined reconstruction shown in red. The height of each red facet is the same as the blue facet from the previous frame. Combined reconstruction results shown in \Cref{app:results} were produced in this way.

\subsection{Computing the vertices of occluded regions}
\label{appendix:occlusion_geometry}
An occluded background region is described by range $\rbg$ measured from the occluding edge at angle $\pmid = (\pmax+\pmin)/2$.
We describe the occluded background region as a subset of the vertical plane facing the occluding edge at range $\rbg$, determined by the position and size of the foreground object as well as the location of the laser spot $\lvec$.\footnote{If the camera pixel and laser spot are not co-located, the camera pixel has a different occluded background region associated with it. In our model, we account for both. The occluded region due to the pixel view is assumed constant across all pixels and is computed for the center of the camera FOV.}
We seek the vertices of the occluded background region so that its response $\sbg$ may be computed using \Cref{alg:forwardModel_computation}. Although in certain geometries the occluded background region may not be exactly rectangular, we make a rectangular approximation for speed. 

Assume we have chosen our coordinate system so that laser spot $\lvec$ is at the origin and the ground is in the $xy$-plane.
The vertical plane that contains the occluded background region may be written in terms of a surface normal vector
$\planenorm = [\planenorm^x,\,\planenorm^y,\,0]^\T$
and a point on the plane surface $\pplane = [\pplane^x,\,\pplane^y,\,\pplane^z]^\T$:\footnote{Plane point $\pplane$ and surface normal $\planenorm$ may be written in terms of parameters $\thetavec$ and $\psivecbg$.
}
\begin{equation}
   \planenorm^x(x-\pplane^x)+\planenorm^y(y-\pplane^y) = 0.
    \label{eq:background_plane}
\end{equation}
A vertex of the background facet $\vbg$ may be written in spherical coordinates:
 \begin{equation}
     \vbg = \left[ \begin{array}{ccc}
     r\cos\alpha\sin\delta\\
     r\sin\alpha\sin\delta\\
     r\cos\delta
     \end{array}\right],
     \label{eq:bg_vertex}
 \end{equation}
 where $\alpha$ is an angle measured down from the positive $z$-axis, $\delta$ is an angle measured from the positive $x$-axis towards the positive $y$-axis, and $r$ is the range from the origin.
 Because the occluded region is projected from the foreground facet onto the back plane, $\alpha$ and $\delta$ may be computed using the corresponding vertex $\vfg$ of the foreground facet:
 \begin{align}
    \alpha &=\tan^{-1}\frac{\vfg^y}{\vfg^x}, \label{eq:alpha}\\
     \delta &= \cos^{-1}\frac{\vfg^z}{\norm{\vfg}_2}.\label{eq:delta}
 \end{align}
The unknown range $r$ of the background vertex $\vbg$ is found by substituting the coordinate expressions of \eqref{eq:bg_vertex} into \eqref{eq:background_plane} and solving for $r$:
\begin{equation}
    r = \frac{\langle \planenorm,\, \pplane\rangle}
             {\planenorm^x\cos\alpha\sin\delta+\planenorm^y\sin\alpha\sin\delta}.
    \label{eq:background_range}
\end{equation}
The background vertex corresponding to foreground vertex $\vfg$ is found by substituting \eqref{eq:alpha}, \eqref{eq:delta}, and \eqref{eq:background_range} into \eqref{eq:bg_vertex}.

\section{Acquisition Time Analysis}
\label{app:acq_time_analysis}
In this work, we use \textit{integration time} to refer to the total time over which the camera collects meaningful data. We use \textit{acquisition time} to refer to the total time it takes the camera to collect, accumulate, and transfer data.
Results have been reported in terms of integration time because it reflects the true capability of our proposed imaging system and not certain limitations of our hardware.
Newer SPAD arrays are capable of achieving acquisition time roughly equal to the integration time, exploiting multi-gates approaches, such as the one implemented in the 32\,$\times$\,32 pixel array presented in \cite{Porta2018}.

For the older camera \cite{villa_2014_32x32SPAD} used in our experiments, the USB 2.0 standard is used to transfer data from the camera to our lab computer, resulting in the loss of 83\% of the measured frames; with the newer USB 3.0 standard, no measured frames would be lost.
The limited gating capabilities of our older camera only allow us to observe 800\,ns of each frame of duration 10\,$\mu$s, resulting in the loss of 92\% of the illumination periods.
Combined, these two acquisition and data-transfer limitations cause our acquisition times to be about $\frac{1}{(0.08)(0.17)} \approx 75$ times longer than they would be on a newer camera.

Although we demonstrate integration times short enough to track objects in motion, the required integration time could be further reduced by increasing the laser power or the SPADs' fill factor and photon detection probability, for instance exploiting 3D stacking technologies or microlens arrays.
In a representative measurement in our system, after removing 39 hot pixels,
we found that the remaining pixels had detections in approximately
0.0025\%
of illumination periods
(see \Cref{table:countrates} in Appendix~\ref{app:acq_time_analysis}).
Thus, we could increase our system detection rates
(through some combination of increases of laser power, fill factors, and detection probability)
by a factor of 
$\sim 2000$
(decreasing our required integration time by the same factor)
before reaching the 5\% threshold at which dead time distortions are conventionally regarded to become significant~\cite{OConnor1984}.
By this calculation,
the results in the paper produced using an integration time of 0.4\,s could be achieved with
200\,$\mu$s
integration time,
enabling tracking at
5000
updates per second.
Furthermore, methods to mitigate dead time effects would facilitate interpretation of data at higher count rates~\cite{RappMa:2019-TSP,Rapp:2021-Optica,Farina2021}.
Note, however, that we are not considering the computational demands of very high-speed tracking.


\section{Additional Experimental Results}
\label{app:results}

Our inversion algorithm has been tested in a variety of conditions, including different hidden scenes, lighting conditions, and frame lengths.
In \Cref{app:single_target}, we demonstrate our reconstruction algorithm on a hidden scene containing a single, moving planar object and explore the effects of frame length and reference measurement integration time.
In \Cref{app:two_target}, we show more detailed reconstruction results for the scene containing two moving planar objects that was presented in the main document.
In \Cref{app:robustness_testing}, we demonstrate that our algorithm continues to work well in challenging conditions.
We show results for scenes with non-planar moving objects, more complicated stationary hidden scenery, and with extreme amounts of ambient light.

\subsection{Single-object demonstrations}
\label{app:single_target}

\begin{figure*}
\centering
\begin{minipage}{.32\textwidth}
        \centering
        \text{0.8\,s frames}\\
        \includegraphics[width=1\linewidth]{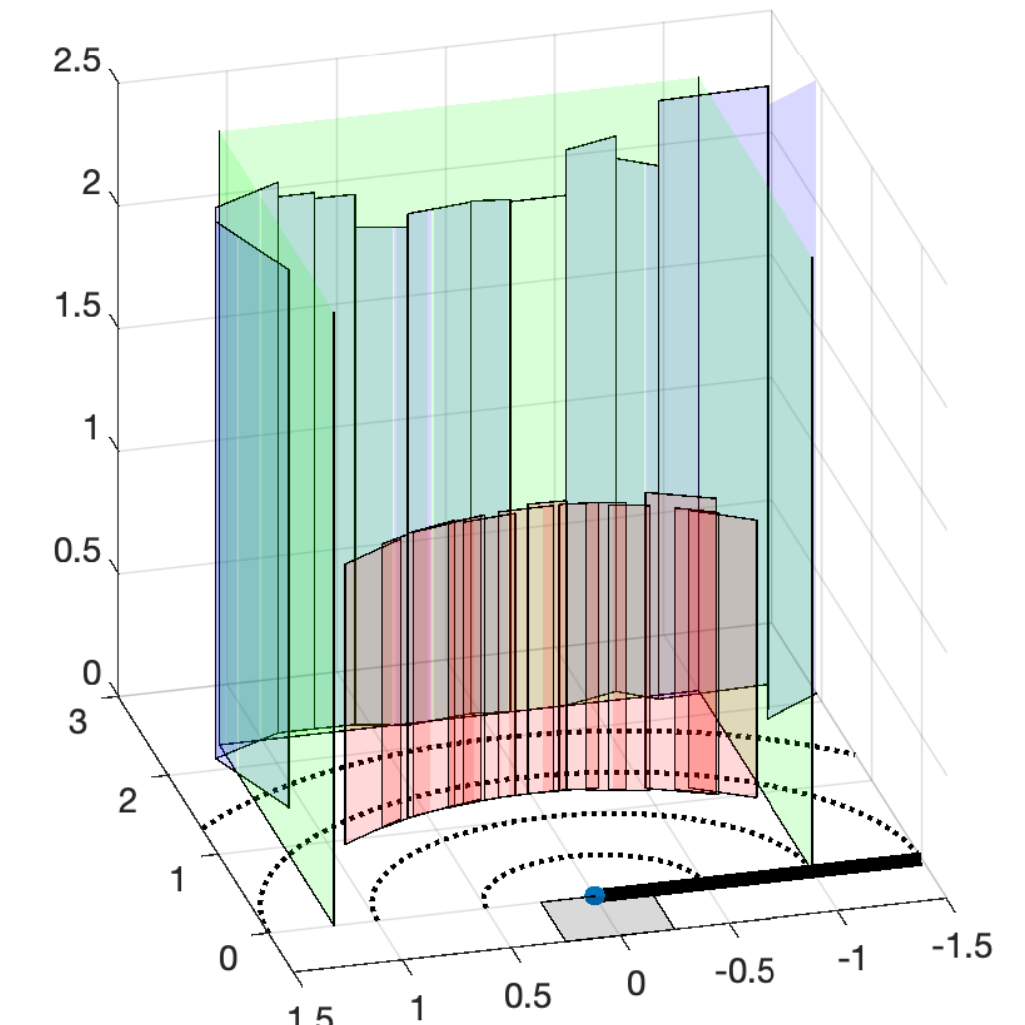}
\end{minipage}
\begin{minipage}{.32\textwidth}
        \centering
        \text{0.4\,s frames}\\
        \includegraphics[width=1\linewidth]{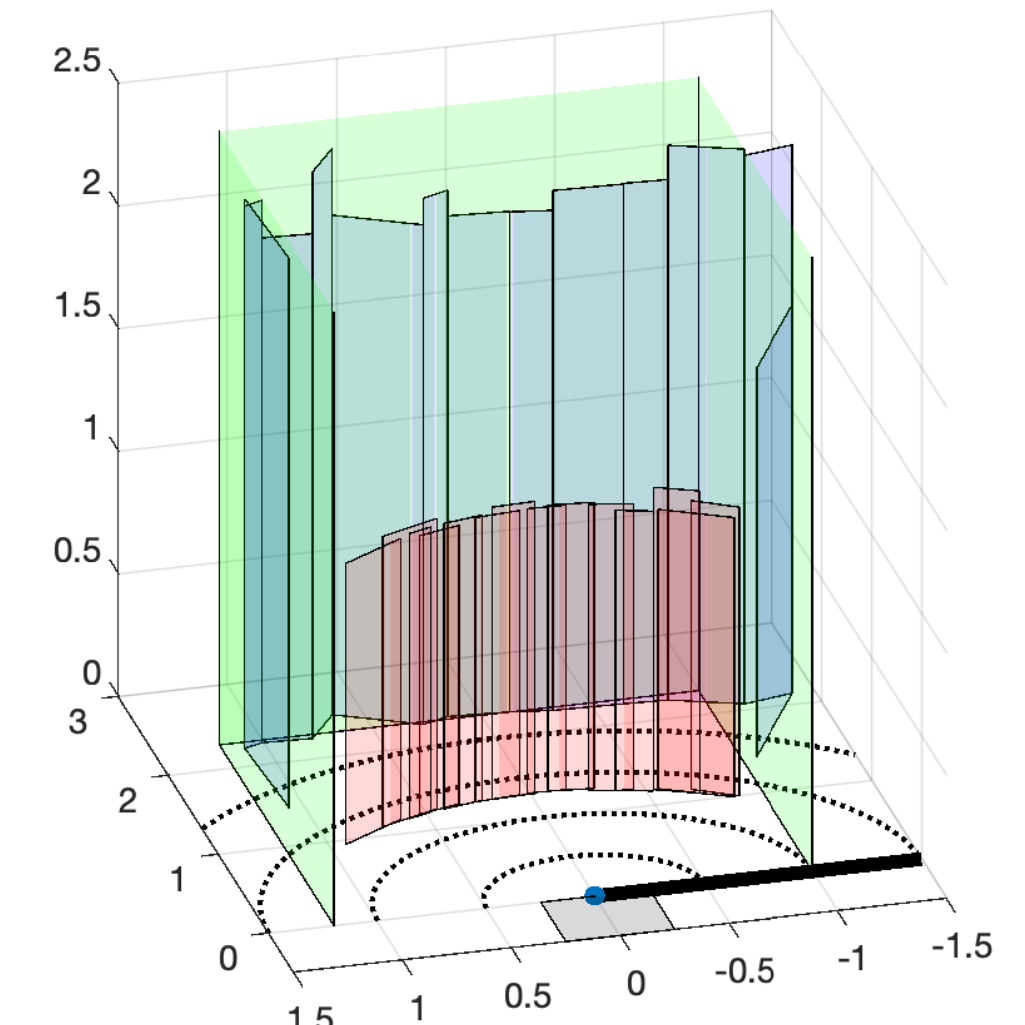}
\end{minipage}
\begin{minipage}{.32\textwidth}
        \centering
        \text{0.14\,s frames}\\
        \includegraphics[width=1\linewidth]{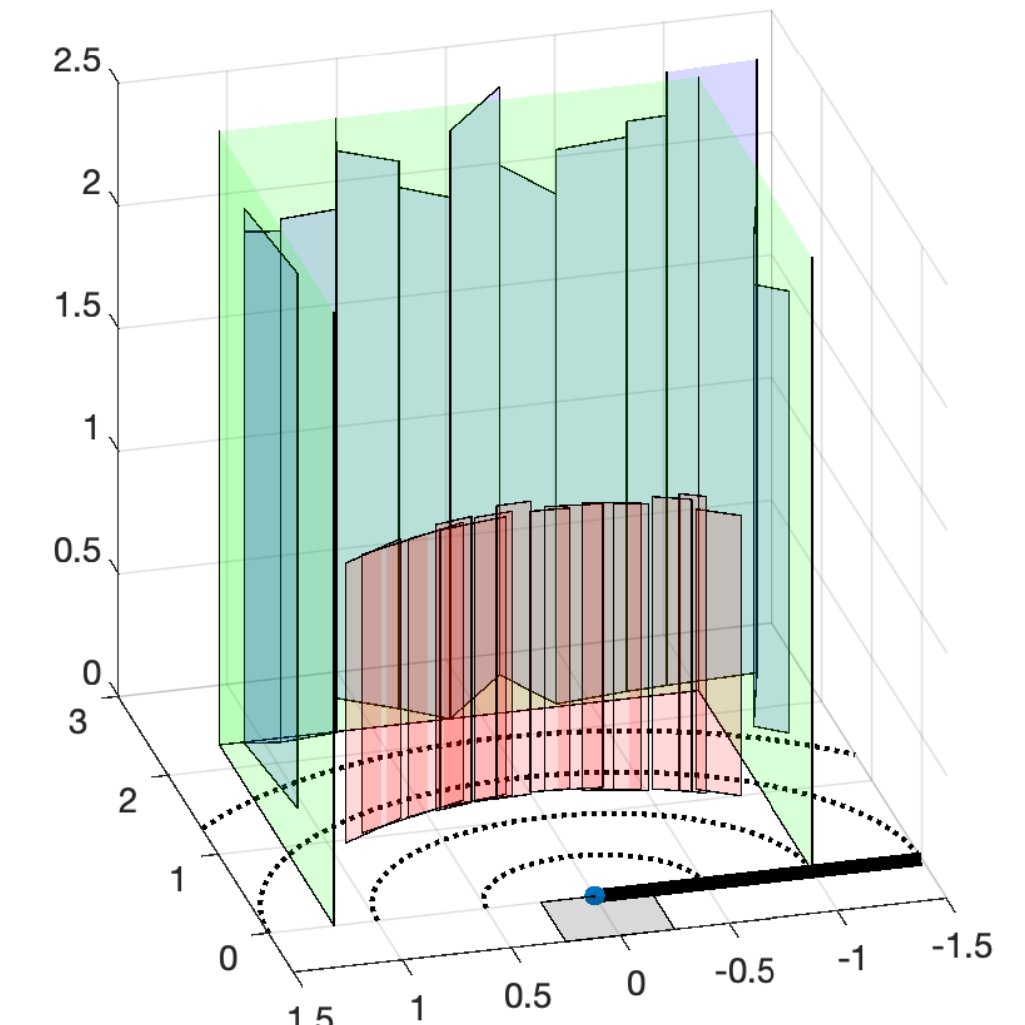}
\end{minipage}
\\
\begin{minipage}{.32\textwidth}
        \centering
        \includegraphics[width=1\linewidth]{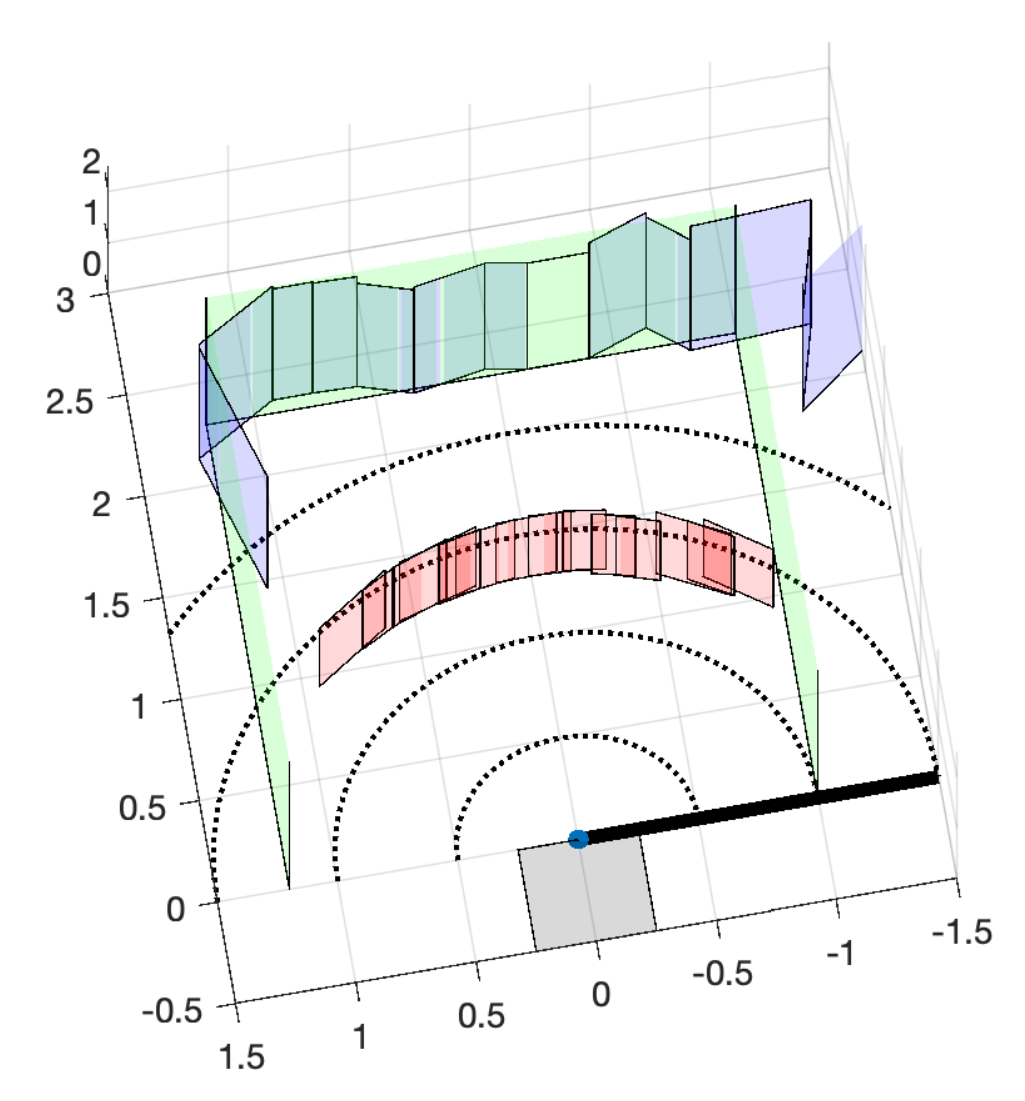}
\end{minipage}
\begin{minipage}{.32\textwidth}
        \centering
        \includegraphics[width=1\linewidth]{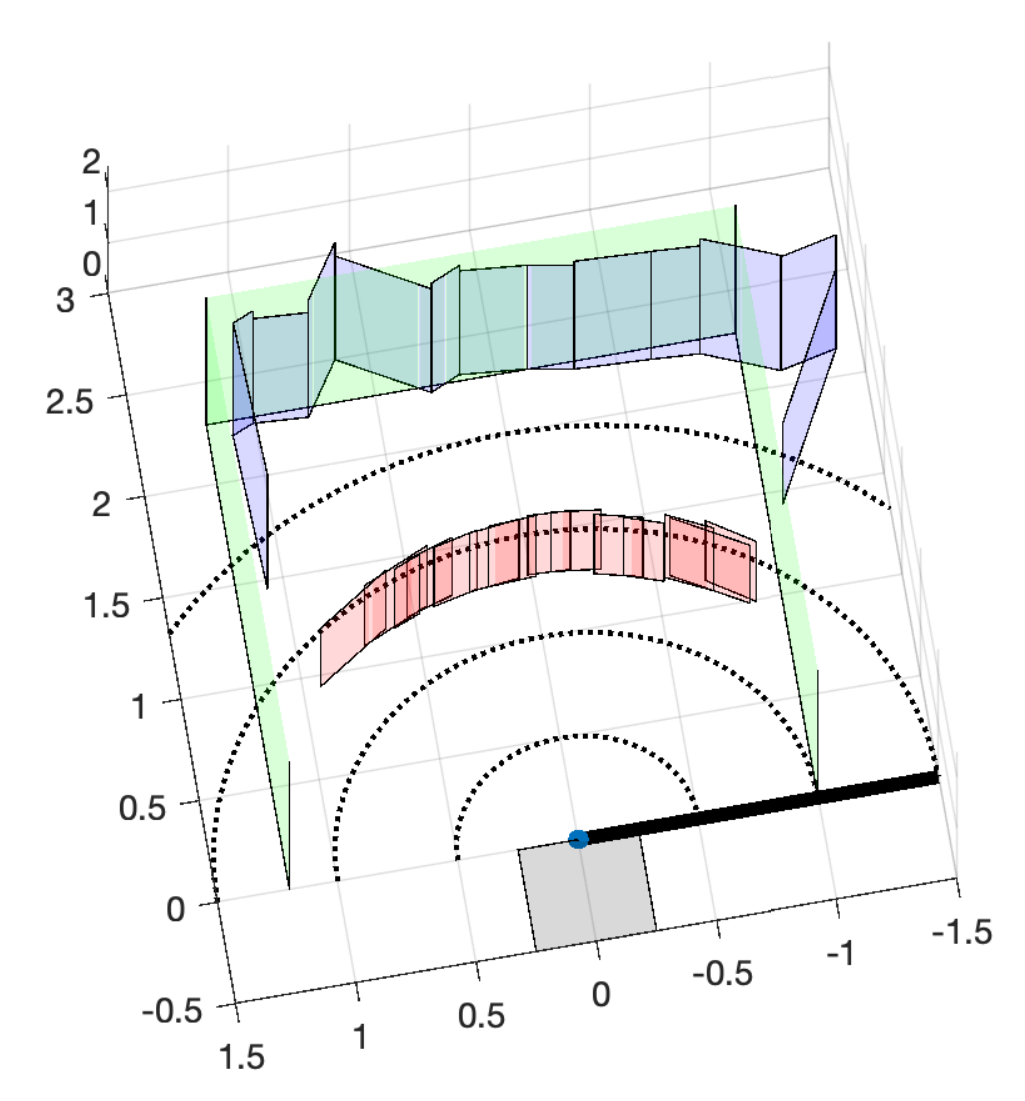}
\end{minipage}
\begin{minipage}{.32\textwidth}
        \centering
        \includegraphics[width=1\linewidth]{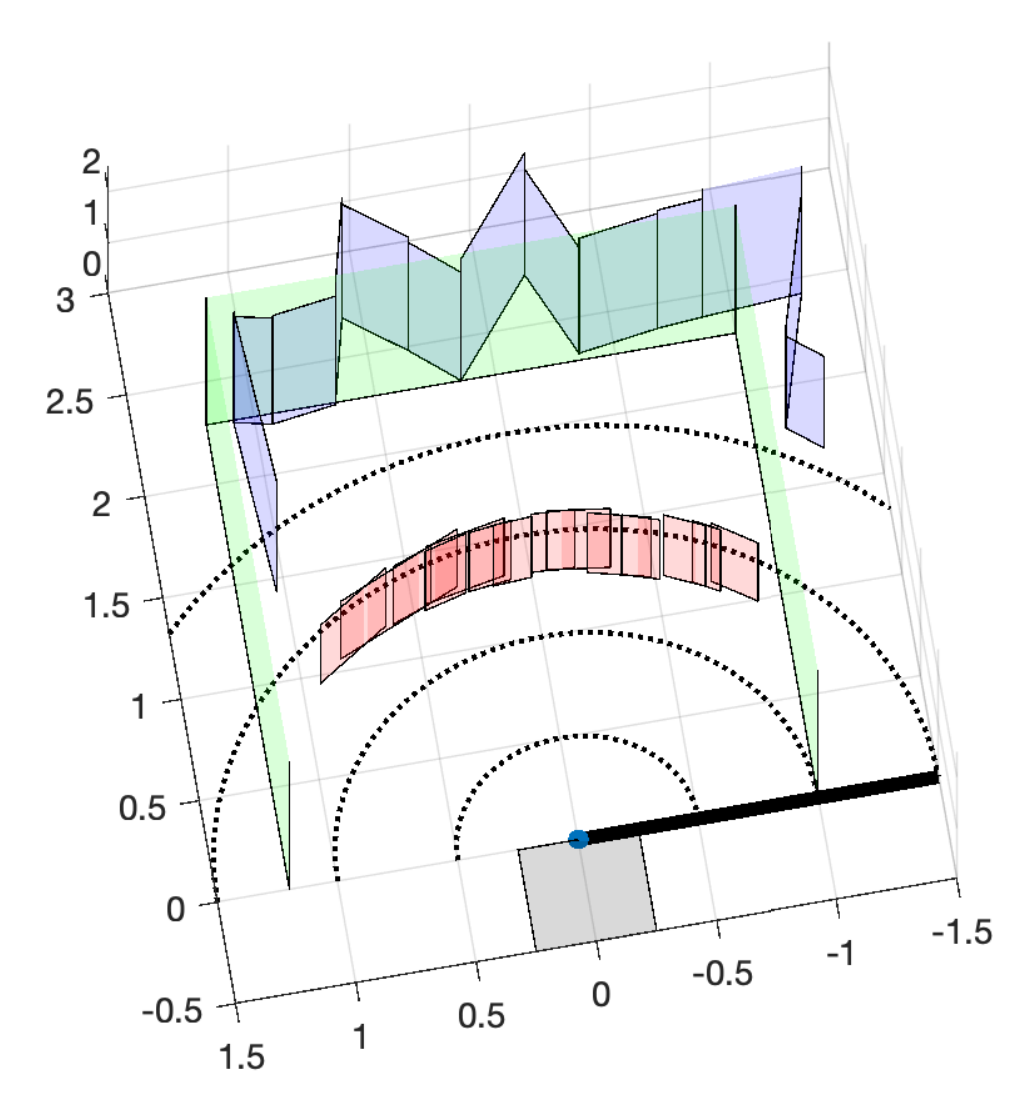}
\end{minipage}
\\
\begin{minipage}{.32\textwidth}
        \centering
        \includegraphics[width=.8\linewidth]{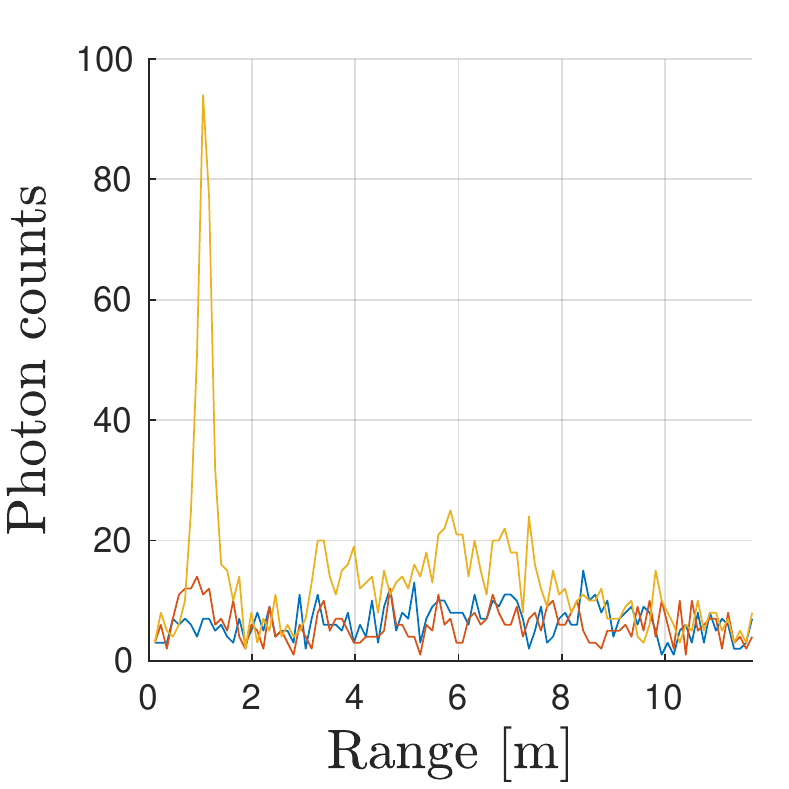}
\end{minipage}
\begin{minipage}{.32\textwidth}
        \centering
        \includegraphics[width=.8\linewidth]{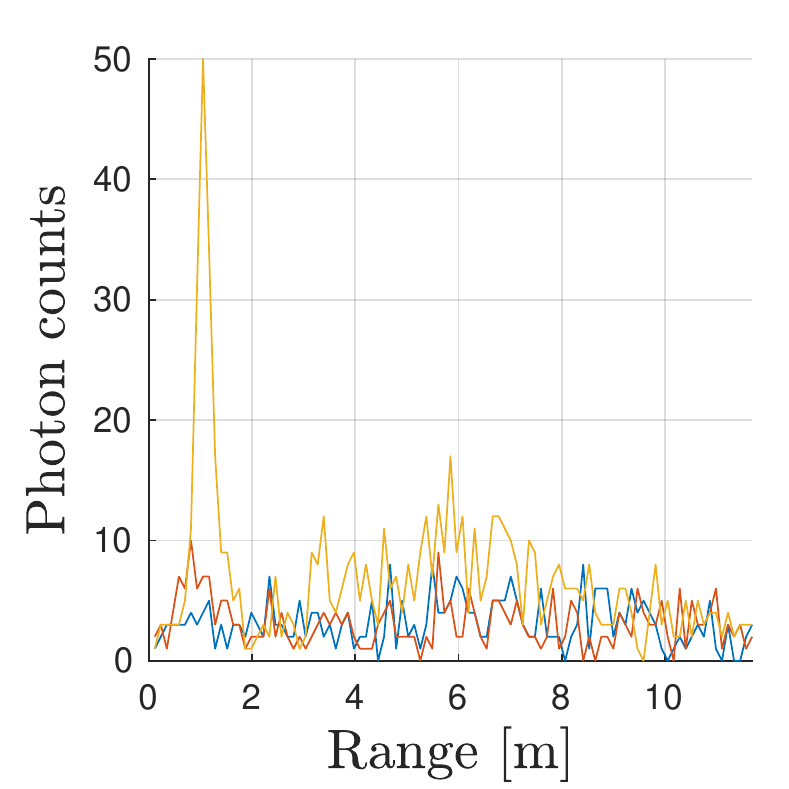}
\end{minipage}
\begin{minipage}{.32\textwidth}
        \centering
        \includegraphics[width=.8\linewidth]{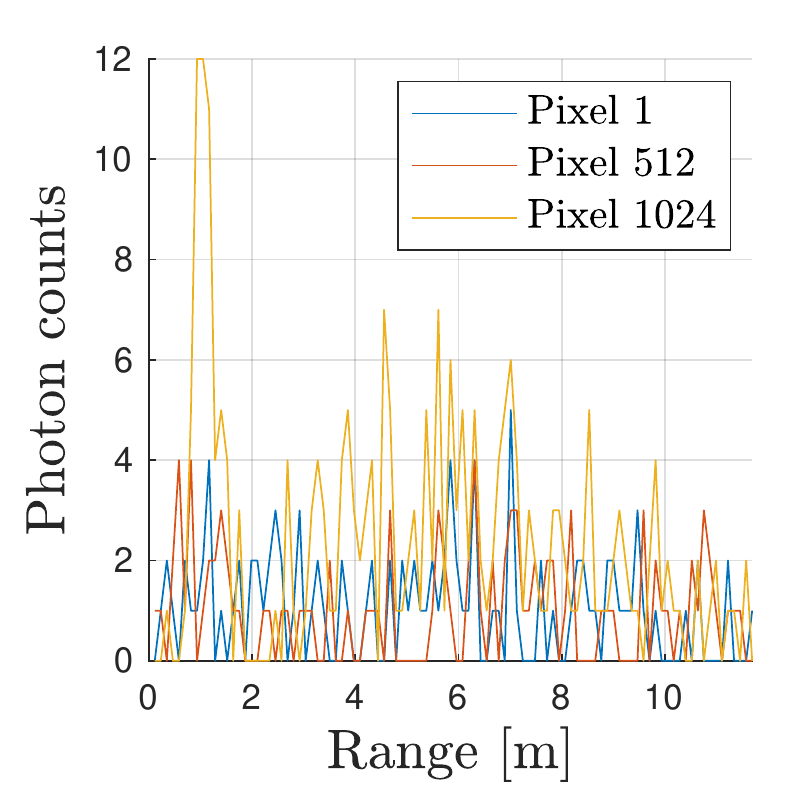}
\end{minipage}
\caption{\label{fig:angleSweep}
Experimental results for when a single planar facet at a range of 1.25\,m from the occluding edge moves in azimuthal angle through 14 positions in the hidden scene. Figure columns correspond to three different frame integration times: 0.8\,s, 0.4\,s, and 0.14\,s. The first and second rows show two different views of the reconstructed hidden scene. The true location of the stationary back walls are shown in green for reference. The composite background reconstruction, accumulated over the 14 frames, is shown in blue. The reconstructed foreground facets at all 14 positions are plotted in red. The SPAD FOV is shaded light gray, the thick dark line on the floor marks the footprint of the occluding wall, and the dotted arcs mark points on the floor that are 0.5\,m, 1\,m, 1.5\,m, and 2\,m from the occluding edge. The third figure row shows measured histograms at three different SPAD pixels. Results were generated using a reference measurement integrated over 33\,s.}
\end{figure*}

In \Cref{fig:angleSweep}, we show reconstruction results for a scenario where a single planar facet, at a range of 1.25\,m from the occluding edge, is moved in angle through 14 positions in the hidden scene.
Figure columns correspond to frame integration times of 0.8\,s, 0.4\,s, and 0.14\,s.
The first two rows show  different views of the reconstructed hidden scene.
The combined background reconstruction, accumulated over the 14 frames, is shown in blue.
The reconstructed foreground facets at all 14 positions are plotted on top of each other in red.
The SPAD FOV is shaded light gray, the thick dark line on the floor marks the footprint of the occluding wall, and the dotted arcs mark points on the floor that are 0.5\,m, 1\,m, 1.5\,m, and 2\,m from the occluding edge.
The third figure row shows sample measured histograms at three different pixels in the camera FOV\@.
Note that in the measured histograms, the peak at about 1\,m corresponds to the first bounce. Due to the location of the laser spot, the first bounce is occluded from view for some SPAD pixels and is thus most pronounced in the histograms for other pixels (e.g., in Pixel 1024 of \Cref{fig:angleSweep}).
The much smaller peak near 3.5\,m in range is due to the foreground object, and the counts observed between 3.5\,m and 8\,m are due to the back walls and ceiling.
For all three frame lengths, the foreground facets are correctly placed on a 1.25\,m arc around the occluding edge.
Foreground facet height estimates are also very close to the true height of 1.1\,m.
This is easily observed in first figure row, where the tops of reconstructed foreground facets clearly align with each other with very little variability.
For all three frame lengths, we observe that the background accuracy is greatest at azimuthal angles closer to the hidden side, with increasing error deeper in angle into the hidden scene.
As expected, scenery deeper into the hidden scene illuminates fewer camera pixels and thus results in lower measurement SNR\@.
We note that although foreground estimates are  quite accurate for all three integration times, error in the background reconstruction increases as the integration time decreases.

\begin{figure*}
\centering
\begin{minipage}{.32\textwidth}
        \centering
        \text{8.2\,s reference frame}\\
        \includegraphics[width=1\linewidth]{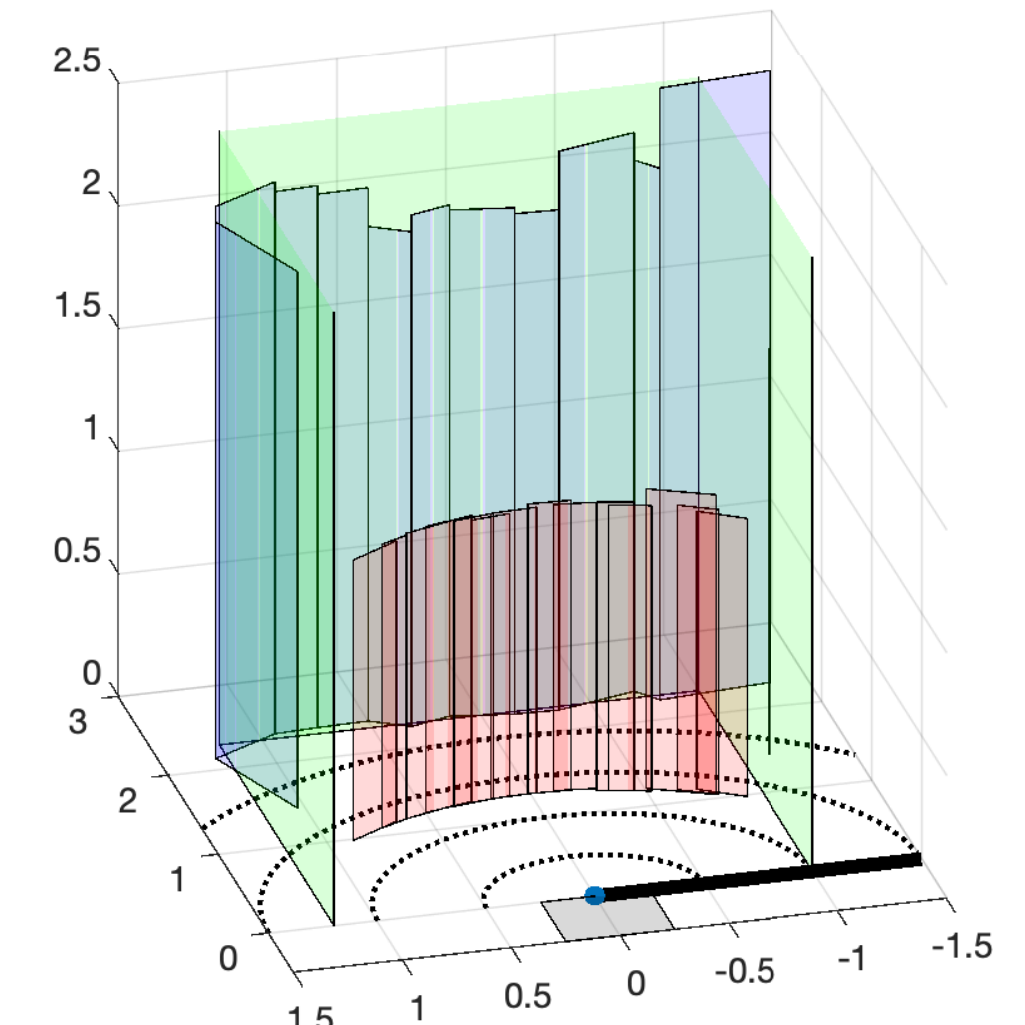}
\end{minipage}
\begin{minipage}{.32\textwidth}
        \centering
        \text{3.3\,s reference frame}\\
        \includegraphics[width=1\linewidth]{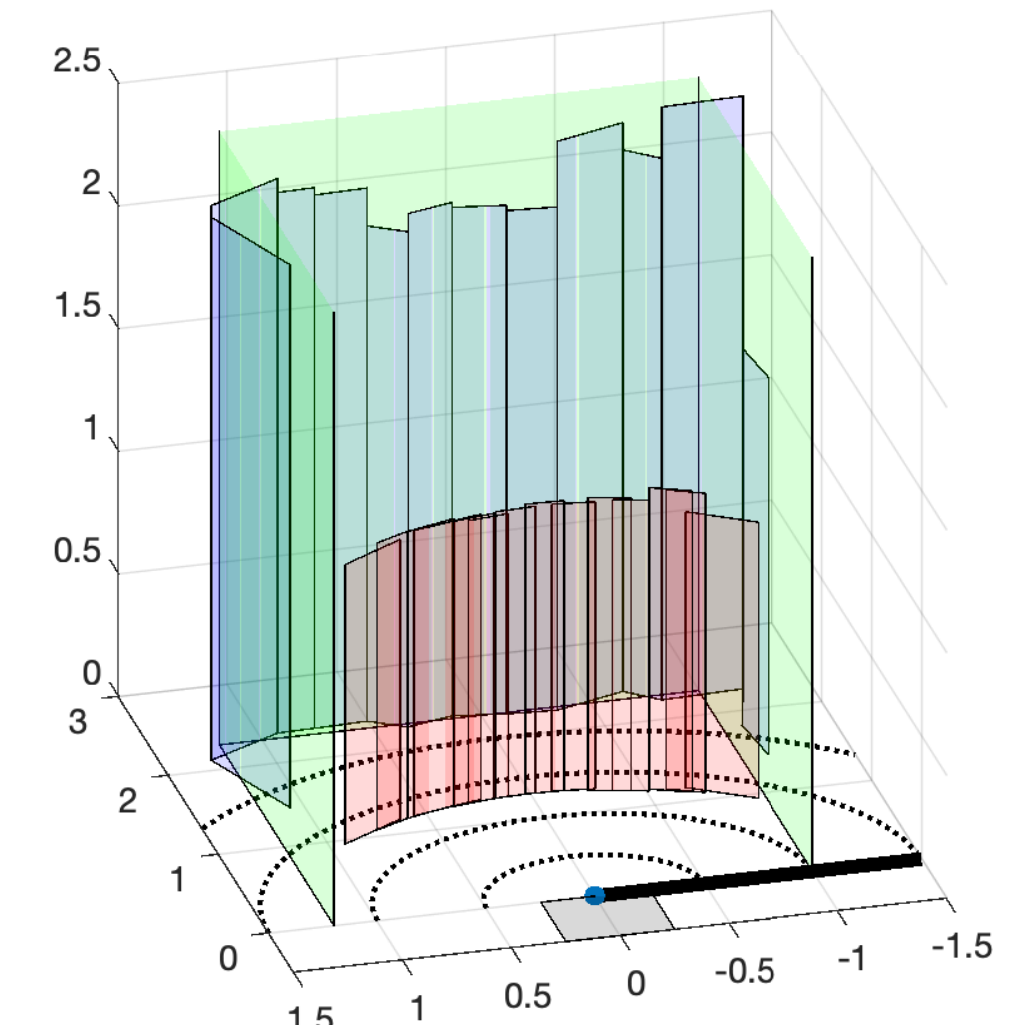}
\end{minipage}
\begin{minipage}{.32\textwidth}
        \centering
        \text{1.6\,s reference frame}\\
        \includegraphics[width=1\linewidth]{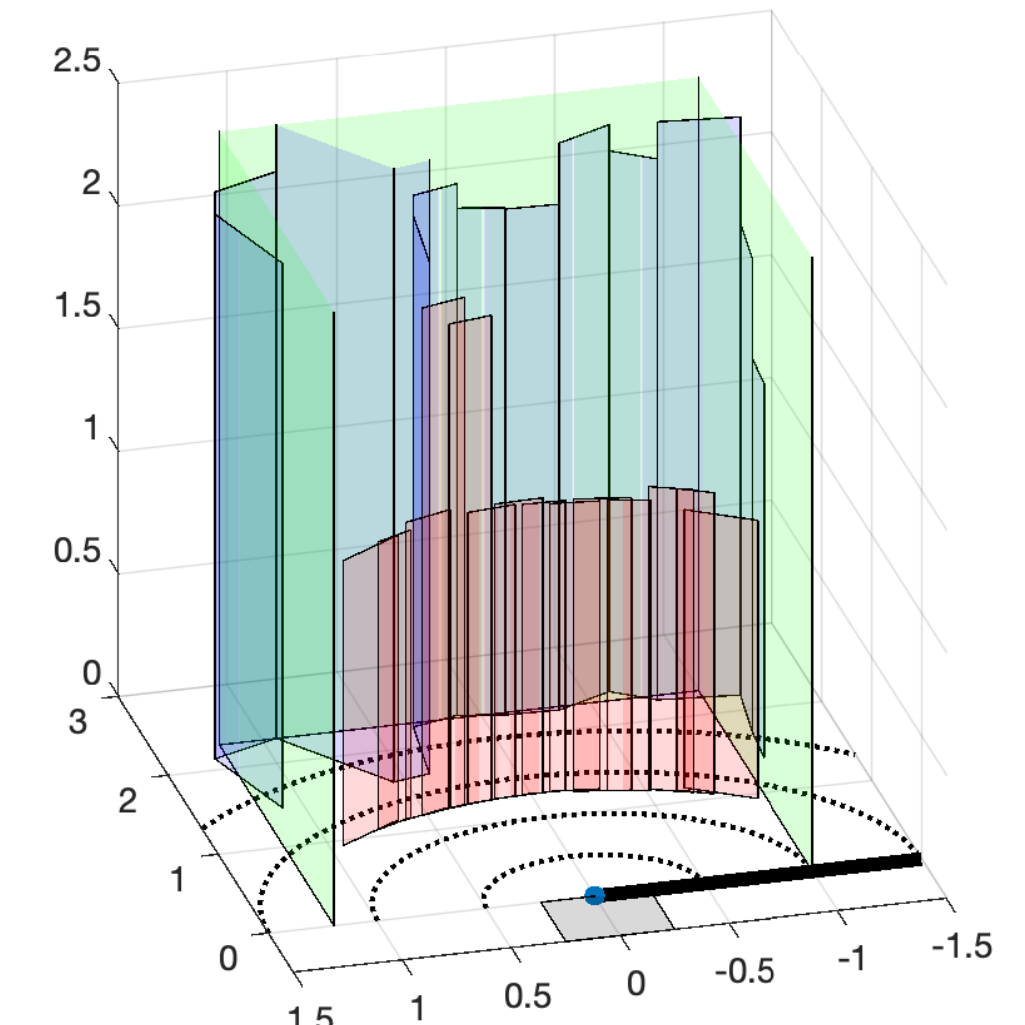}
\end{minipage}
\\
\begin{minipage}{.32\textwidth}
        \centering
        \includegraphics[width=1\linewidth]{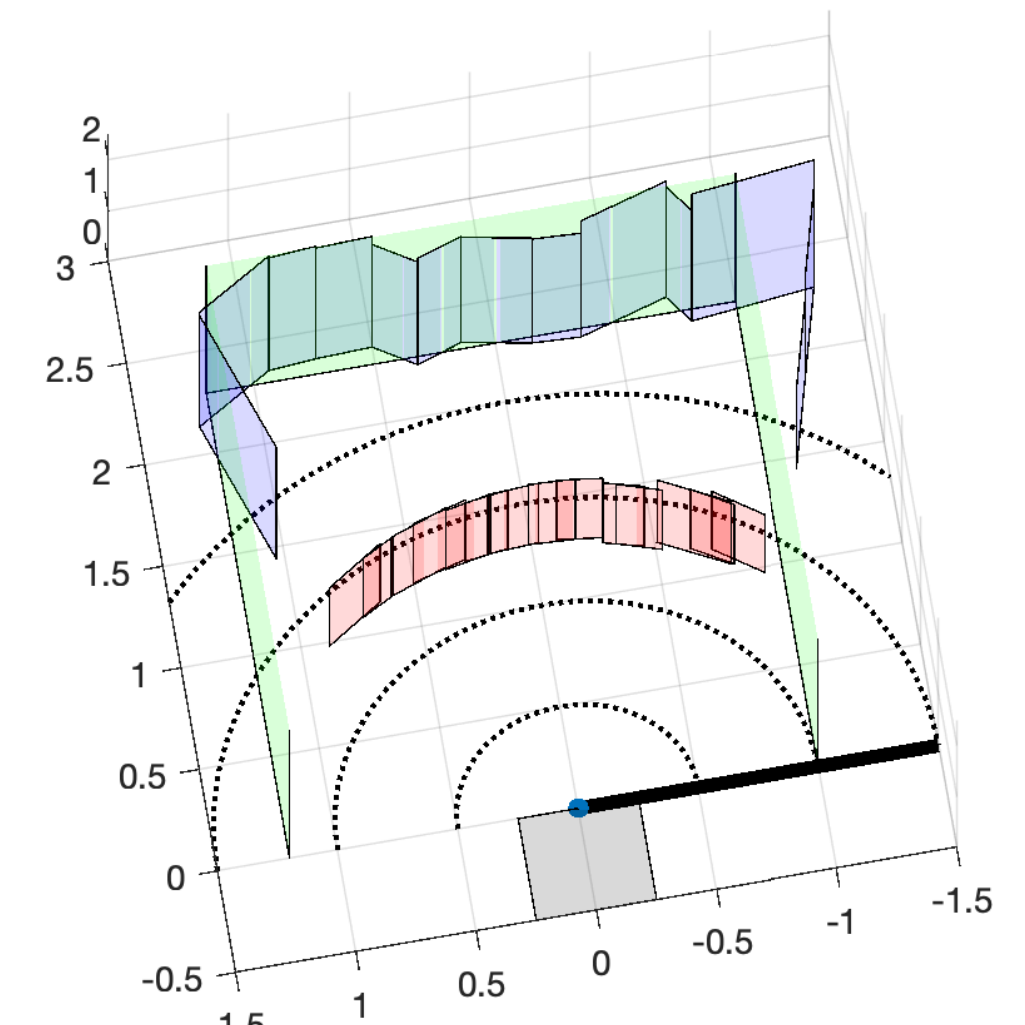}
\end{minipage}
\begin{minipage}{.32\textwidth}
        \centering
        \includegraphics[width=1\linewidth]{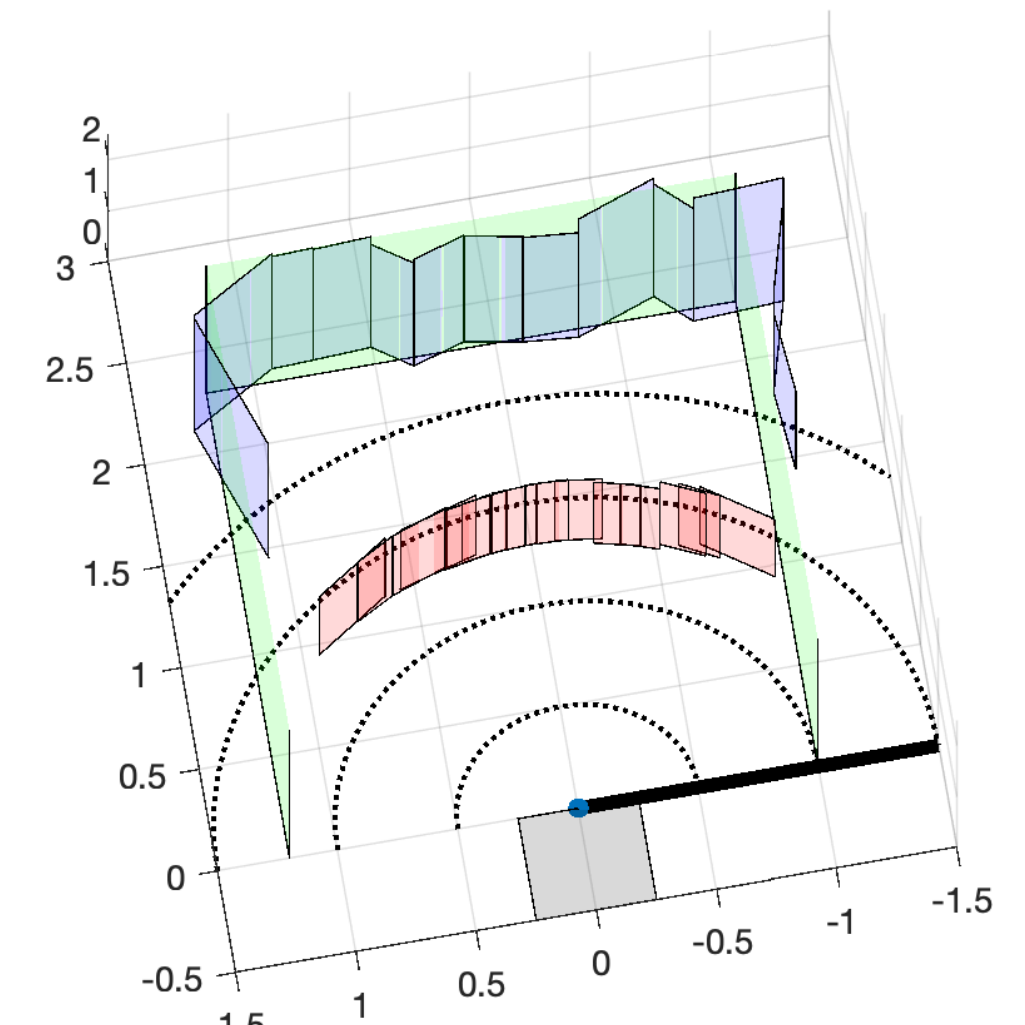}
\end{minipage}
\begin{minipage}{.32\textwidth}
        \centering
        \includegraphics[width=1\linewidth]{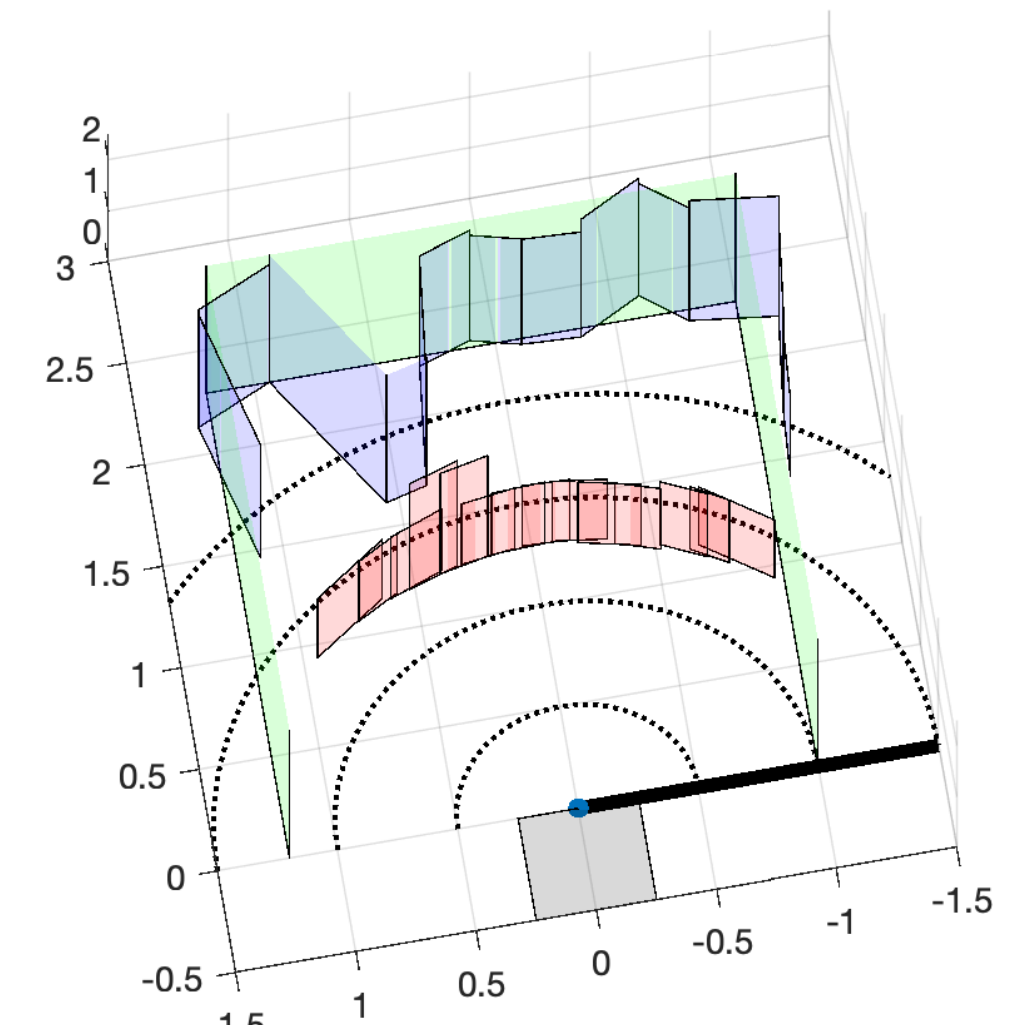}
\end{minipage}
\caption{\label{fig:angleSweepIntegrationTime} Hidden scene reconstructions for reference measurement integration times of 8.2\,s, 3.3\,s, and 1.6\,s.
The scenario is the same as in \Cref{fig:angleSweep}, with a single planar facet fixed at a range of 1.25\,m and swept in angle through 14 positions.
Each new measurement frame is integrated over 0.8\,s.
Figure rows show two different views of the hidden scene reconstruction.
The true location of the stationary back walls are shown in green for reference.
The composite background reconstruction, accumulated over the 14 frames, is shown in blue.
The reconstructed foreground facets at all 14 positions are plotted in red.
The SPAD FOV is shaded light gray, the thick dark line on the floor marks the footprint of the occluding wall, and the dotted arcs mark points on the floor that are 0.5\,m, 1\,m, 1.5\,m, and 2\,m from the occluding edge.}
\end{figure*}

In \Cref{fig:angleSweepIntegrationTime}, we use the same single-object scenario as in \Cref{fig:angleSweep} to explore the effect of reference measurement integration time on reconstruction accuracy.
Here, each new frame has an integration time of 0.8\,s
while testing reference measurement integration times of 8.1\,s, 3.3\,s, and 1.6\,s seconds.
Figure columns show results for different integration times; figure rows show two different reconstruction views.
As in \Cref{fig:angleSweep}, foreground facets are correctly placed in range at 1.25\,m.
However, when stationary scene integration time is decreased to 1.6\,s, we observe more error in foreground facet height estimates.
Similarly, background reconstructions for 8.1\,s and 3.3\,s are comparable to the results in \Cref{fig:angleSweep}, where an integration time of 33\,s was used for the reference measurement.
Increased error is seen in the background reconstruction when background integration time is reduced to 1.6\,s.
Although some of the results in this paper use the longest available background integration times, the results in \Cref{fig:angleSweepIntegrationTime} suggest that there is little cost to using background integration times as short as 3.3\,s.

\begin{figure}
\begin{minipage}{0.015\textwidth}
        \begin{sideways}
            \text{\normalsize \textbf{ground truth}}
        \end{sideways}
\end{minipage}
\begin{minipage}{0.155\textwidth}
        \centering
        \text{0.5\,m}\\
        \includegraphics[width=1\linewidth]{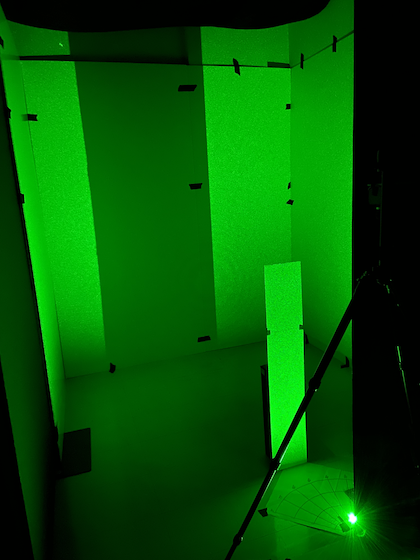}
\end{minipage}
\begin{minipage}{0.155\textwidth}
        \centering
        \text{0.75\,m}\\
        \includegraphics[width=1\linewidth]{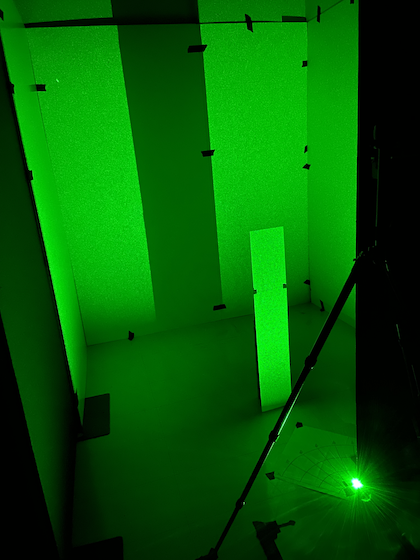}
\end{minipage}
\begin{minipage}{0.155\textwidth}
        \centering
        \text{1\,m}\\
        \includegraphics[width=1\linewidth]{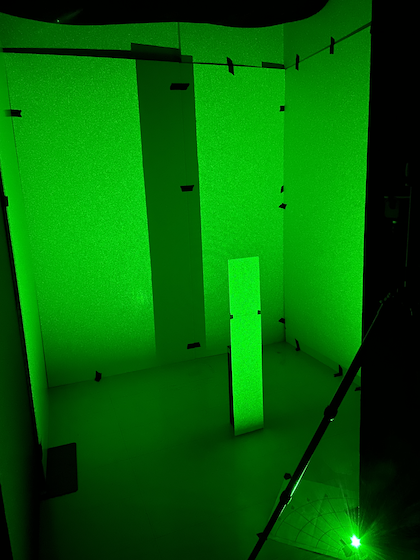}
\end{minipage}
\begin{minipage}{0.155\textwidth}
        \centering
        \text{1.25\,m}\\
        \includegraphics[width=1\linewidth]{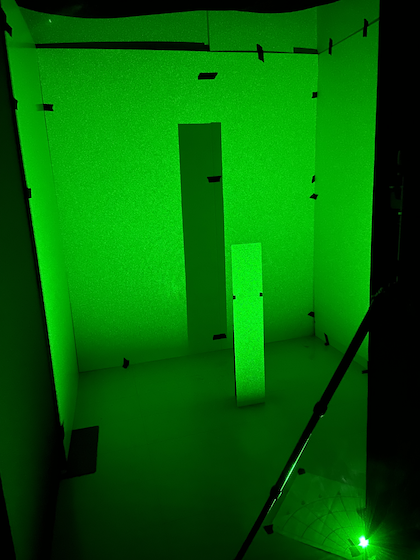}
\end{minipage}
\begin{minipage}{0.155\textwidth}
        \centering
        \text{1.5\,m}\\
        \includegraphics[width=1\linewidth]{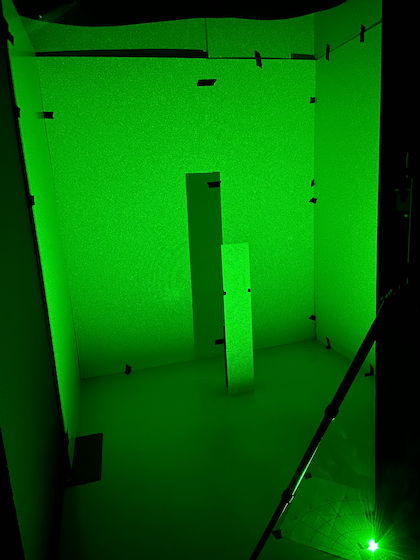}
\end{minipage}
\begin{minipage}{0.155\textwidth}
        \centering
        \text{1.75\,m}\\
        \includegraphics[width=1\linewidth]{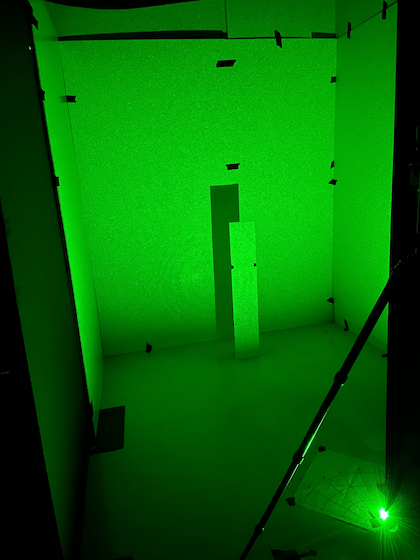}
\end{minipage}
\\
\begin{minipage}{0.015\textwidth}
        \begin{sideways}
            \text{\normalsize \textbf{reconstruction}}
        \end{sideways}
\end{minipage}
\begin{minipage}{0.155\textwidth}
        \centering
        \includegraphics[width=1\linewidth]{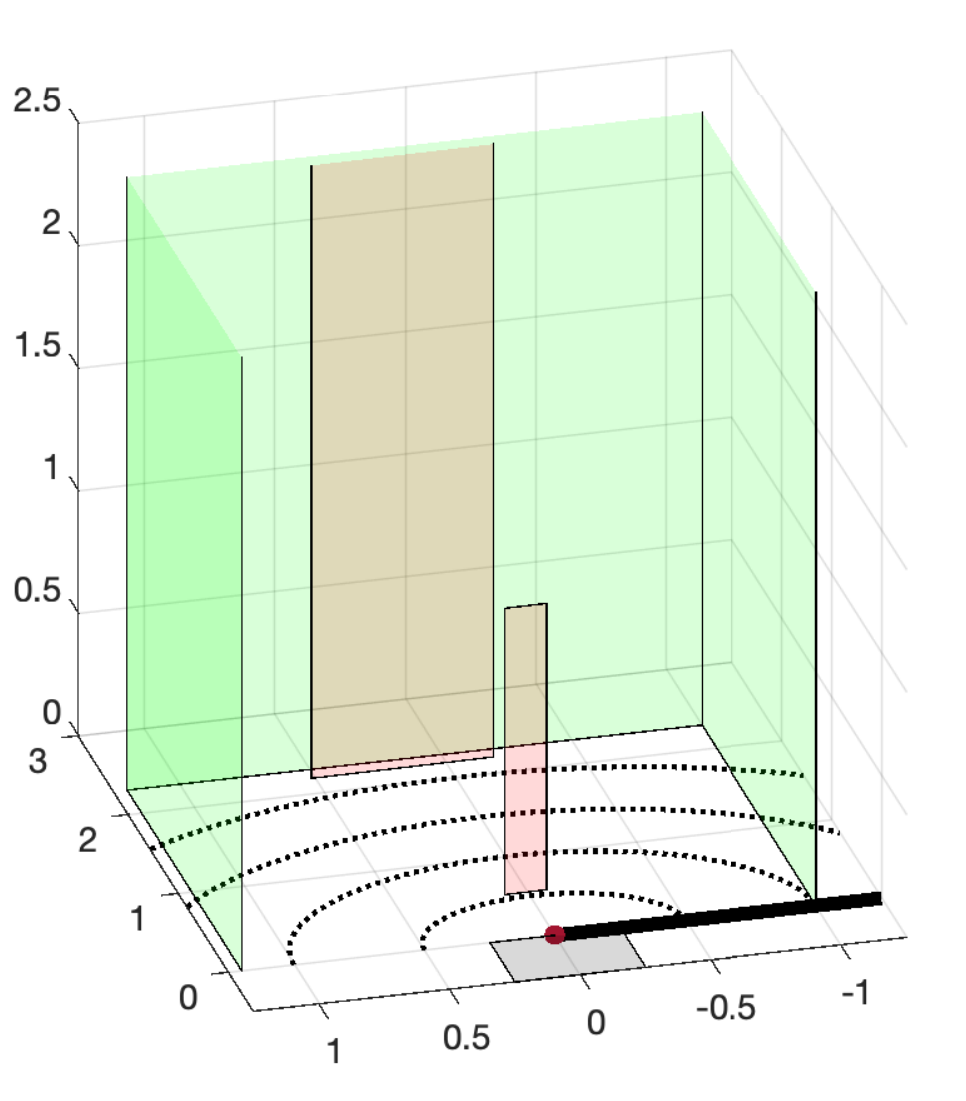}
\end{minipage}
\begin{minipage}{0.155\textwidth}
        \centering
        \includegraphics[width=1\linewidth]{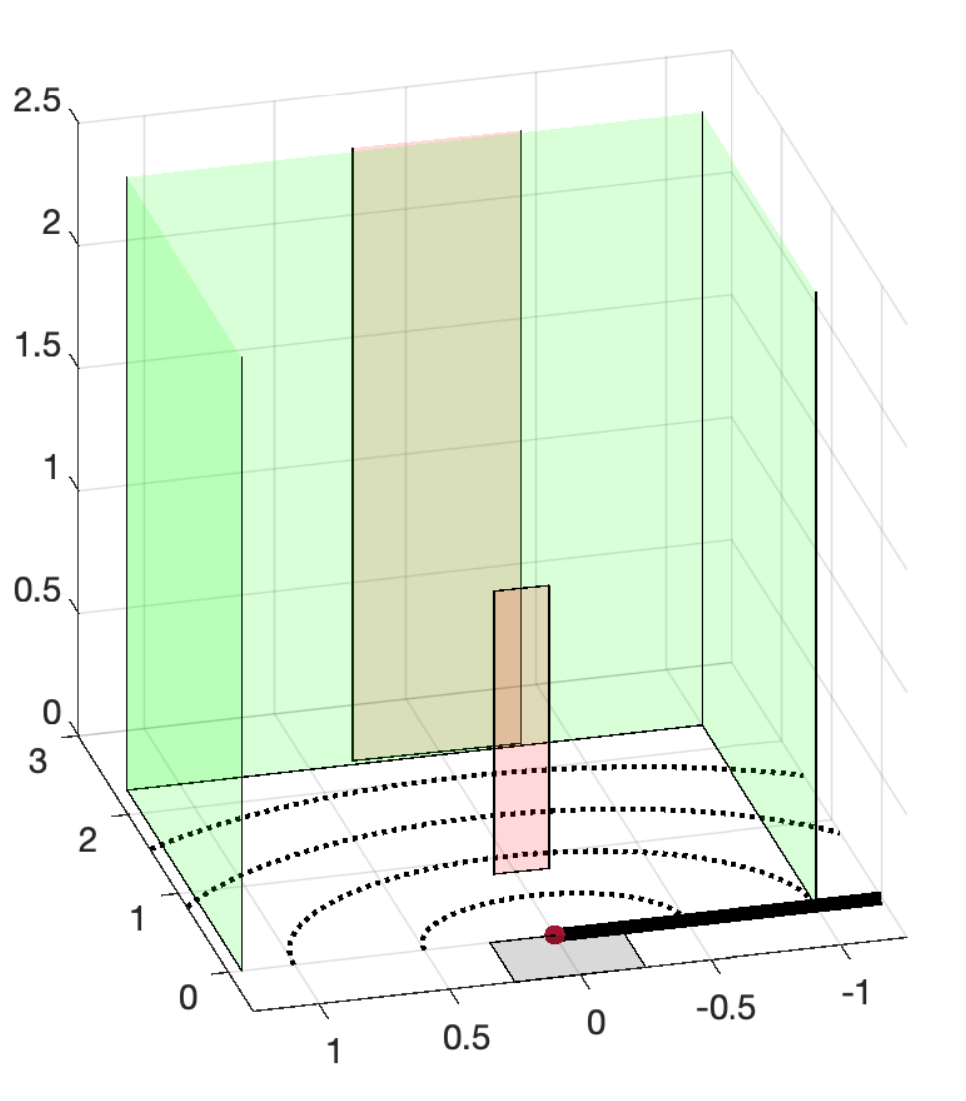}
\end{minipage}
\begin{minipage}{0.155\textwidth}
        \centering
        \includegraphics[width=1\linewidth]{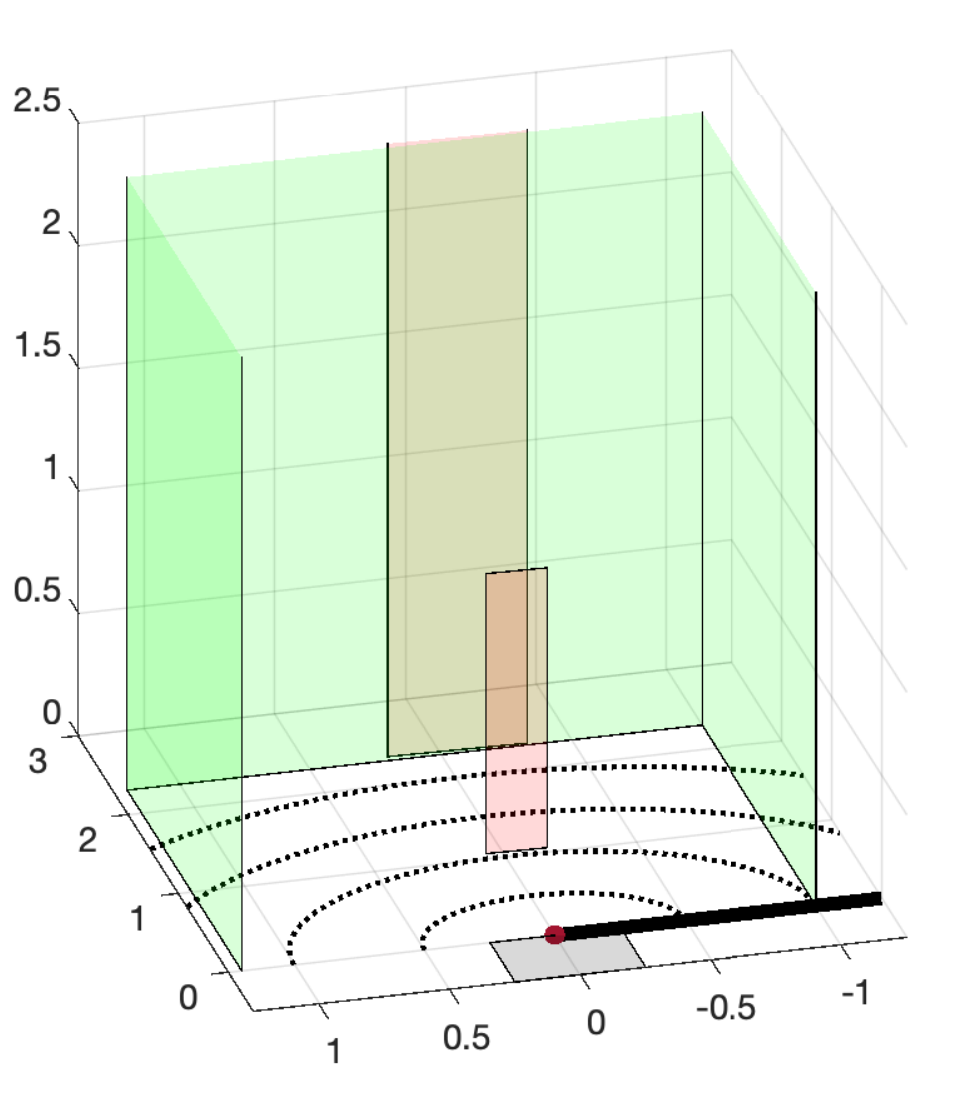}
\end{minipage}
\begin{minipage}{0.155\textwidth}
        \centering
        \includegraphics[width=1\linewidth]{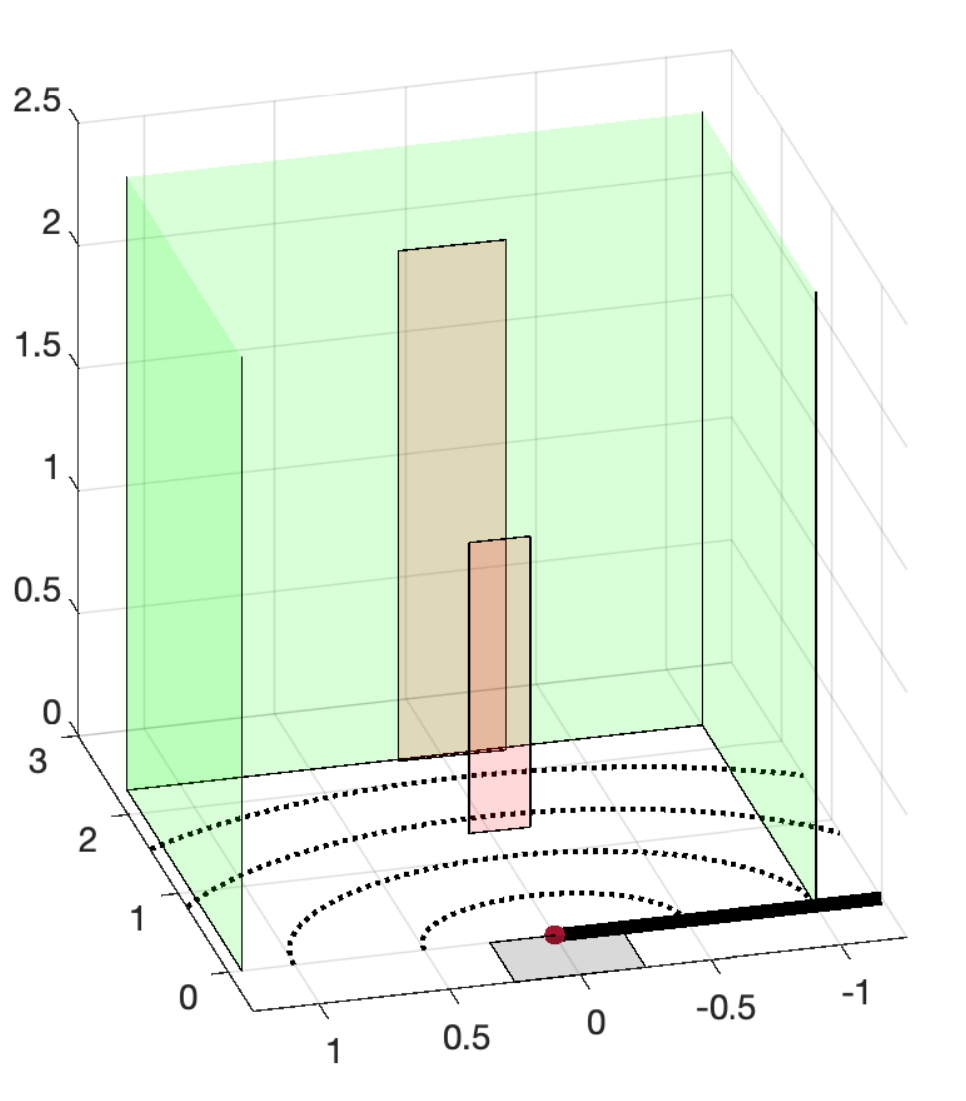}
\end{minipage}
\begin{minipage}{0.155\textwidth}
        \centering
        \includegraphics[width=1\linewidth]{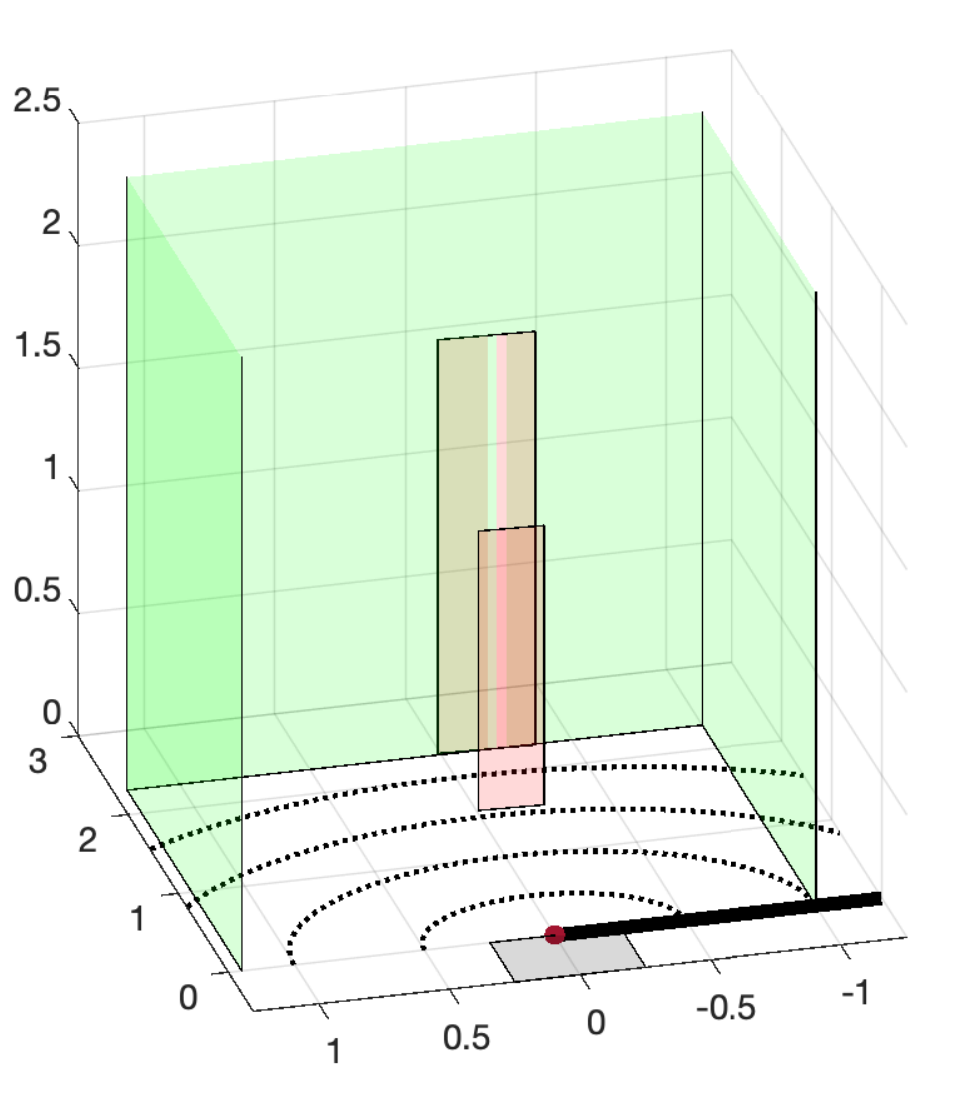}
\end{minipage}
\begin{minipage}{0.155\textwidth}
        \centering
        \includegraphics[width=1\linewidth]{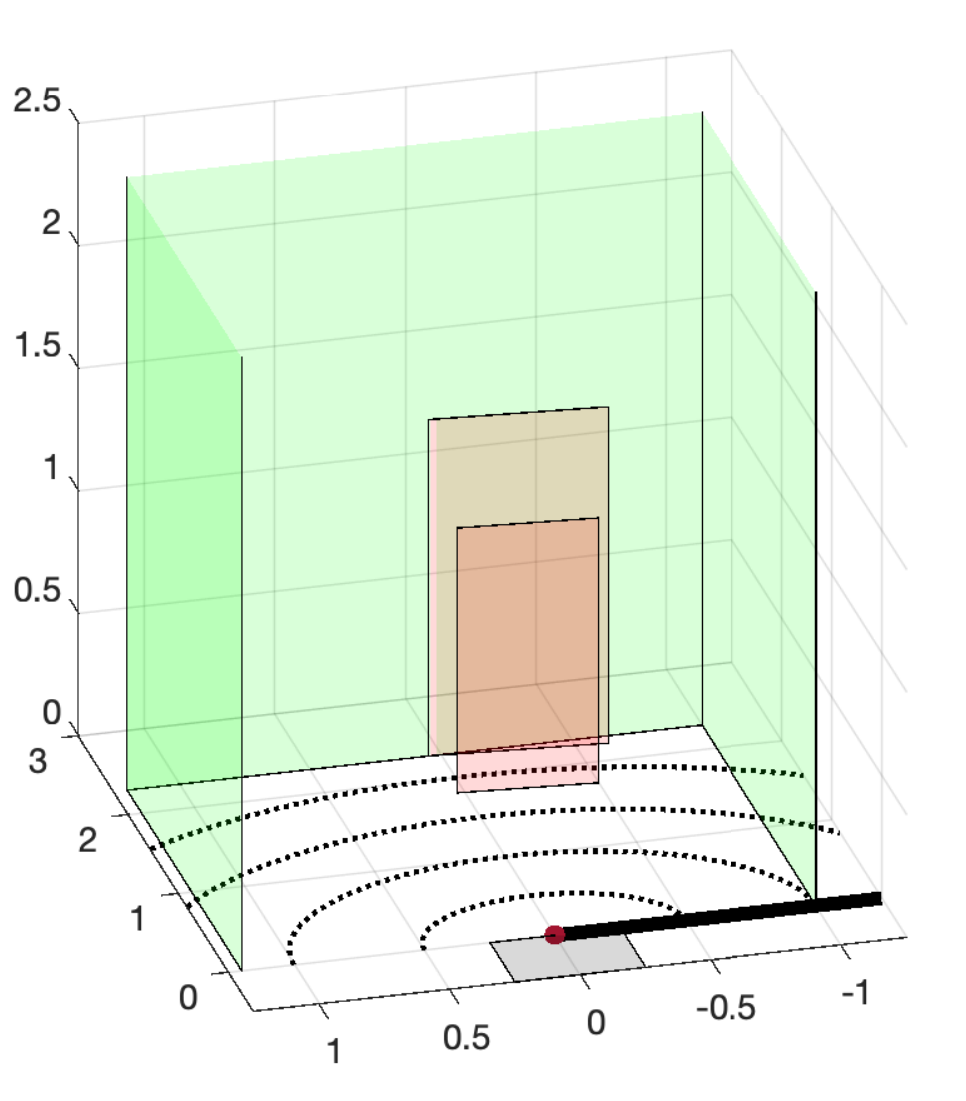}
\end{minipage}

\caption{\label{fig:differentRangeFacet} Hidden scene reconstructions for a object fixed in angle at $\pi/2$ (measured around the occluding edge and into the hidden scene) and moved through six different positions in range.
The first figure row shows a reference LOS photograph of the hidden scene; the second row shows the single-frame reconstruction results.
In the reconstructions, the true location of the back wall is shown in green with foreground and background reconstructions shown in red.
Foreground and background reconstructions closely match the ground truth in all cases, with more error in the last column when the moving object is very close to the back wall.
Each measurement frame had a 0.4\,s integration period, with the reference measurement integrated over 32.6\,s.
}
\end{figure}

The results in \Cref{fig:angleSweep,fig:angleSweepIntegrationTime} were for a single object fixed in range, and moving in angle around the occluding edge.
In \Cref{fig:differentRangeFacet}, we fix a object in angle at $\pi/2$ and move it in range away from the occluding edge to demonstrate our algorithm on a variety of object positions within the hidden scene.
Each figure column corresponds to a different object range (labeled above).
The first row shows LOS photographs of the hidden scene in each case, and the second row shows our corresponding reconstructions.
Because the object is not moving in angle, we do not attempt to process multiple frames into a combined background reconstruction.
Instead, we plot the single-frame reconstruction results for each object position.
Reconstruction results for the first five object positions closely match the ground truth.
When the object reaches 1.75\,m in range from from the edge, it is extremely close to the back wall and the reconstruction quality starts to degrade slightly, although the ranges of both foreground and background reconstructions are still very accurate.
Because parts of the planar facet at 1.75\,m share round-trip travel times with the back wall, separately estimating foreground and background parameters is less justified and there is less total change in the measurement due to the object.

\subsection{Two-object demonstration}
\label{app:two_target}

\begin{figure}
\begin{minipage}{0.015\textwidth}
        \begin{sideways}
            \text{\normalsize \textbf{ground truth}}
        \end{sideways}
\end{minipage}
\begin{minipage}{.135\textwidth}
        \centering
        \text{\textbf{Frame 1}}\\
        \includegraphics[width=1\linewidth]{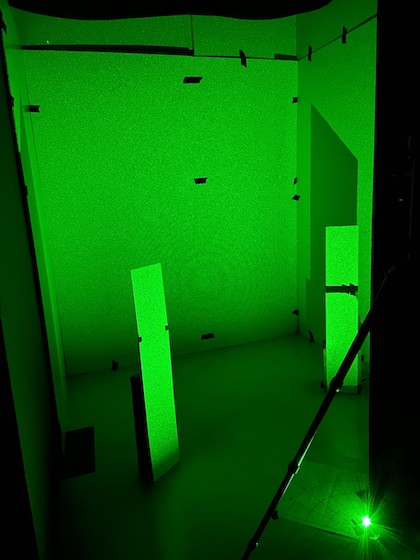}
\end{minipage}
\begin{minipage}{.135\textwidth}
        \centering
        \text{\textbf{Frame 2}}\\
        \includegraphics[width=1\linewidth]{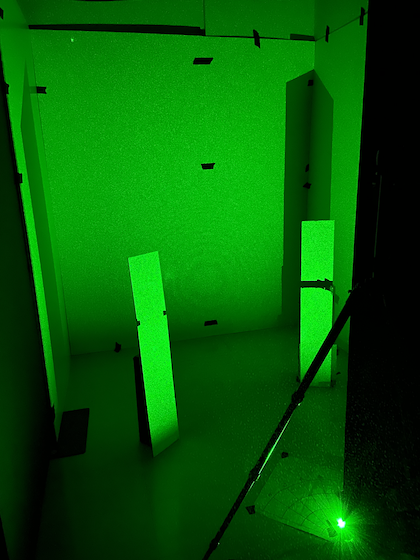}
\end{minipage}
\begin{minipage}{.135\textwidth}
        \centering
        \text{\textbf{Frame 3}}\\
        \includegraphics[width=1\linewidth]{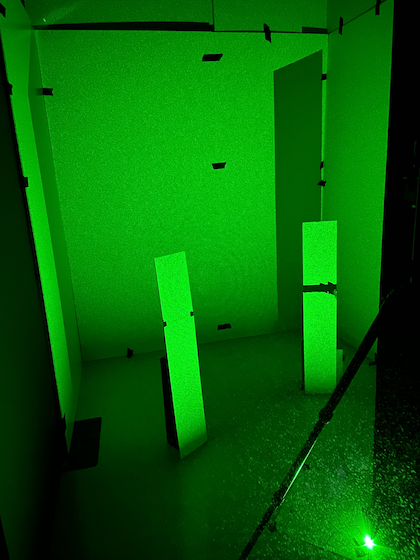}
\end{minipage}
\begin{minipage}{.135\textwidth}
        \centering
        \text{\textbf{Frame 4}}\\
        \includegraphics[width=1\linewidth]{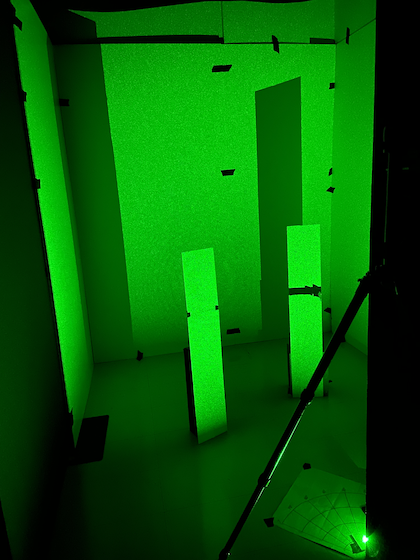}
\end{minipage}
\begin{minipage}{.135\textwidth}
        \centering
        \text{\textbf{Frame 5}}\\
        \includegraphics[width=1\linewidth]{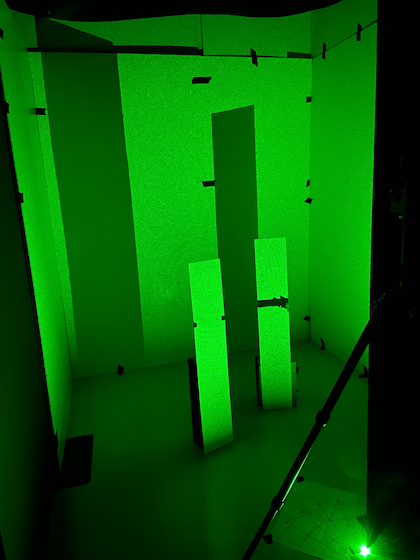}
\end{minipage}
\begin{minipage}{.135\textwidth}
        \centering
        \text{\textbf{Frame 6}}\\
        \includegraphics[width=1\linewidth]{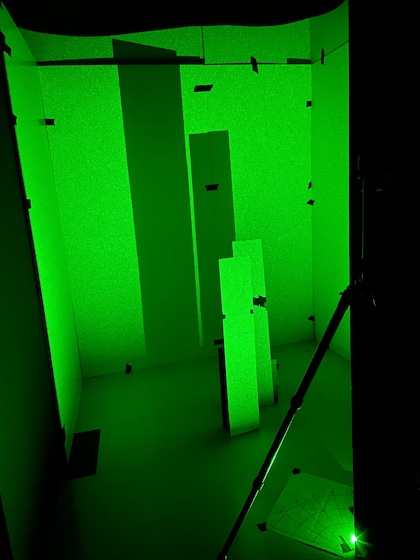}
\end{minipage}
\begin{minipage}{.135\textwidth}
        \centering
        \text{\textbf{Frame 7}}\\
        \includegraphics[width=1\linewidth]{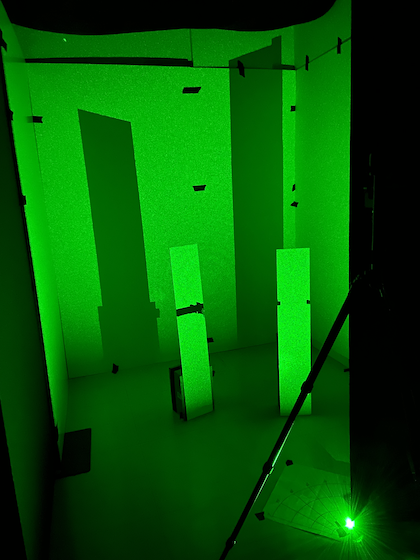}
\end{minipage}
\\
\begin{minipage}{0.015\textwidth}
        \begin{sideways}
            \text{\normalsize \textbf{reconstruction}}
        \end{sideways}
\end{minipage}
\begin{minipage}{.135\textwidth}
        \centering
        \includegraphics[width=1\linewidth]{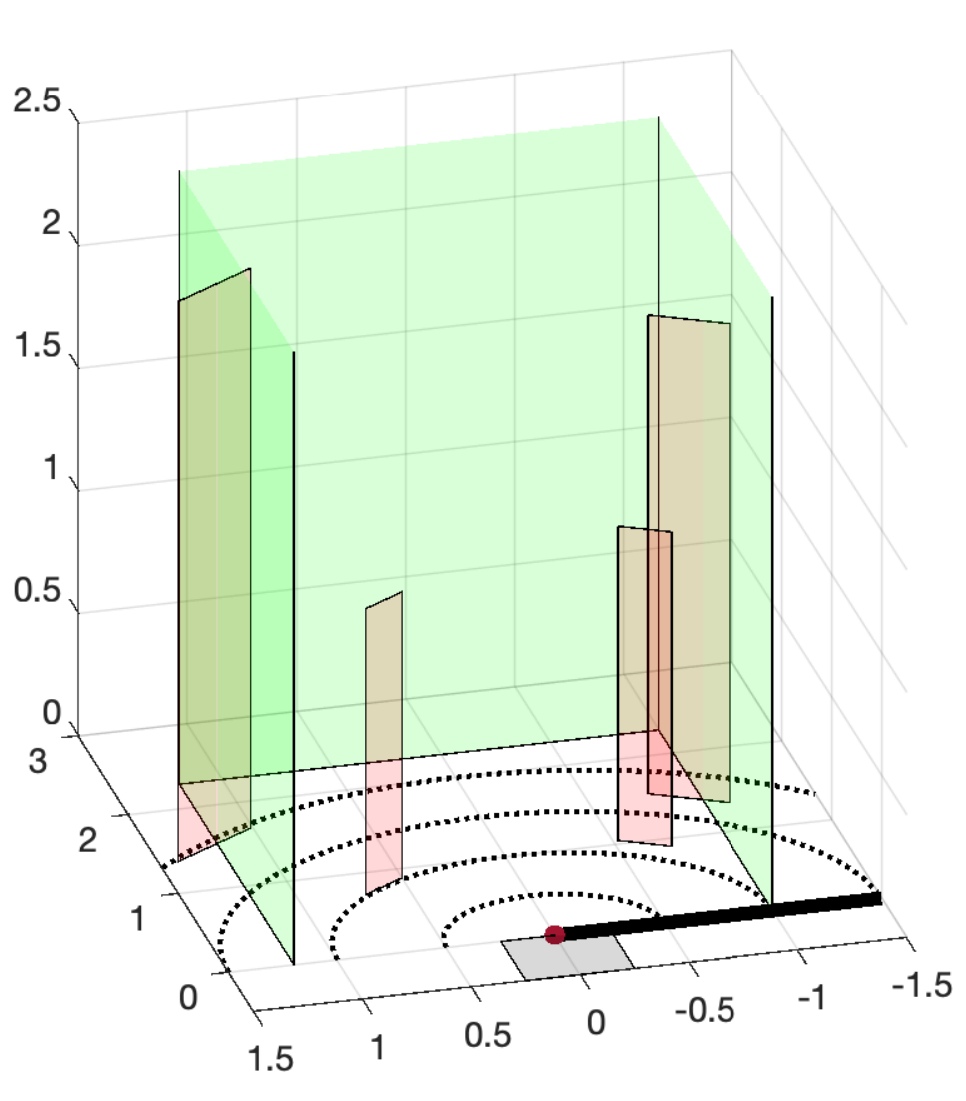}
\end{minipage}
\begin{minipage}{.135\textwidth}
        \centering
        \includegraphics[width=1\linewidth]{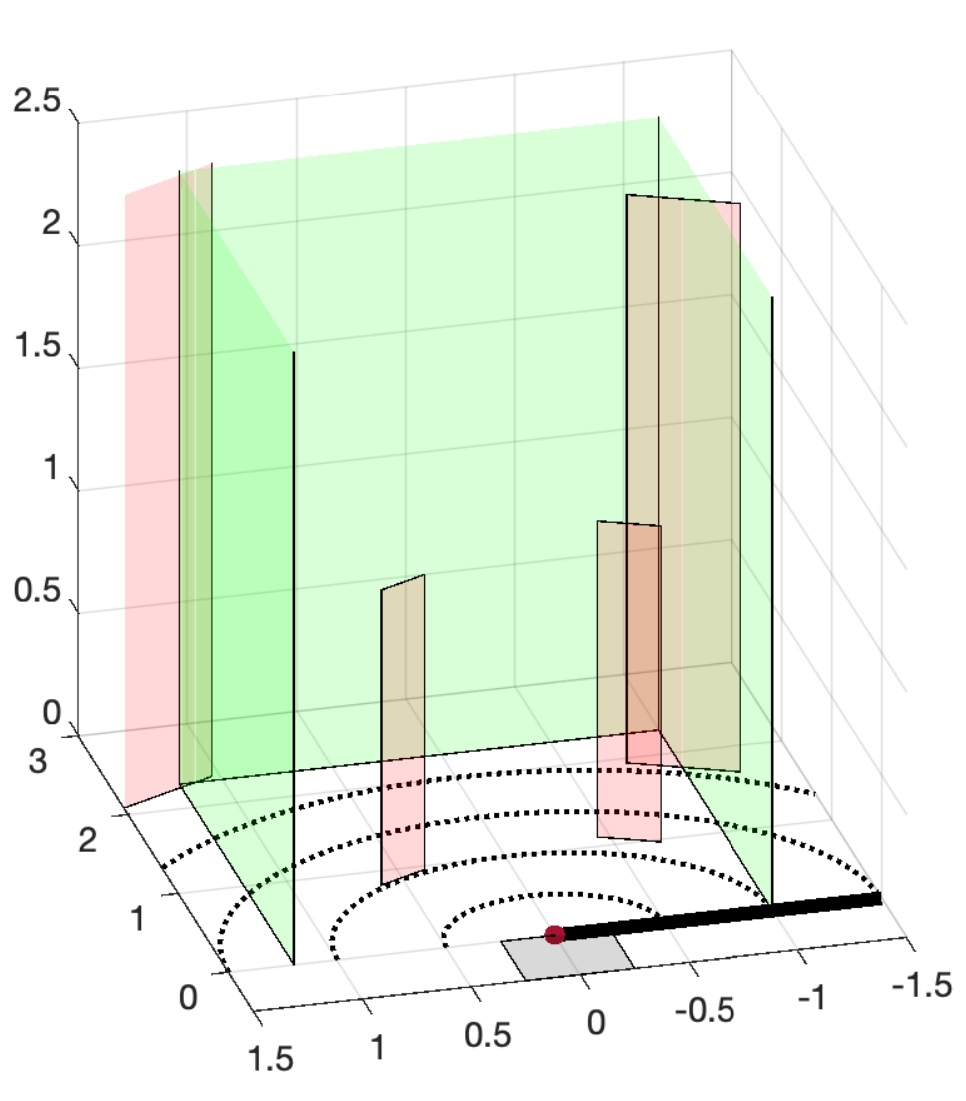}
\end{minipage}
\begin{minipage}{.135\textwidth}
        \centering
        \includegraphics[width=1\linewidth]{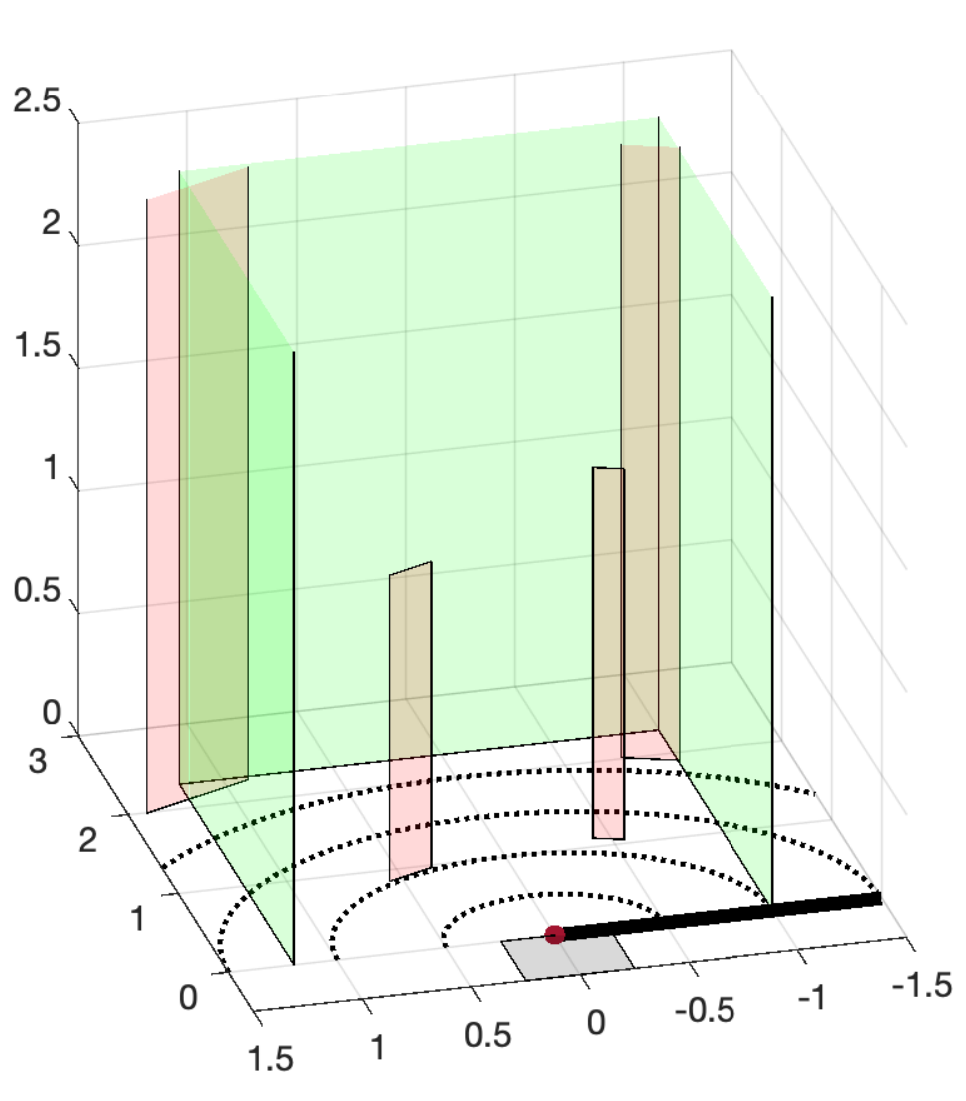}
\end{minipage}
\begin{minipage}{.135\textwidth}
        \centering
        \includegraphics[width=1\linewidth]{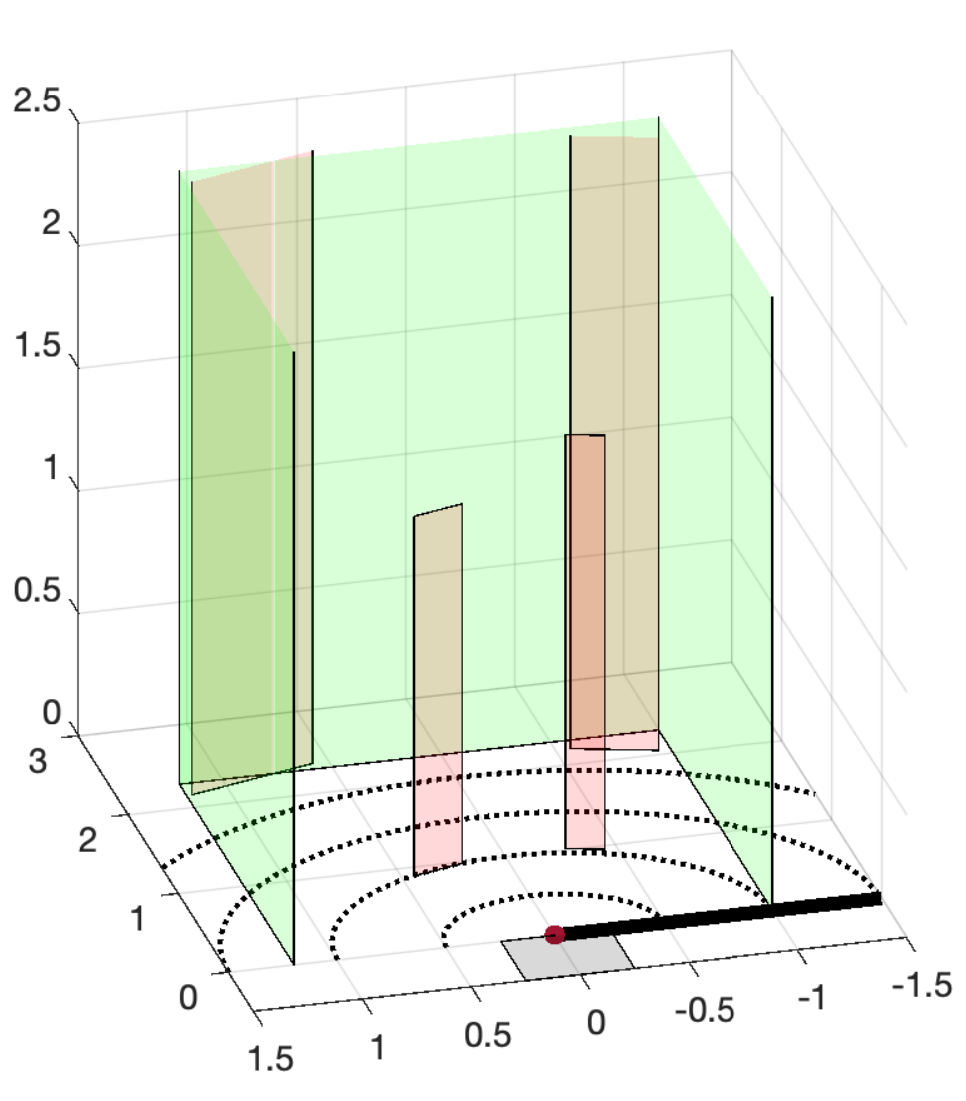}
\end{minipage}
\begin{minipage}{.135\textwidth}
        \centering
        \includegraphics[width=1\linewidth]{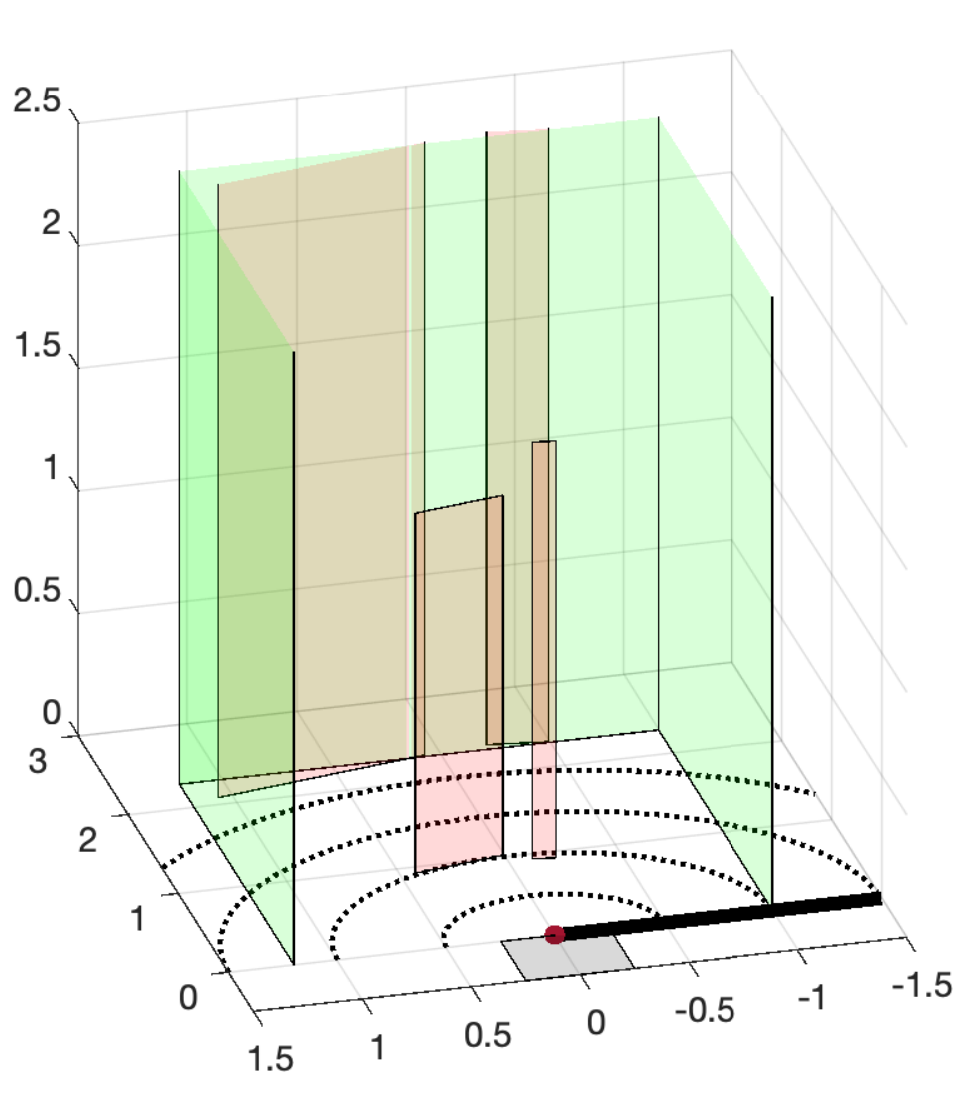}
\end{minipage}
\begin{minipage}{.135\textwidth}
        \centering
        \includegraphics[width=1\linewidth]{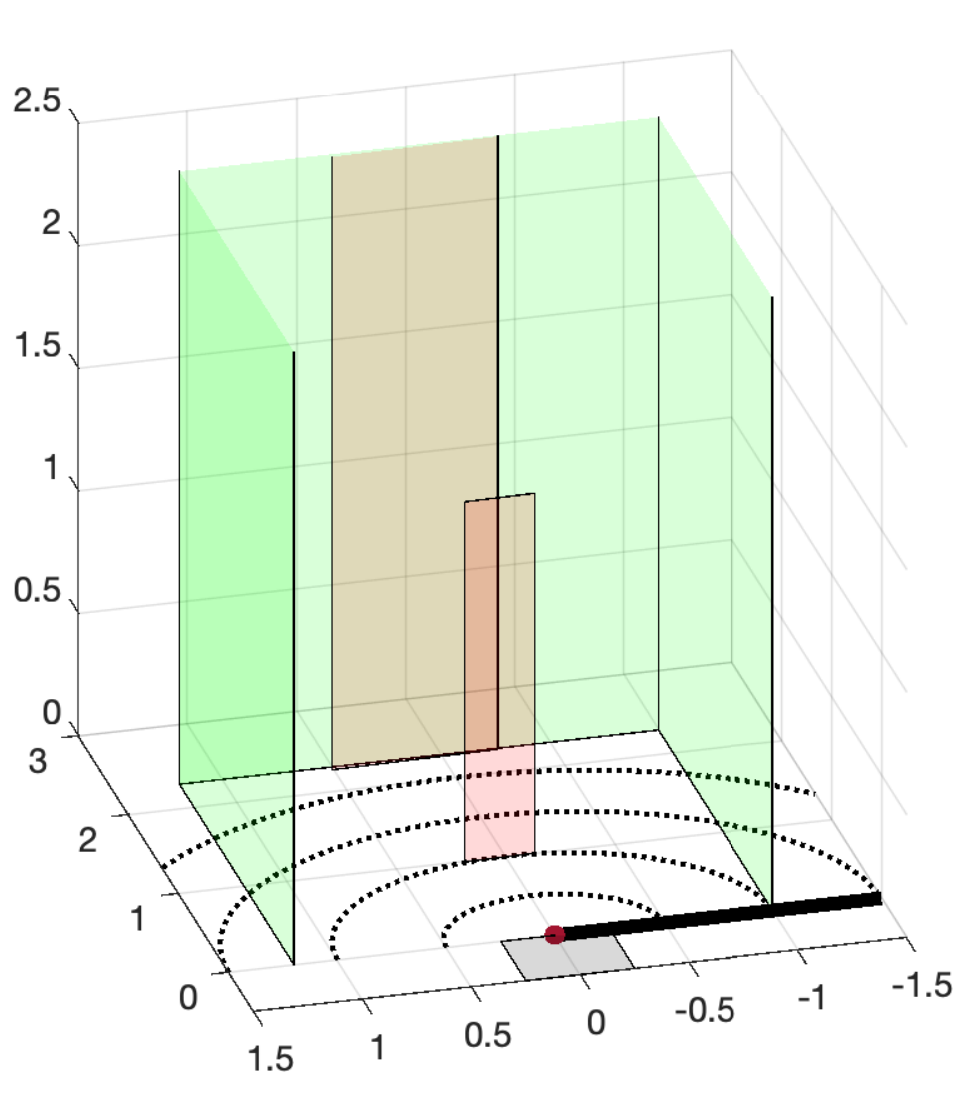}
\end{minipage}
\begin{minipage}{.135\textwidth}
        \centering
        \includegraphics[width=1\linewidth]{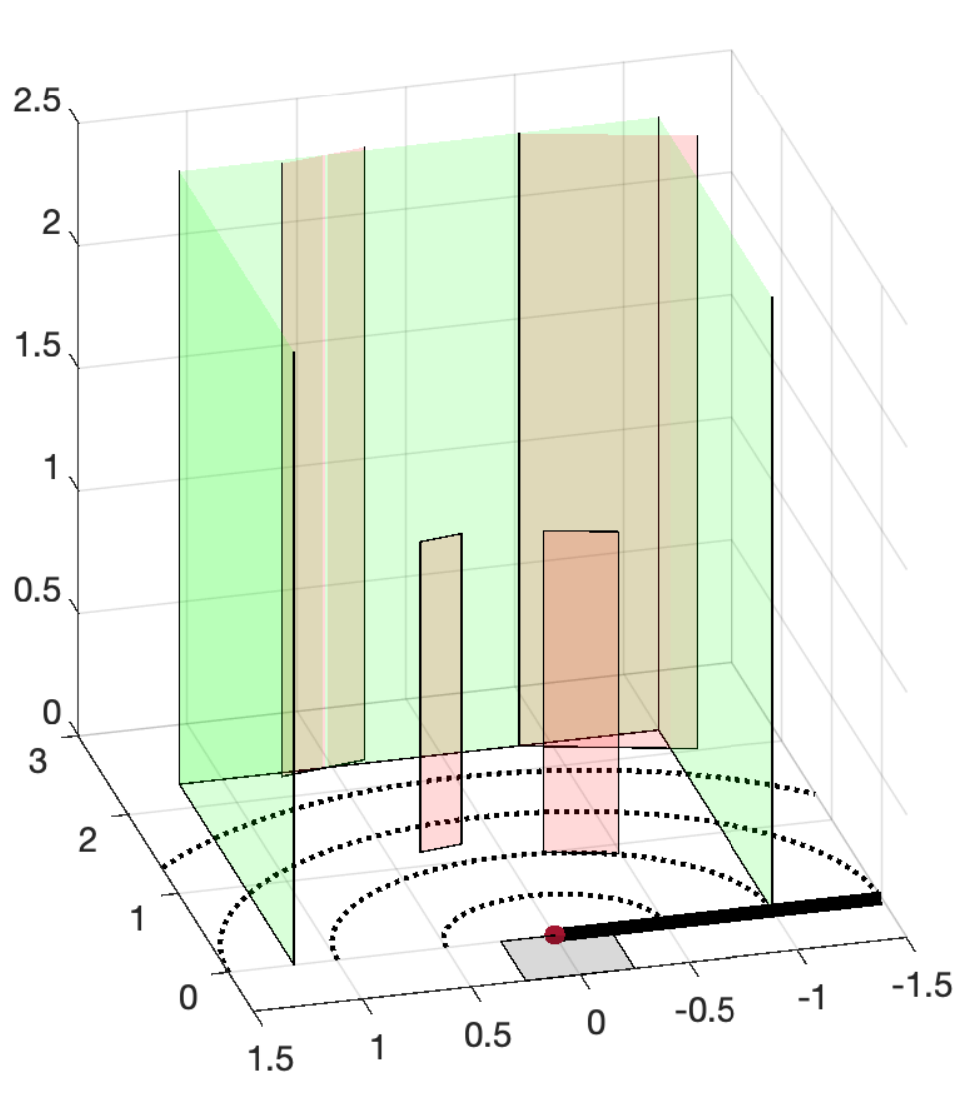}
\end{minipage}
\caption{Single-frame reconstruction results for seven different frames of a two-object scenario.
The first figure row shows LOS photographs of the ground truth; the second row shows single-frame reconstruction results.
The true location of the back wall is shown in green with foreground and background reconstructions shown in red.
Each measurement frame had a 0.4\,s integration period, with the reference measurement integrated over 32.6\,s.
}
\label{fig:two_target_scenario}
\end{figure}

\begin{figure}
    \centering
    \begin{subfigure}{0.35\textwidth}
    \includegraphics[width=1\linewidth]{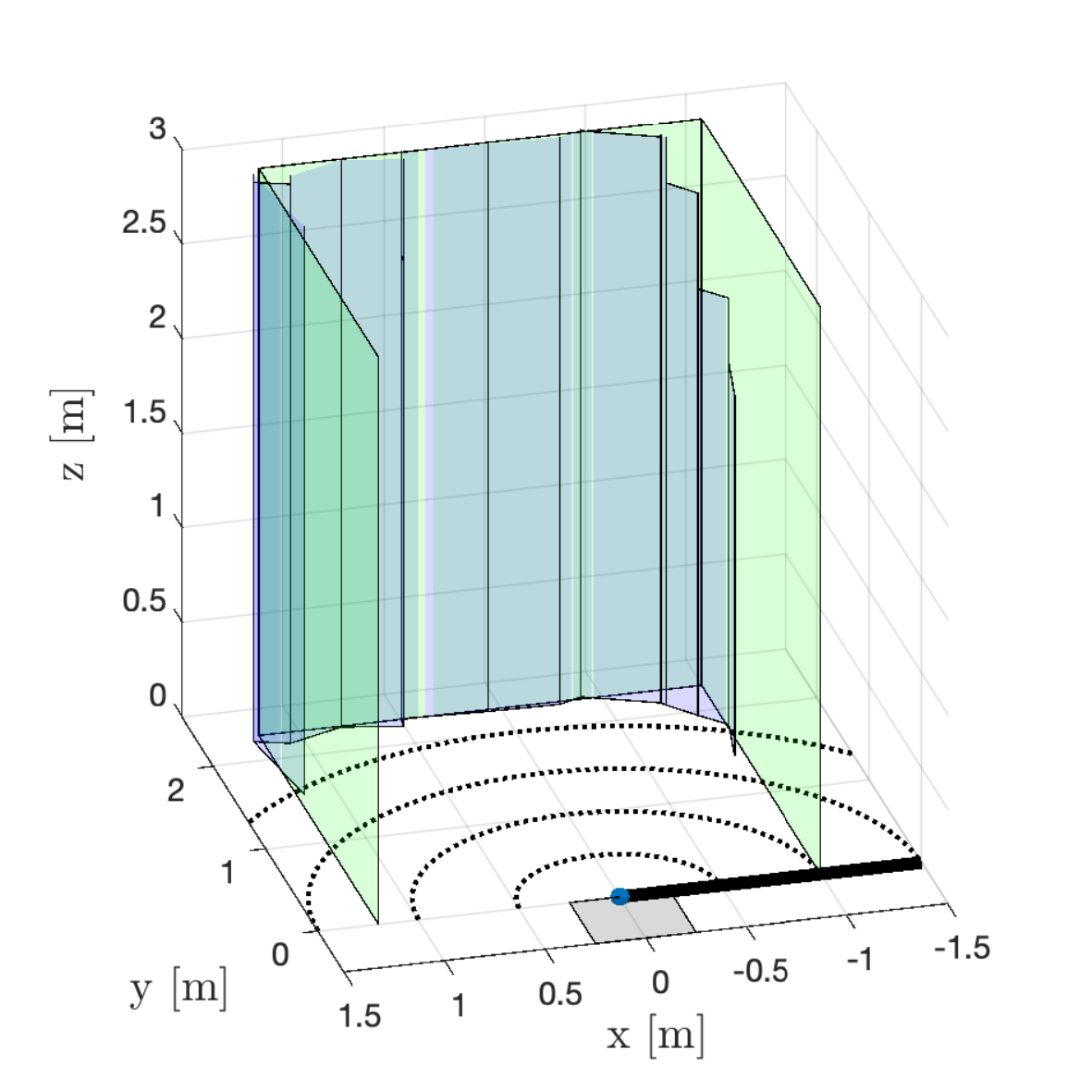}
    \caption{\label{fig:twoTarg_wallCompositeView1} Side view}
    \end{subfigure}
    \begin{subfigure}{0.35\textwidth}
    \includegraphics[width=1\linewidth]{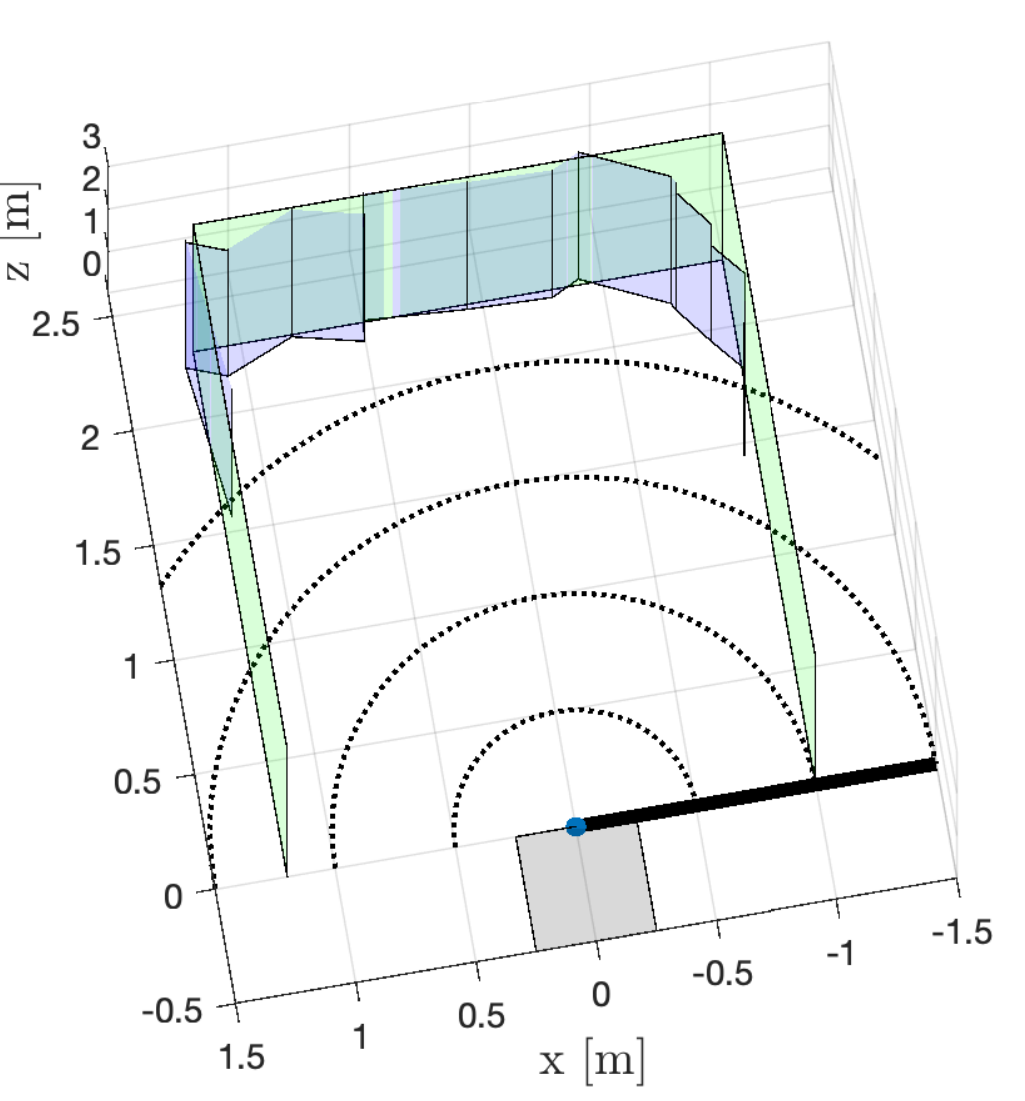}
    \caption{\label{fig:twoTarg_wallCompositeView2} Bird's eye view}
    \end{subfigure}
    \caption{\label{fig:twoTarg_wallComposite} Two views of the \textit{combined} wall reconstruction formed using the seven two-object frames in \Cref{fig:two_target_scenario}. The true location of the back wall is shown in green with stationary scene reconstructions shown in blue.
    Each measurement frame had a 0.4\,s integration period, with the reference measurement integrated over 32.6\,s.}
\end{figure}

In \Cref{fig:two_target_scenario,fig:twoTarg_wallComposite},
we show more detailed reconstruction results for the two-object scenario in the main document.
Columns of \Cref{fig:two_target_scenario} correspond to seven measurement frames as the objects move towards and then past each other.
The object that starts on the left in Frame~1 is fixed at a range of 1\,m;
the object that starts on the right side is slightly further away at 1.25\,m.
LOS ground truth photographs are shown in the first row, with single-frame reconstruction results shown in the second row.
For increased legibility, the $z$-axis in the reconstructed frames is cropped at 2.5\,m, although the true ceiling height is 3\,m.
In all but Frame~6, two objects are correctly resolved and placed with very little error in range.
There is some variability in object height estimates, most notably in Frames 4~and~5\@.
In Frame~6, when the left object begins to cross in front of the right object, our algorithm resolves a single object.
In all result frames, the background estimates closely match the ground truth.
In \Cref{fig:twoTarg_wallComposite}, we show two views of the combined background reconstruction, formed by accumulating a total of 13 background estimates (two each from Frames 1--5 and 7, and one from Frame~6).
The estimate (blue) is extremely close to the measured ground truth (green).

\subsection{Robustness demonstrations}
\label{app:robustness_testing}
\begin{figure*}
\centering
\begin{minipage}{0.015\textwidth}
        \begin{sideways}
            \text{\normalsize \textbf{ground truth}}
        \end{sideways}
\end{minipage}
\begin{minipage}{0.22\textwidth}
        \centering
        \text{\textbf{\normalsize white facet}}\\
        \includegraphics[width=.75\textwidth]{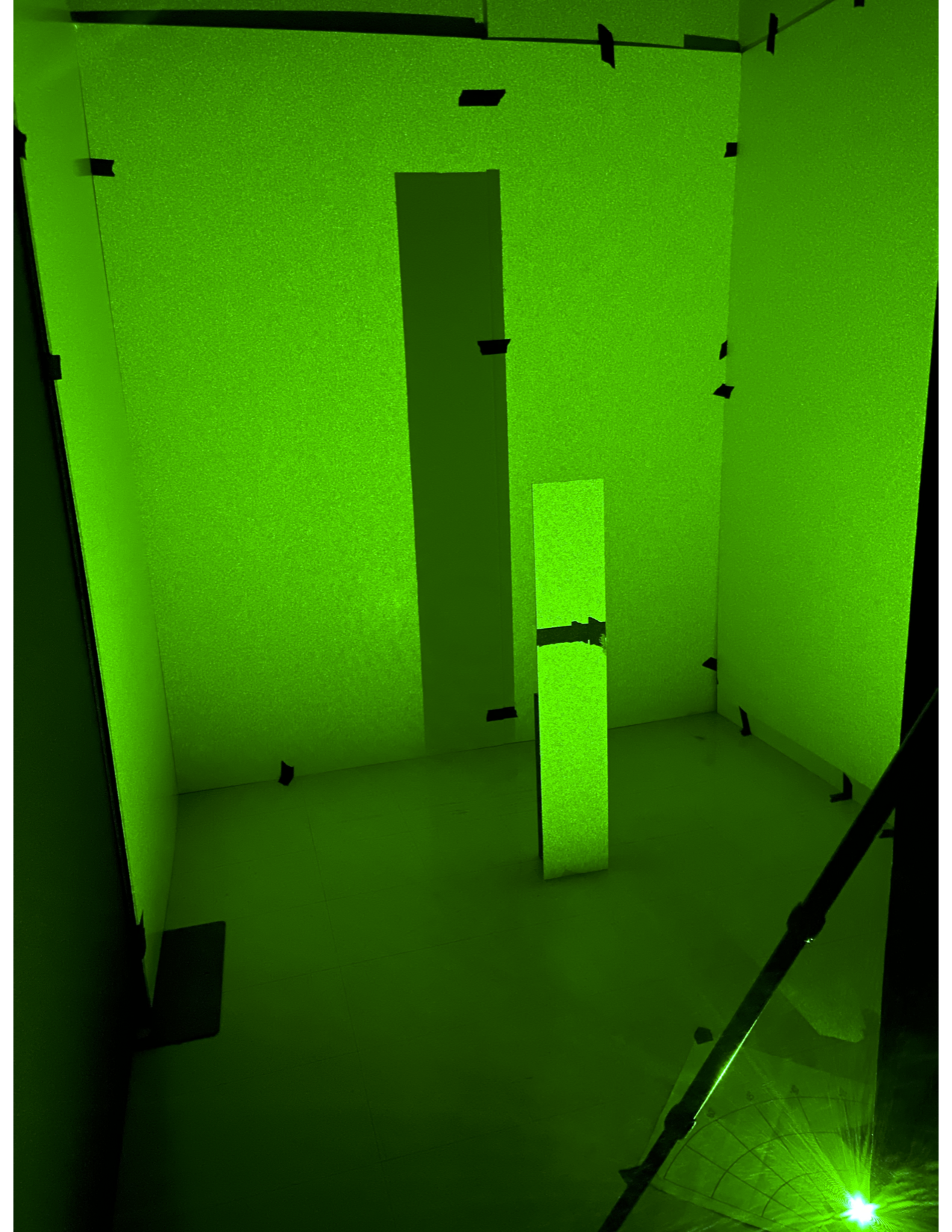}
\end{minipage}
\begin{minipage}{0.22\textwidth}
        \centering
        \text{\textbf{\normalsize gray facet}}\\
        \includegraphics[width=.75\textwidth]{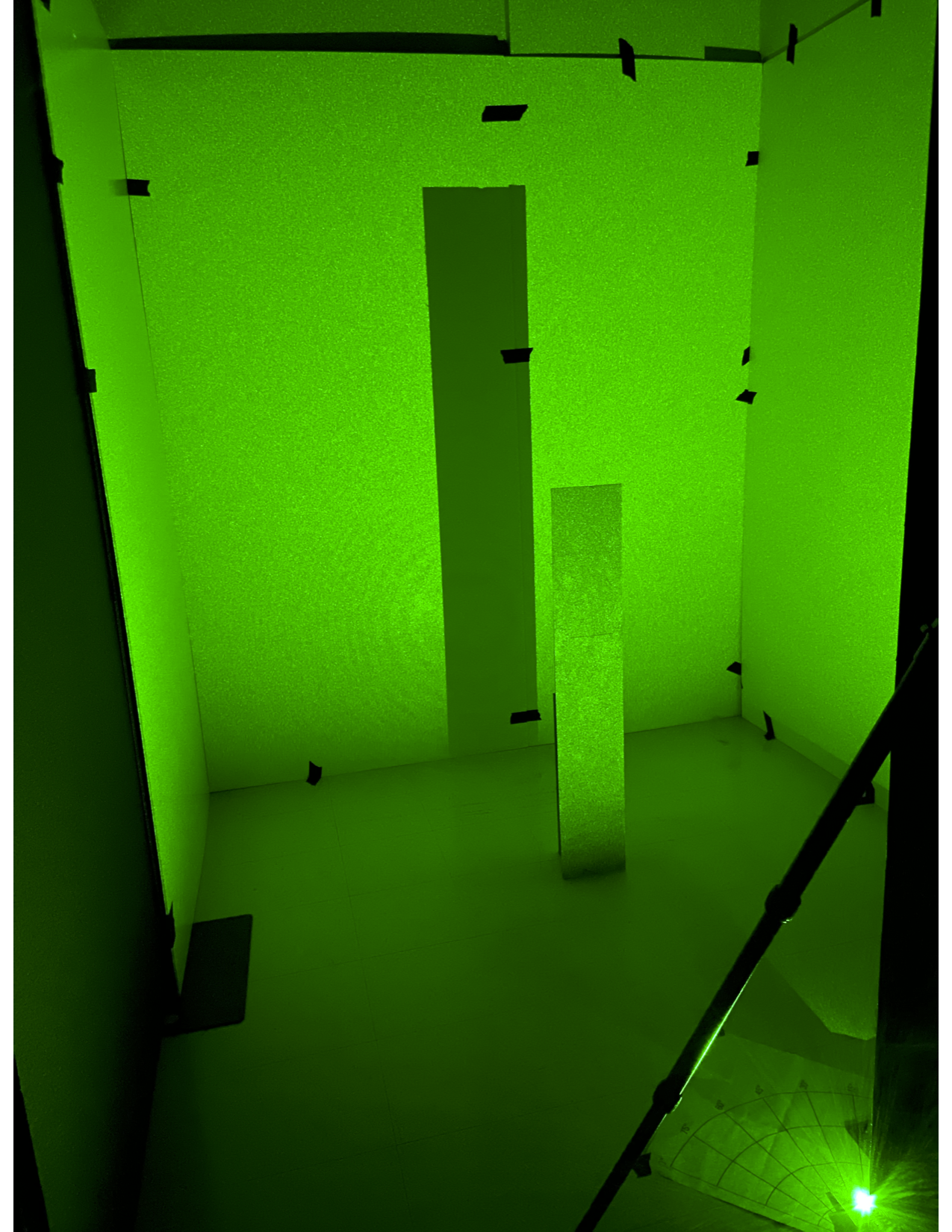}
\end{minipage}
\begin{minipage}{0.22\textwidth}
        \centering
        \text{\textbf{\normalsize mannequin}}\\
        \includegraphics[width=.75\textwidth]{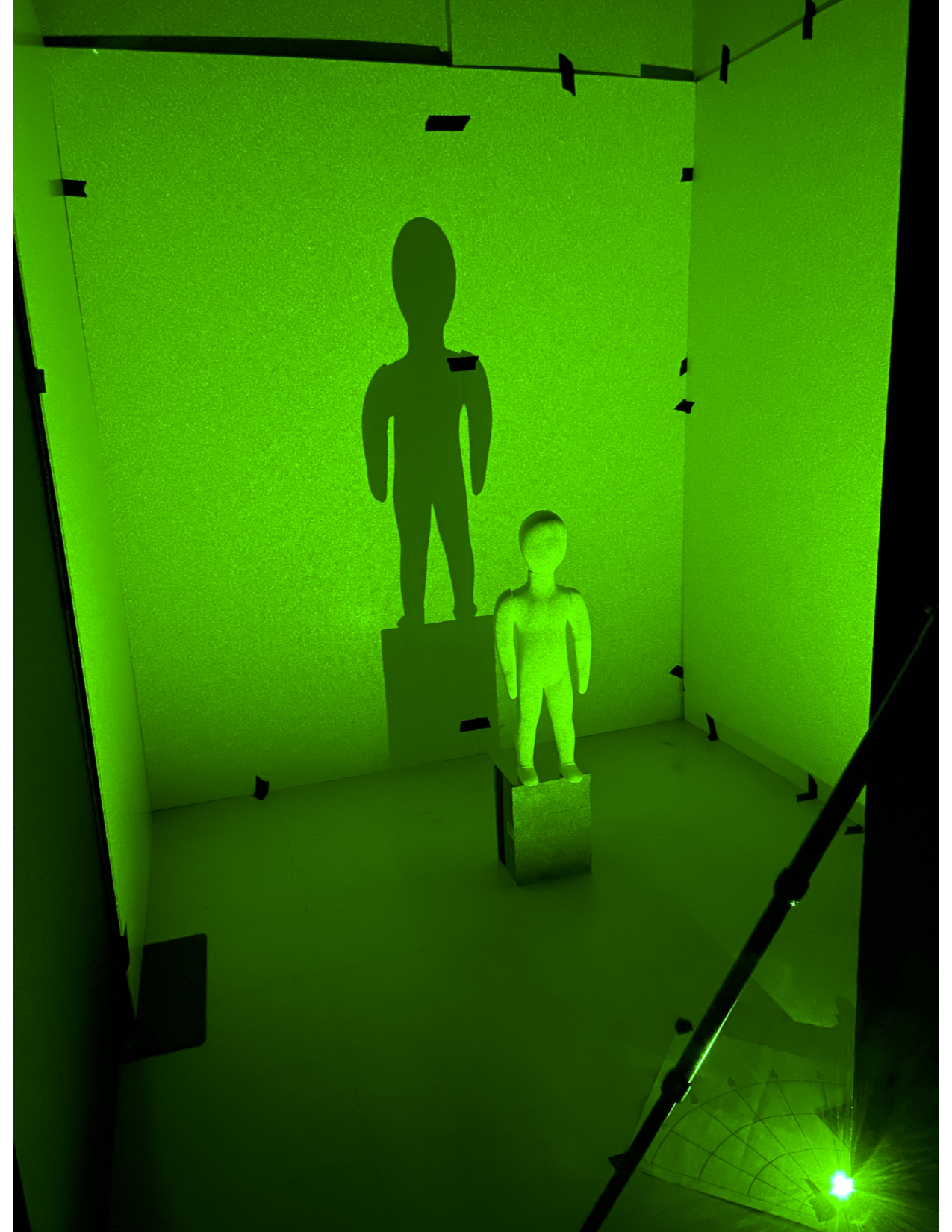}
\end{minipage}
\begin{minipage}{0.22\textwidth}
        \centering
        \text{\textbf{\normalsize stairs}}\\
        \includegraphics[width=.75\textwidth]{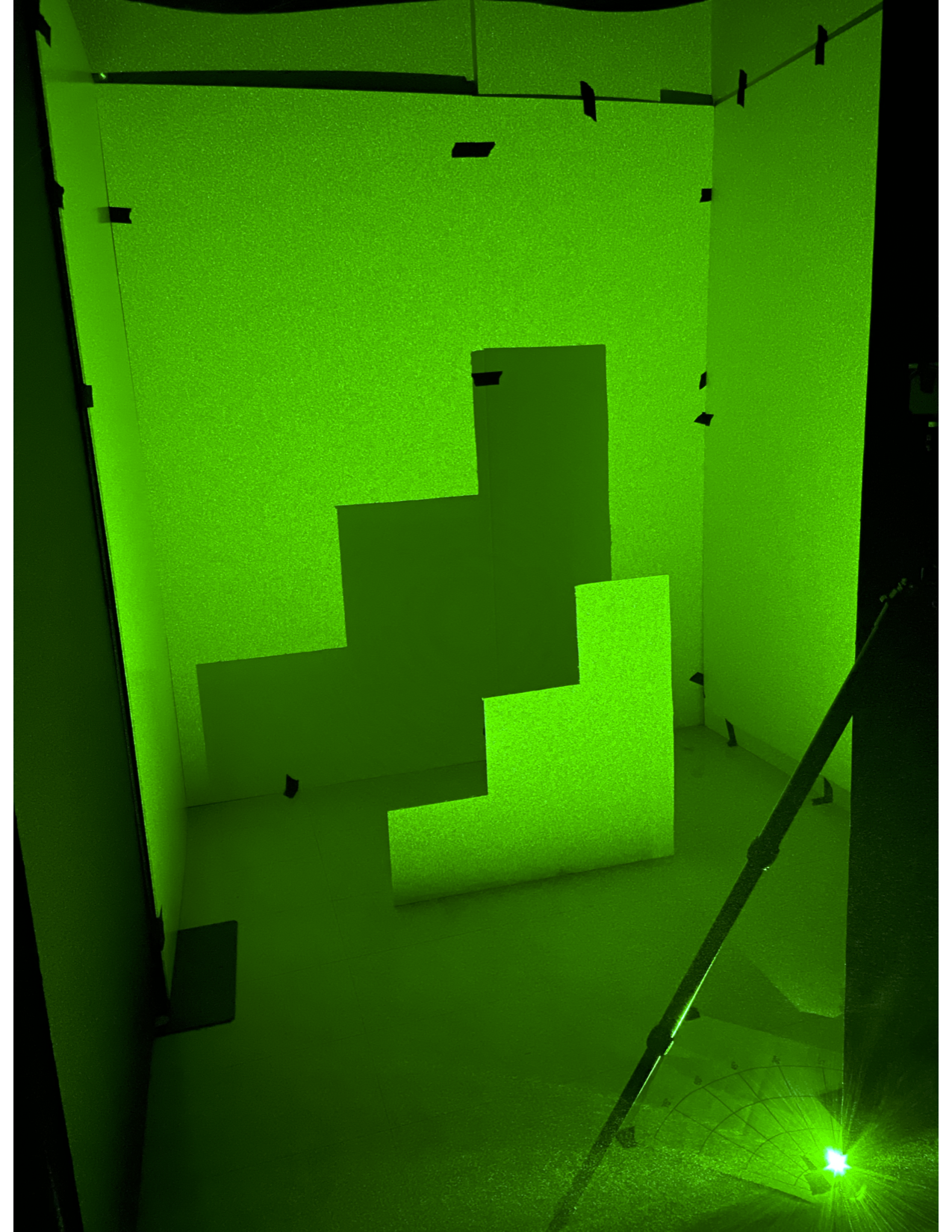}
\end{minipage}\\
\begin{minipage}{0.015\textwidth}
        \begin{sideways}
            \text{\normalsize \textbf{reconstruction}}
        \end{sideways}
\end{minipage}
\begin{minipage}{0.22\textwidth}
        \centering
        \includegraphics[width=1\linewidth]{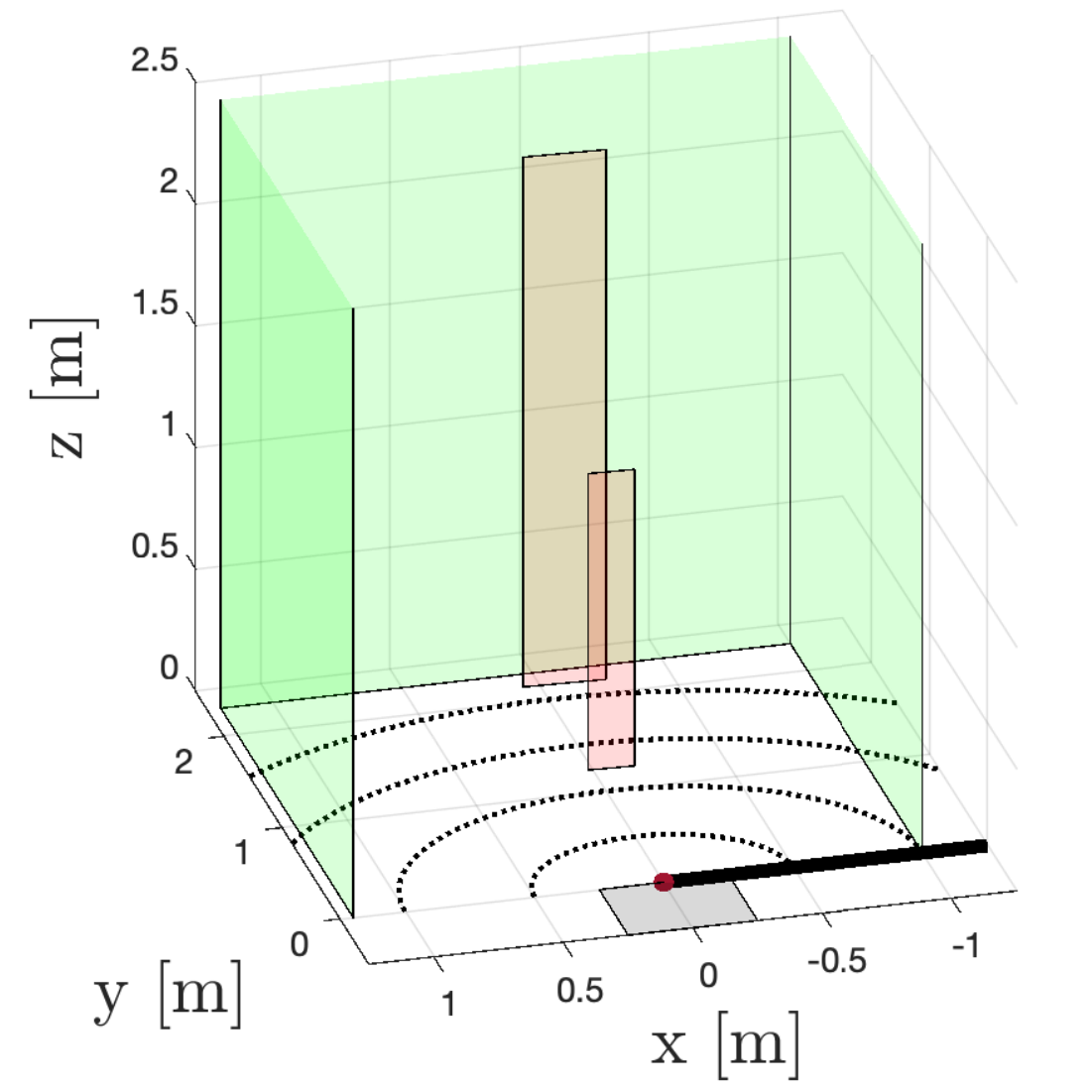}
\end{minipage}
\begin{minipage}{0.22\textwidth}
        \centering
        \includegraphics[width=1\linewidth]{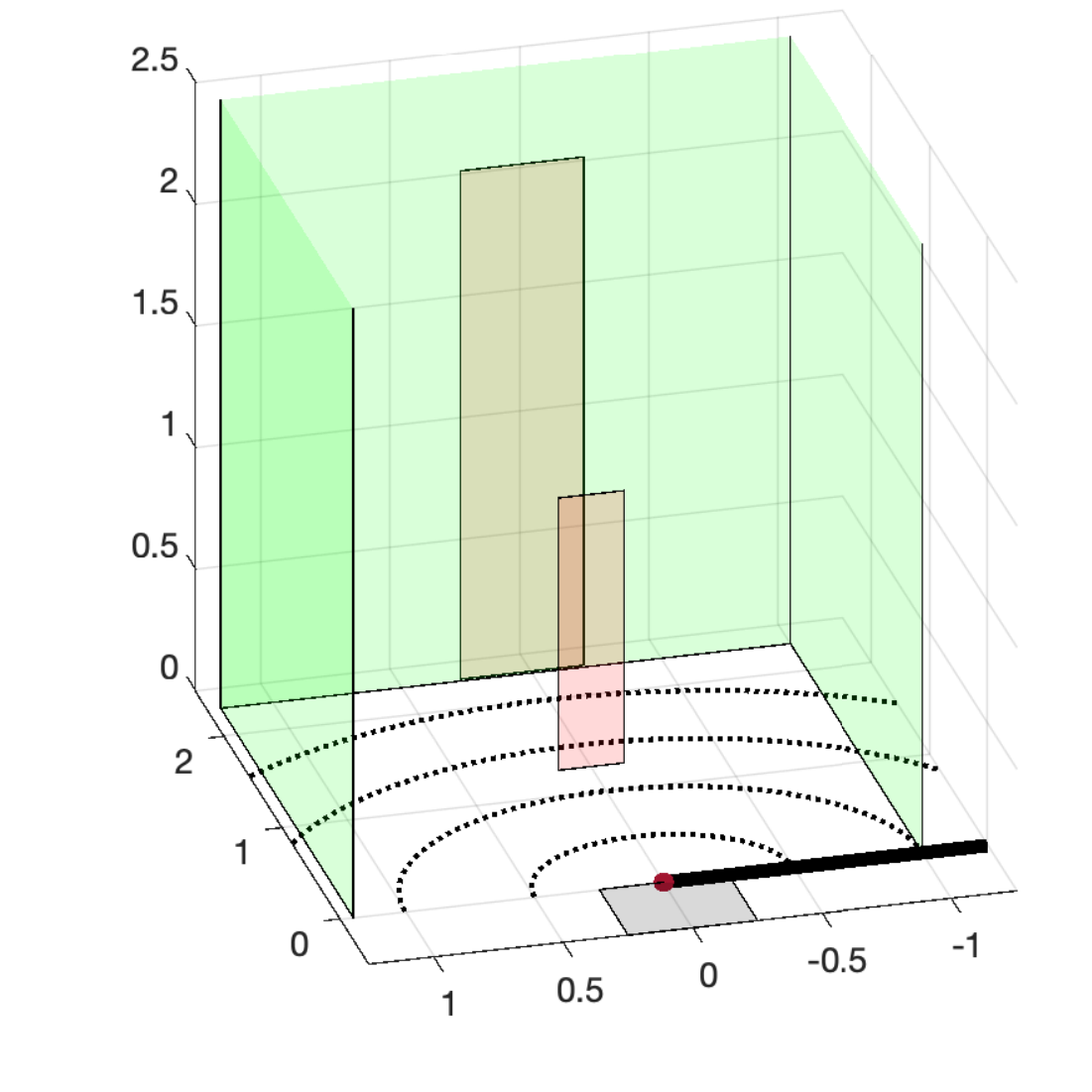}
\end{minipage}
\begin{minipage}{0.22\textwidth}
        \centering
        \includegraphics[width=1\linewidth]{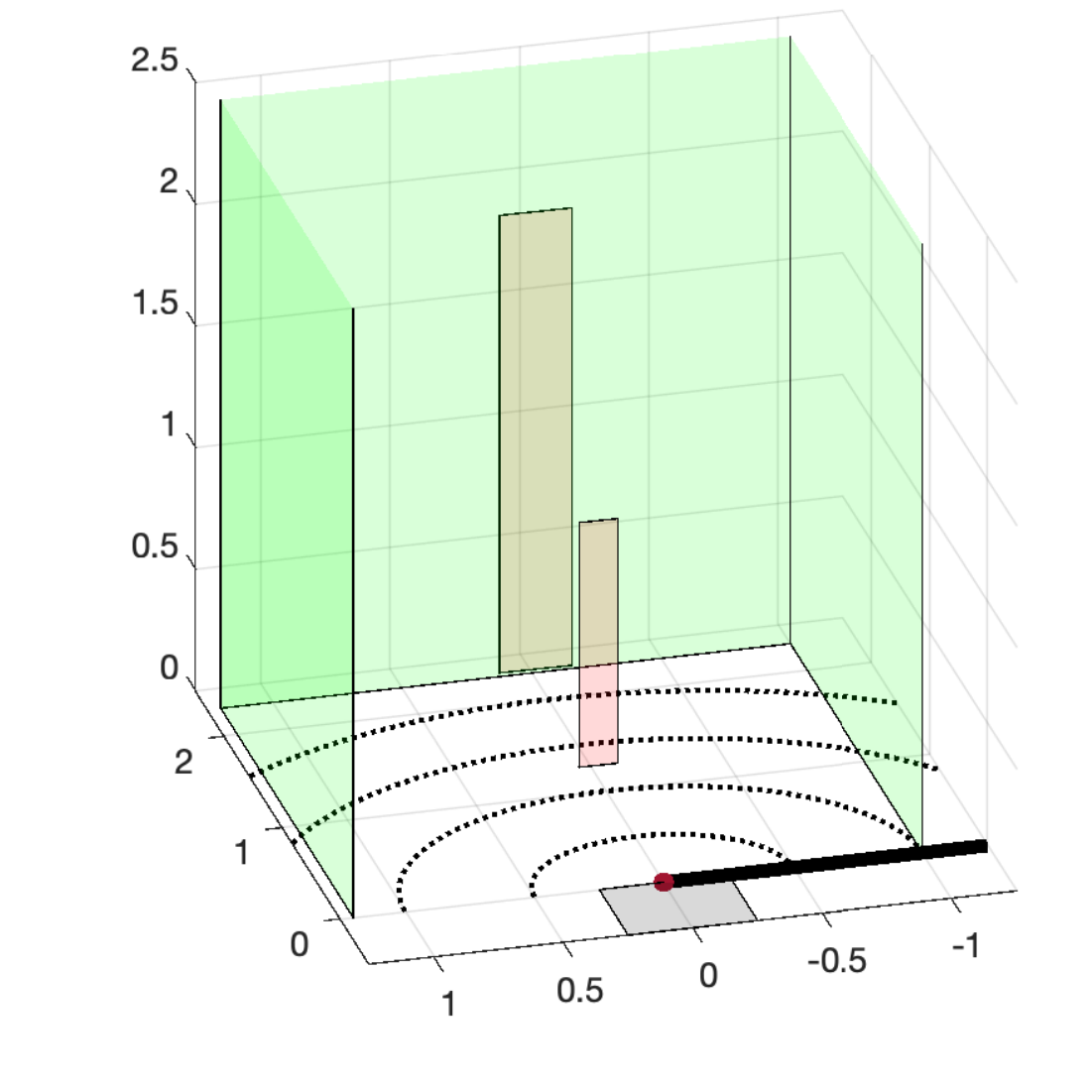}
\end{minipage}
\begin{minipage}{0.22\textwidth}
        \centering
        \includegraphics[width=1\linewidth]{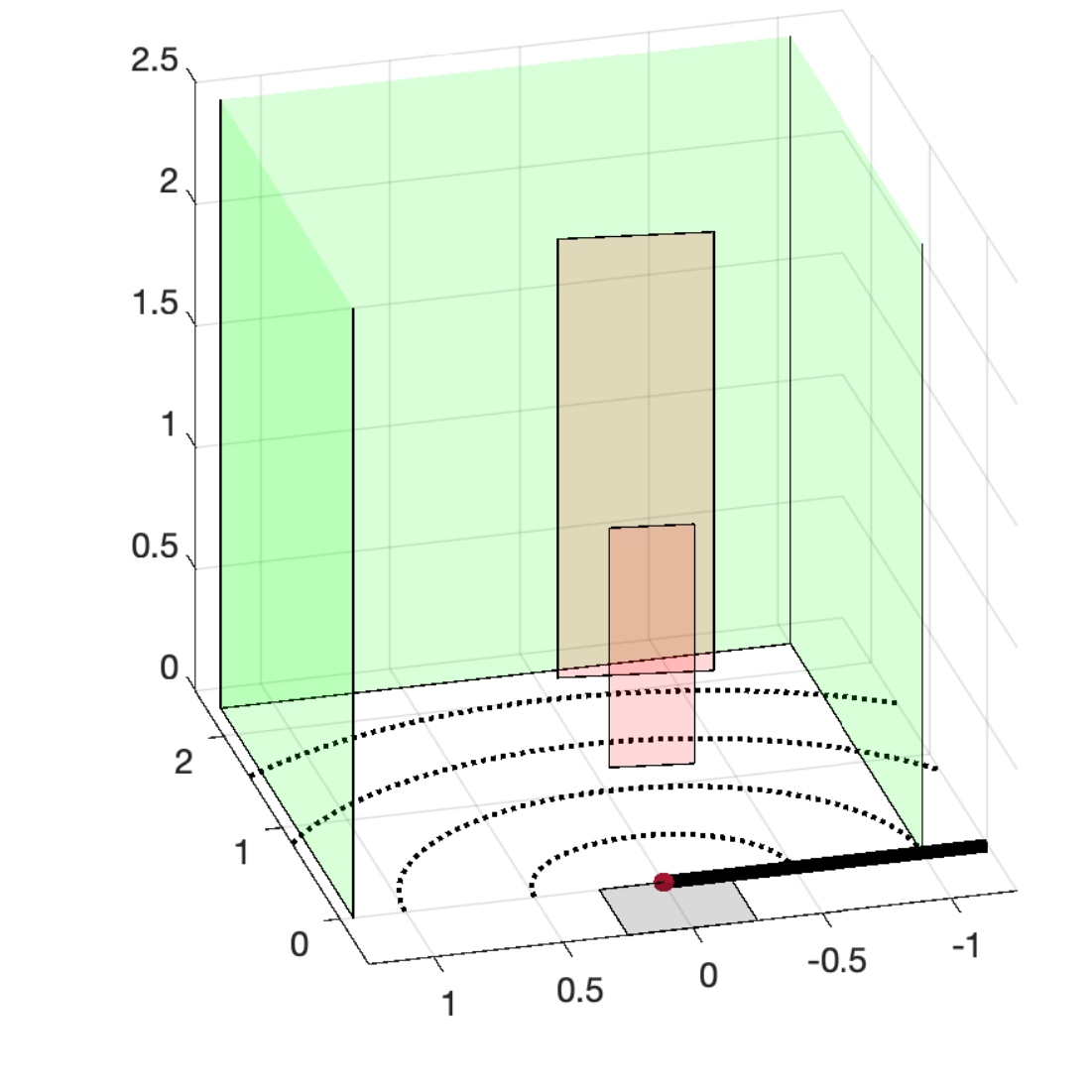}
\end{minipage}
\caption{\label{fig:differentTargets} A demonstration of algorithm robustness to different object types at a distance of 1.25\,m from the occluding edge.
The four columns correspond to different object types: a white facet, a dark gray facet, a mannequin, and a stair-shaped facet.
The first row shows LOS photographs of the hidden scene; the second row shows our reconstructions.
The true location of the back wall is shown in green with foreground and background reconstructions shown in red.
Although the first two objects perfectly match the rectangular, planar-facet model, the dark gray facet returns far fewer photons, and the mannequin and stairs are not rectangular facets at all.
 In all four cases, the foreground object is well fit by our facet model and the occluded background region is placed correctly in range.
Each measurement frame had a 0.4\,s integration period, with the reference measurement integrated over 33\,s.
}
\end{figure*}

In \Cref{fig:differentTargets}, we explore the effects of model mismatch on our inversion algorithm. Each figure column corresponds to a different hidden object type while rows show ground truth LOS photographs and single-frame reconstruction results.
The white and gray facet objects fit the rectangular, planar facet model, although the gray facet reflects fewer photons resulting in lower SNR\@.
The mannequin is not a facet at all, and it is meant to test the common scenario where the hidden object is a person moving through the hidden scene.
The stairs object is planar, but it is also wide and not rectangular, similar perhaps to a piece of furniture that has been relocated within the hidden scene or a moving car.
The white facet reconstruction is the most accurate of the four examples. We note that the gray facet reconstruction is slightly wider than the ground truth (and the white facet reconstruction), with error likely due to the SNR reduction.
In all four cases, the hidden object was fit by a rectangular planar facet placed correctly in range, and the range of the occluded background region is accurately recovered.
We note that although our model does not allow us to reconstruct the varying height profile of the stairs, we correctly reconstruct it to be wider and more to the right than the other hidden objects.
These results indicate that our rectangular, planar-facet model does not prevent our inversion algorithm from giving useful results under significant model mismatch.
Instead, we expect to correctly locate and roughly describe (in width and height) a variety of hidden objects while also reconstructing the hidden scene behind them.

\begin{figure*}
\centering
\begin{minipage}{0.015\textwidth}
        \begin{sideways}
            \text{\normalsize \textbf{ground truth}}
        \end{sideways}
\end{minipage}
\begin{minipage}{0.2\textwidth}
        \centering
        \text{\textbf{stationary scene}}\\
        \includegraphics[width=1\linewidth]{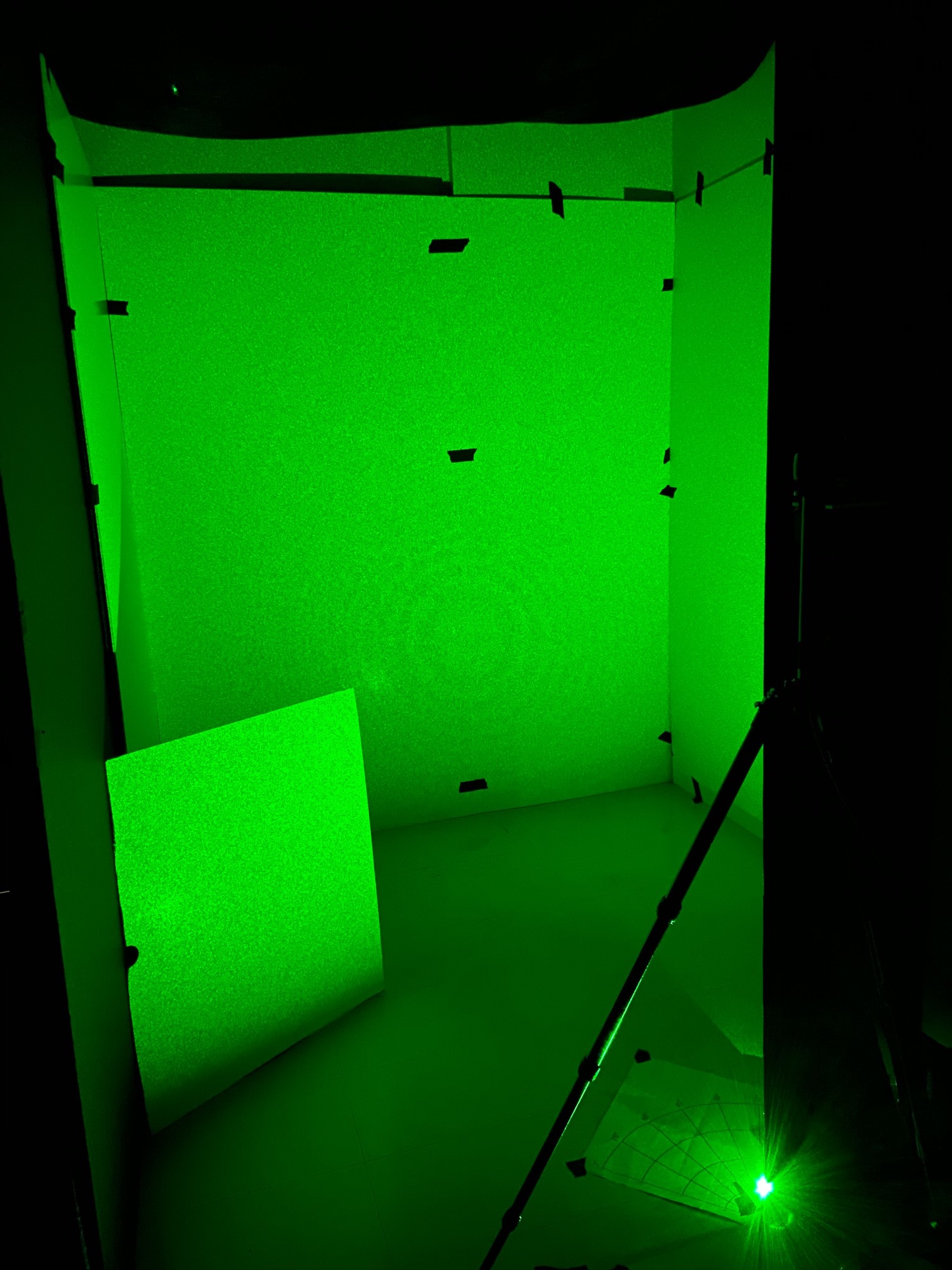}
\end{minipage}
\begin{minipage}{0.2\textwidth}
        \centering
        \text{1 meter}\\
        \includegraphics[width=1\linewidth]{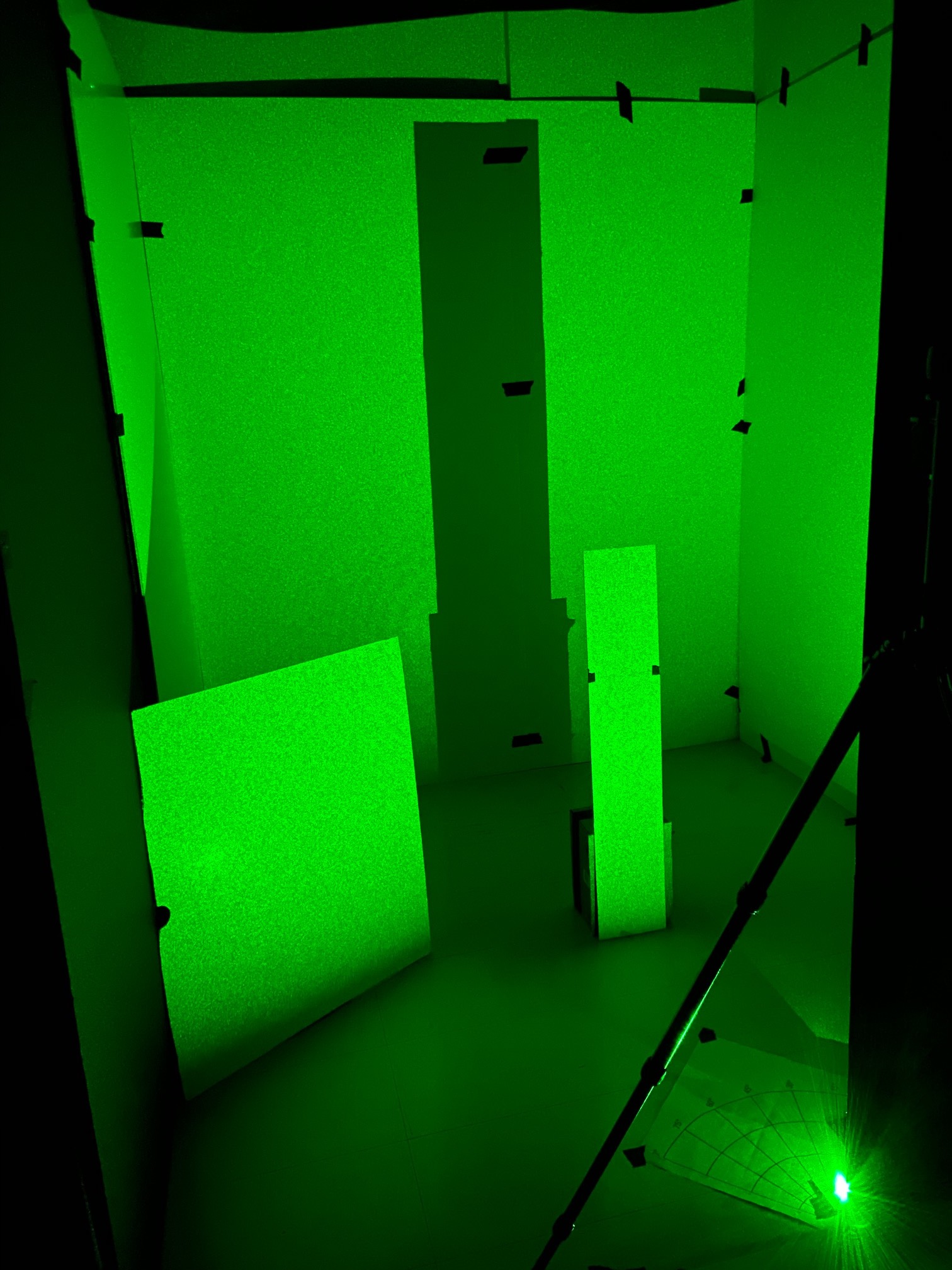}
\end{minipage}
\begin{minipage}{0.2\textwidth}
        \centering
        \text{1.25 meter}\\
        \includegraphics[width=1\linewidth]{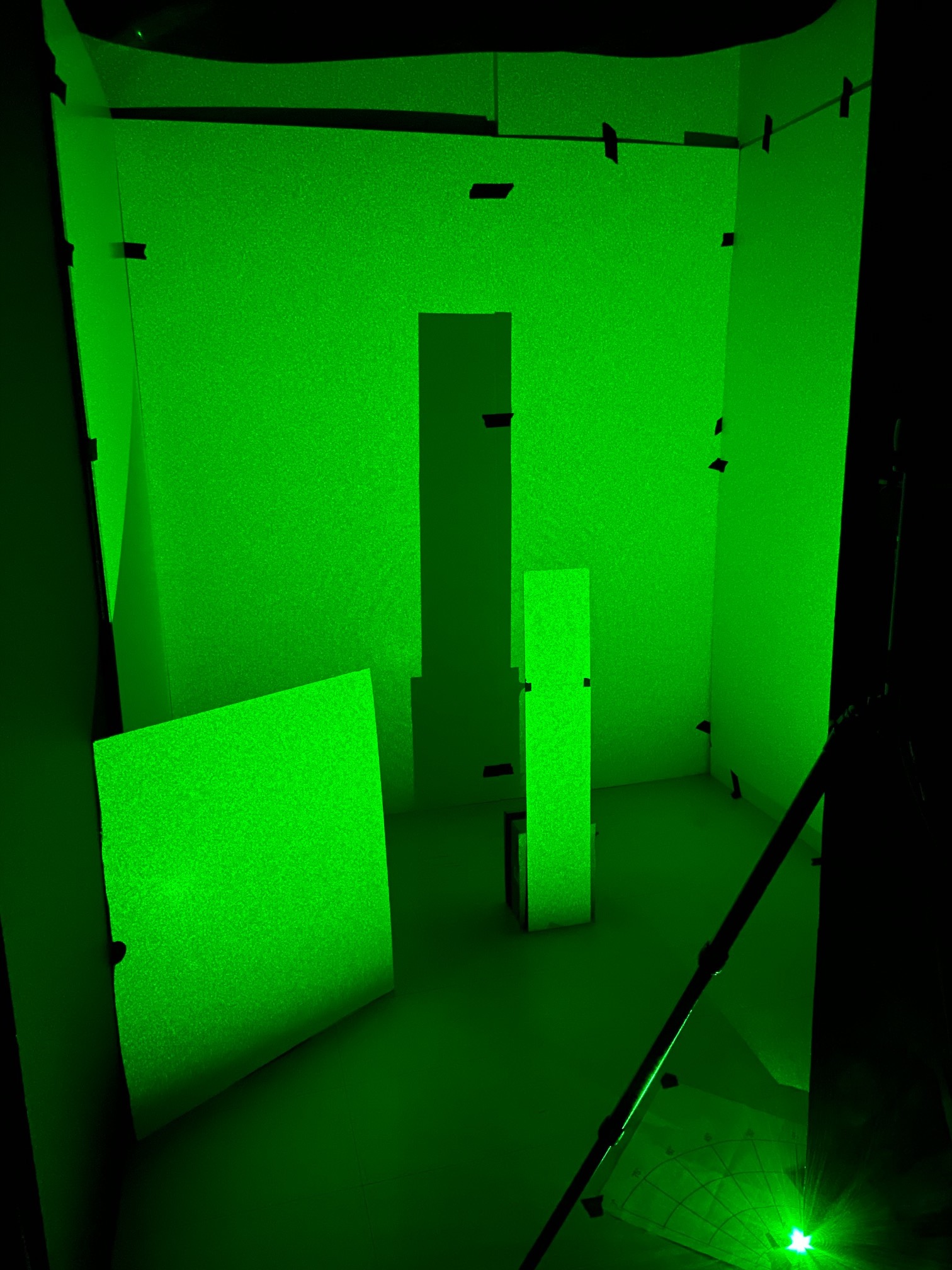}
\end{minipage}
\begin{minipage}{0.2\textwidth}
        \centering
        \text{1.5 meter}\\
        \includegraphics[width=1\linewidth]{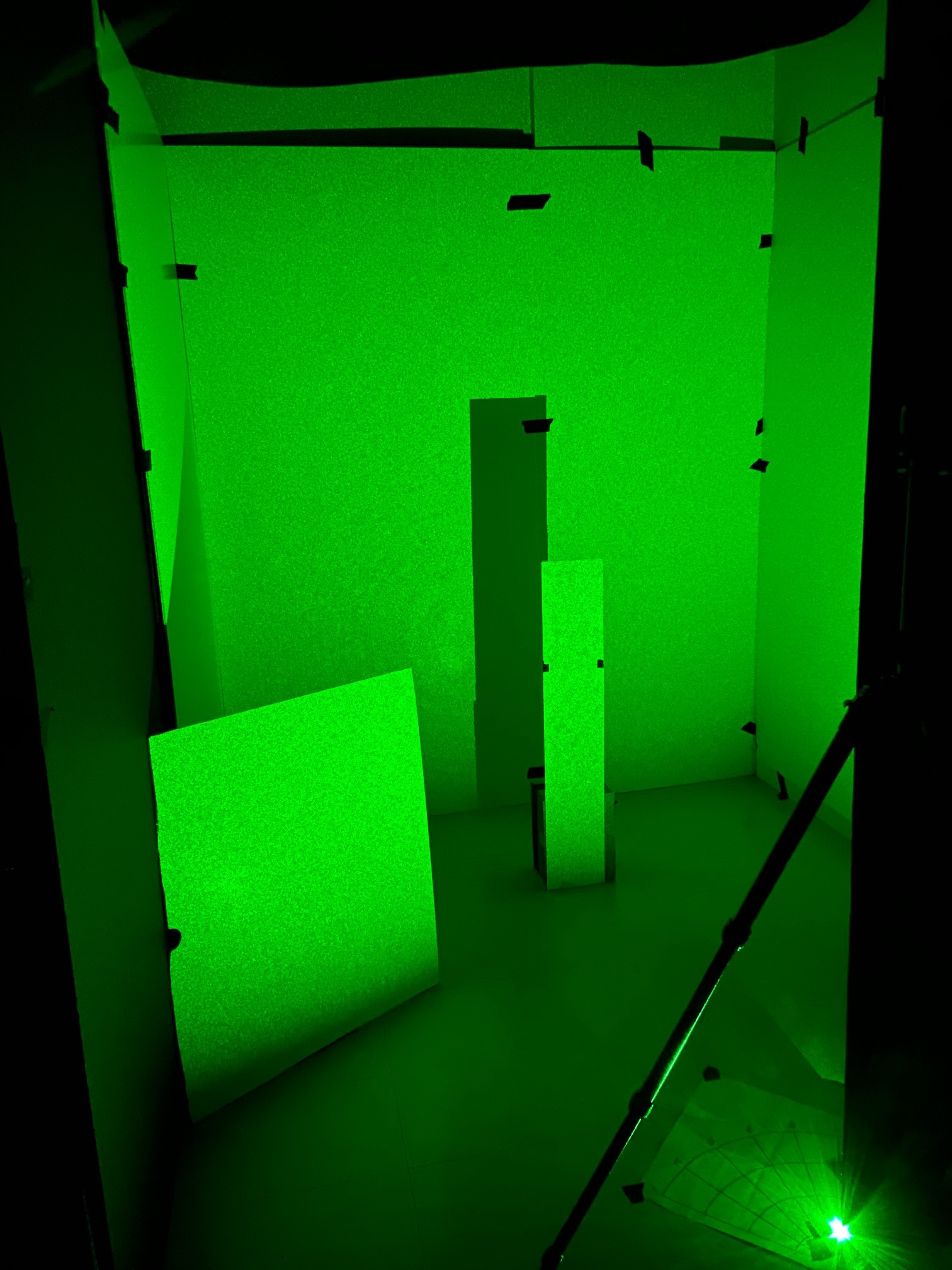}
\end{minipage}\\
\begin{minipage}{0.015\textwidth}
        \begin{sideways}
            \text{\normalsize \textbf{reconstruction}}
        \end{sideways}
\end{minipage}
\hspace{37mm}
\begin{minipage}{0.2\textwidth}
        \centering
        \includegraphics[width=1\linewidth]{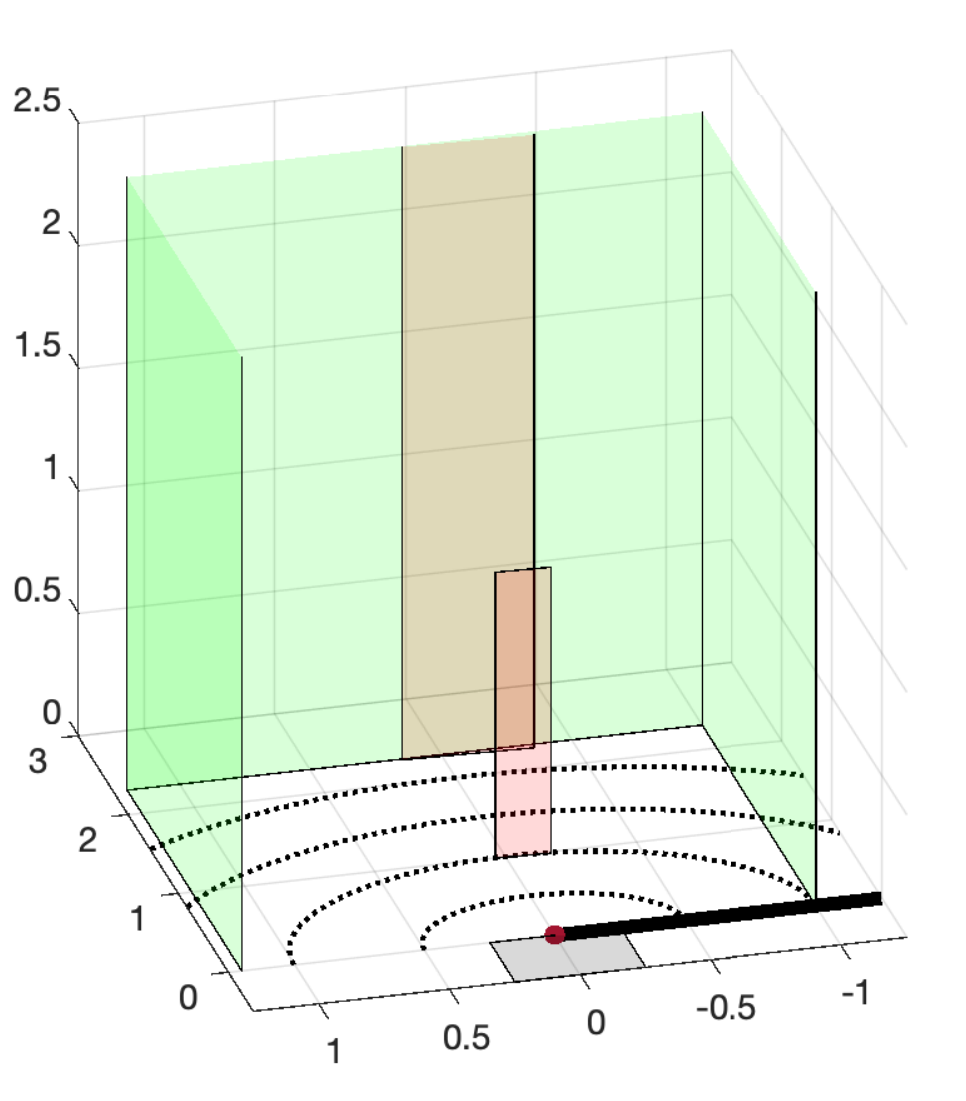}
\end{minipage}
\begin{minipage}{0.2\textwidth}
        \centering
        \includegraphics[width=1\linewidth]{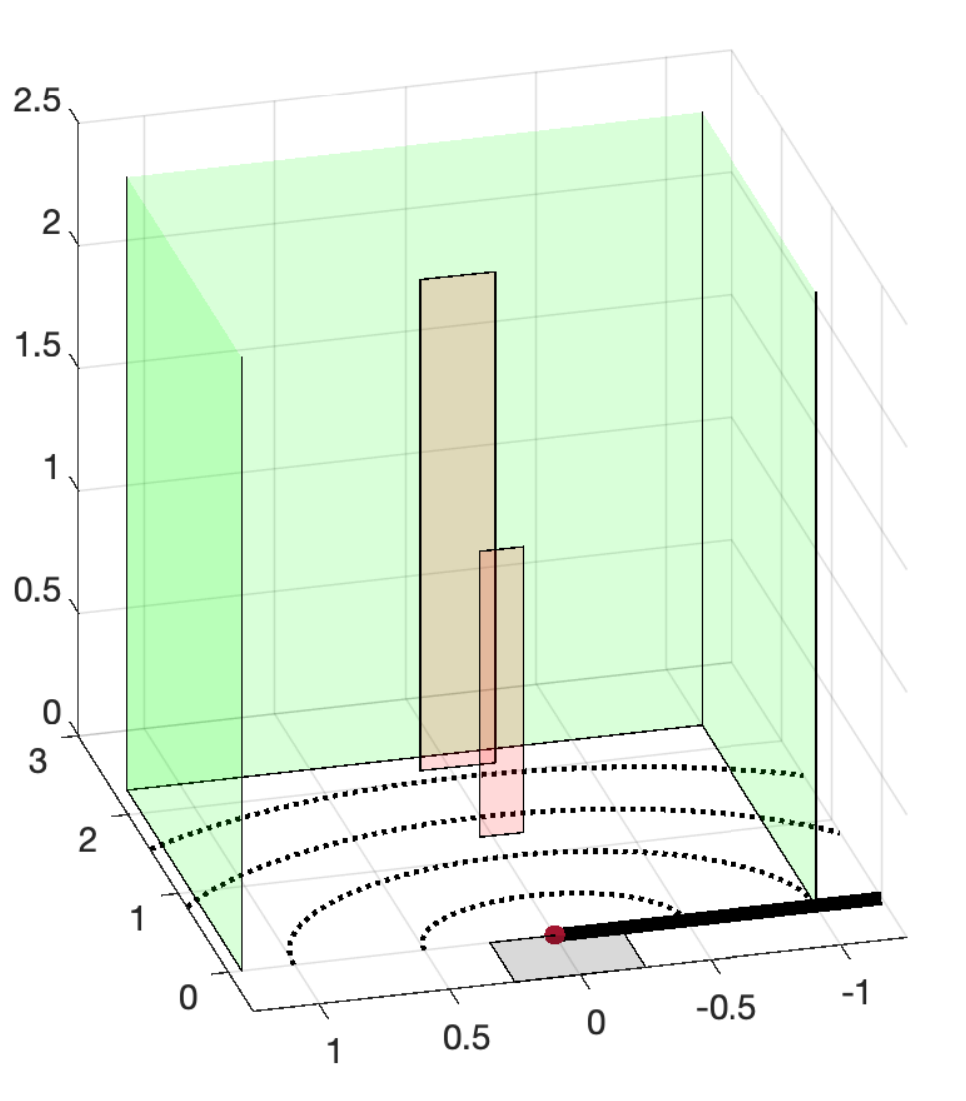}
\end{minipage}
\begin{minipage}{0.2\textwidth}
        \centering
        \includegraphics[width=1\linewidth]{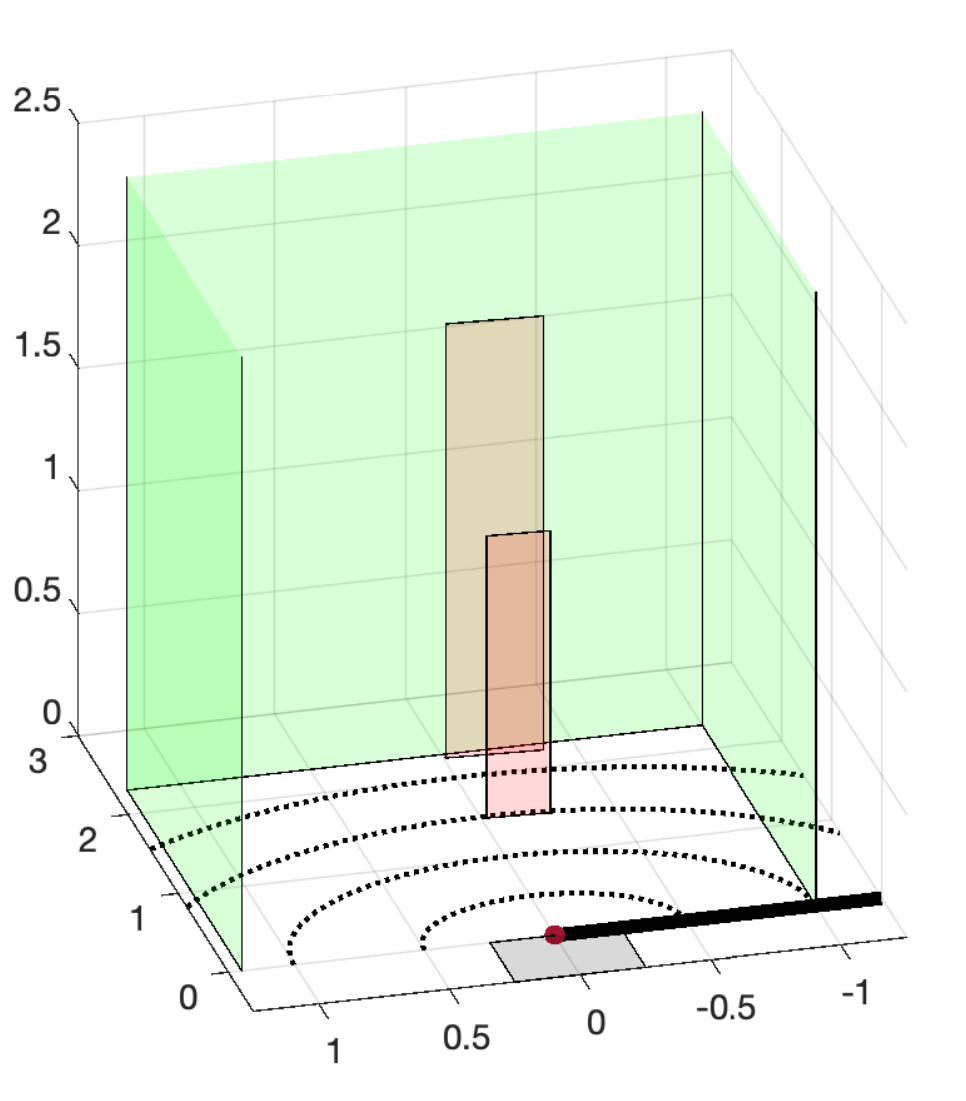}
\end{minipage}
\caption{\label{fig:stationaryFacet} Reconstruction results for a challenging hidden scene with a large stationary foreground object. The first row shows LOS photographs of the hidden scene; the second row shows the corresponding reconstructions. The true location of the back wall is shown in green with foreground and background reconstructions shown in red. The first column shows a photograph of the stationary scene, acquired before the moving object enters the hidden scene. Subsequent columns correspond to three different positions of the moving object, which is fixed in azimuthal angle at $\pi/2$ (measured around the edge and into the hidden scene) and moved in range to 1\,m, 1.25\,m, and 1.5\,m from the occluding edge. Note that even though the stationary foreground object and the moving object occupy similar range bins, reconstructions closely match the ground truth in all test cases. These results were produced using a 0.4\,s integration time per frame with an 24.5\,s reference measurement.}
\end{figure*}

Our reconstruction approach does not require any knowledge of the stationary hidden scene, as long as its \textit{response} can be well characterized by a reference measurement.
As a result, our inversion algorithm is not constrained to work only on simple hidden scenes containing just a few walls.
In \Cref{fig:stationaryFacet}, we demonstrate our inversion algorithm on a stationary hidden scene that contains a large stationary object in the foreground.
In the first figure row, we show LOS photographs of the hidden scene.
The first photograph, taken of the stationary scene before the moving object enters, shows a large white foreground object in addition to side and back walls.
After the object enters the scene, it remains at an azimuthal angle of $\pi/2$ around the corner and moves to 1\,m, 1.25\,m, and 1.5\,m ranges.
At these positions, the moving object shares some round-trip travel times with the large stationary facet.
The second figure row shows single-frame reconstruction results for the three moving object positions.
In all cases, an accurate reconstruction of both the moving object and the occluded background region is formed.

\begin{figure*}
\centering
\begin{subfigure}{0.3\textwidth}
    \centering
    \includegraphics[width=.8\linewidth]{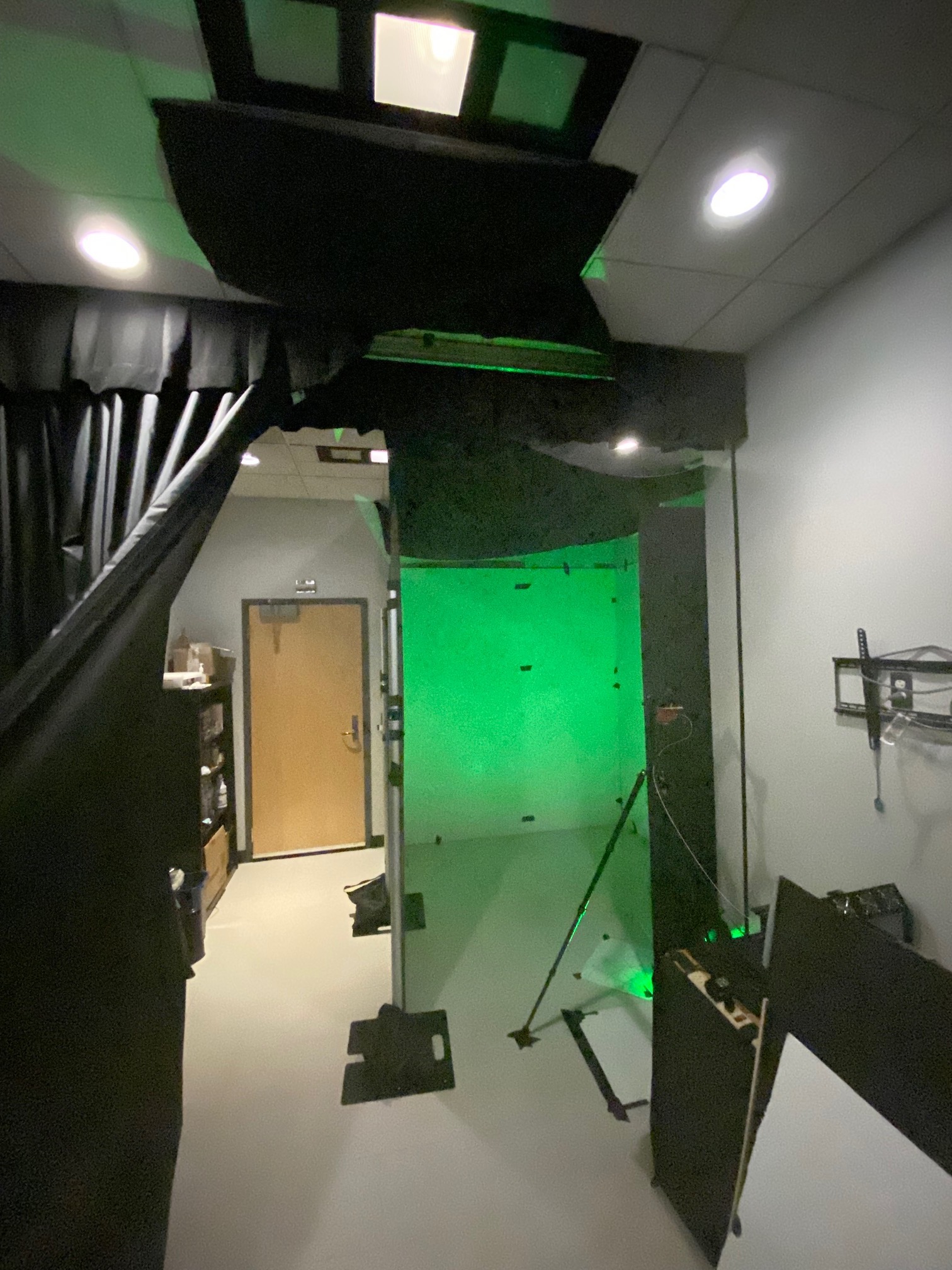}
    \caption{Laboratory photo}
    \label{fig:lights_on_lab_photo}
\end{subfigure}
\begin{subfigure}{0.3\textwidth}
        \centering
        \includegraphics[width=.8\linewidth]{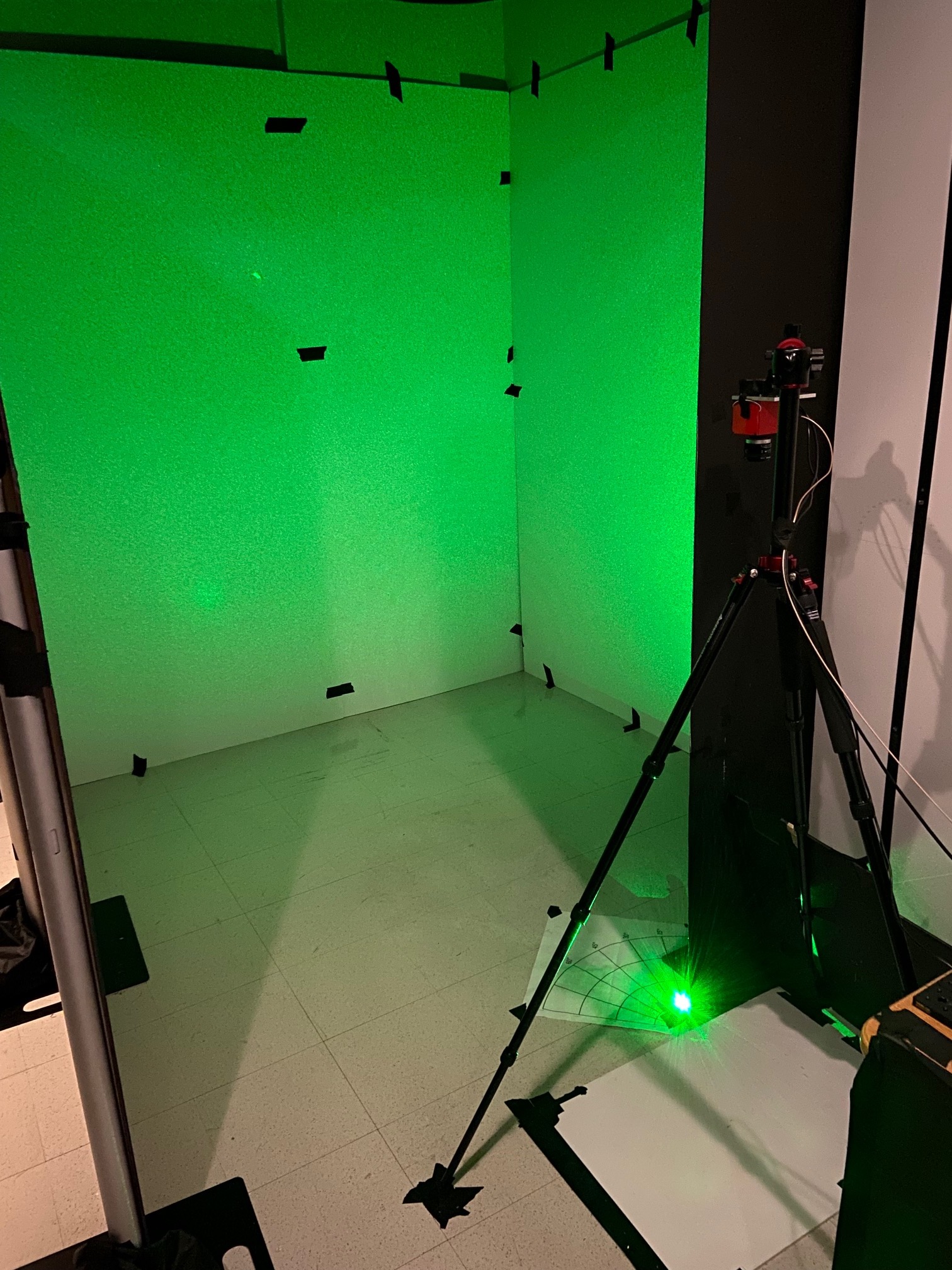}
        \caption{Stationary hidden scene}
        \label{fig:lights_on_stationary_scene}
\end{subfigure}
\begin{subfigure}{0.3\textwidth}
        \centering
        \includegraphics[width=.8\linewidth]{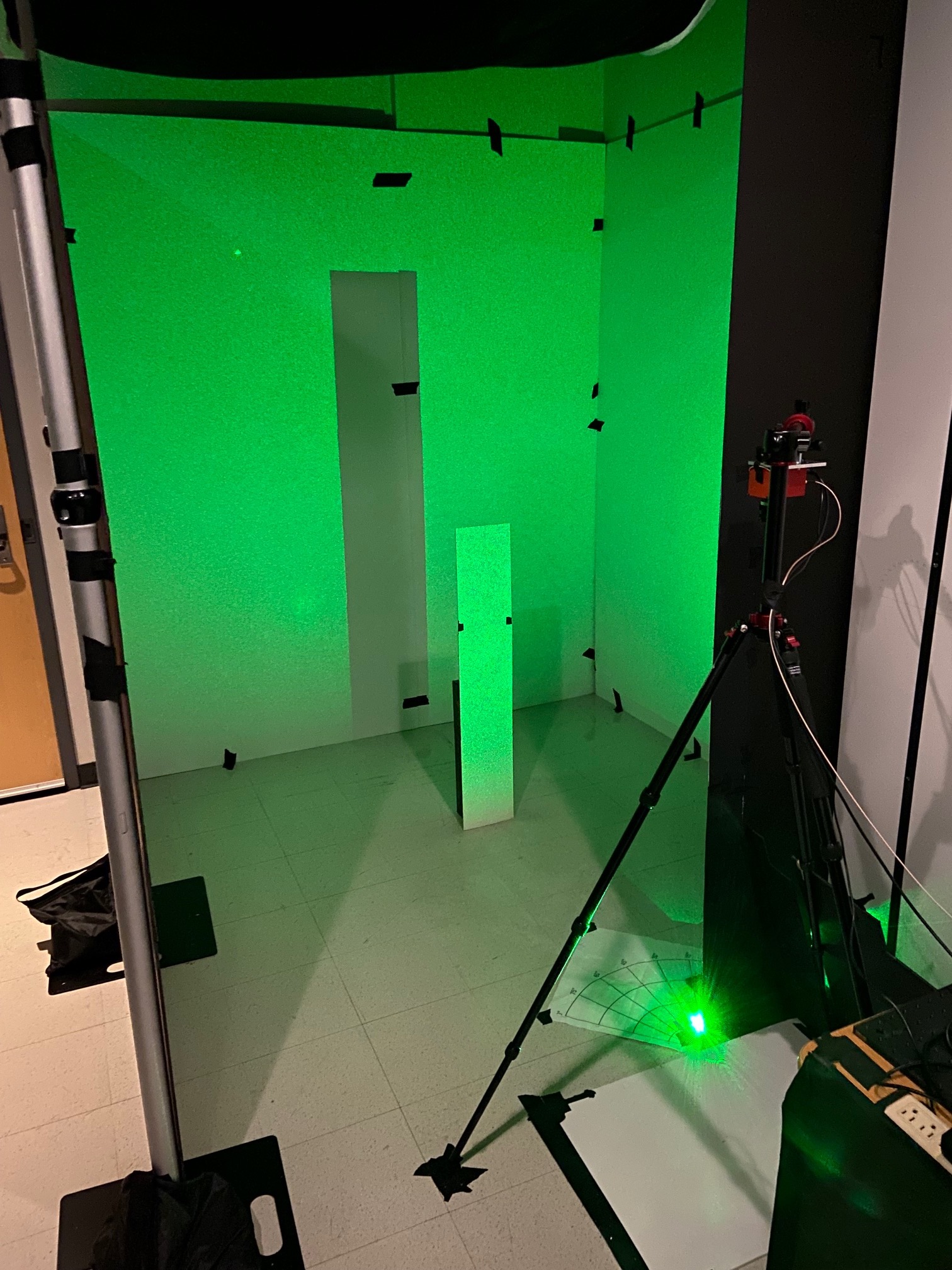}
        \caption{Hidden scene with object}
        \label{fig:lights_on_target}
\end{subfigure}
\\
\begin{subfigure}{0.3\textwidth}
        \centering
        \includegraphics[width=1\linewidth]{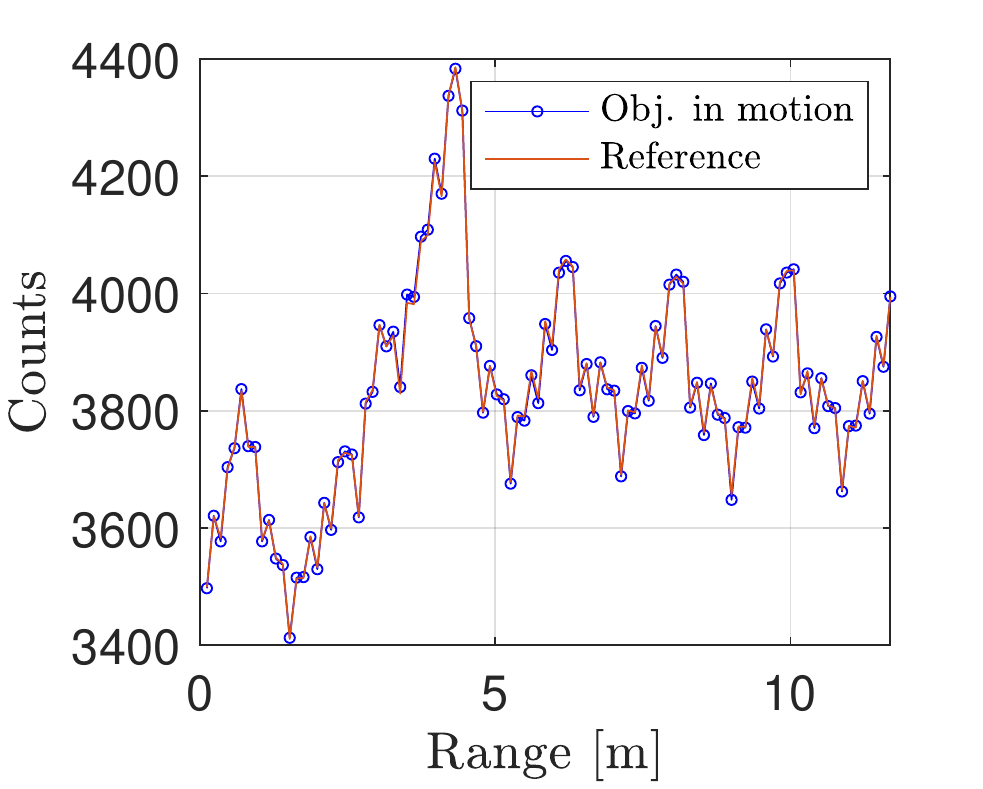}
        \caption{Spatially averaged measurements}
        \label{fig:lights_on_bg_counts}
\end{subfigure}
\begin{subfigure}{0.3\textwidth}
        \centering
        \includegraphics[width=1\linewidth]{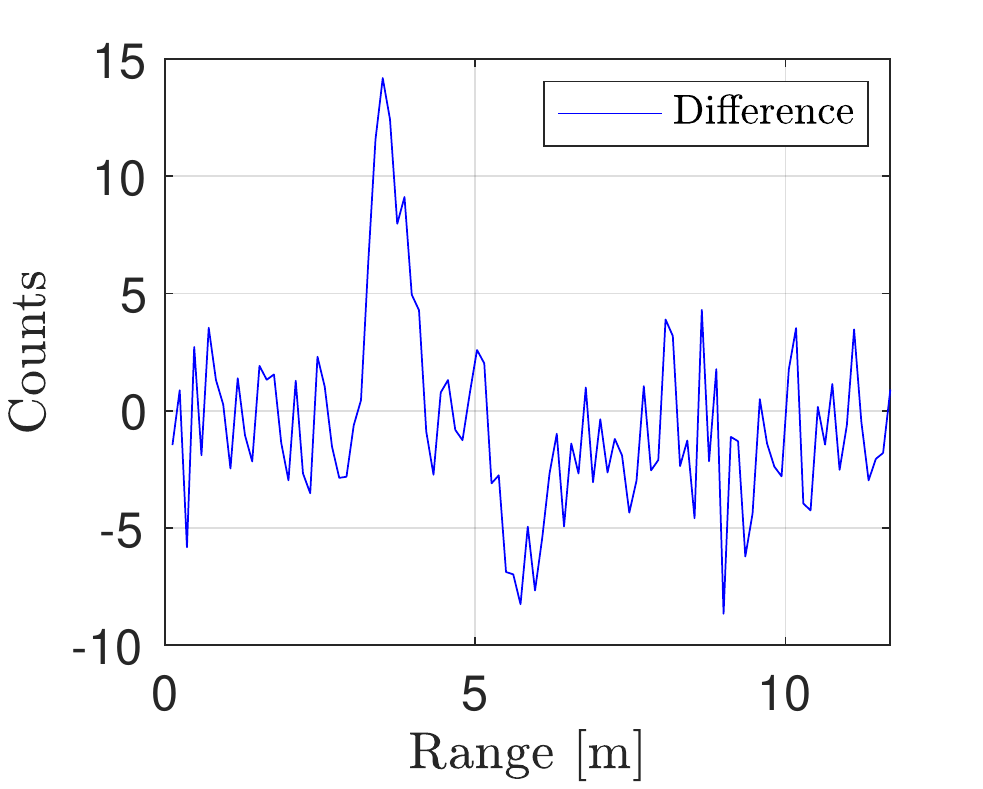}
        \caption{Spatially averaged difference measurement}
        \label{fig:lights_on_new_frame_counts}
\end{subfigure}
\begin{subfigure}{0.3\textwidth}
        \centering
        \includegraphics[width=1\linewidth]{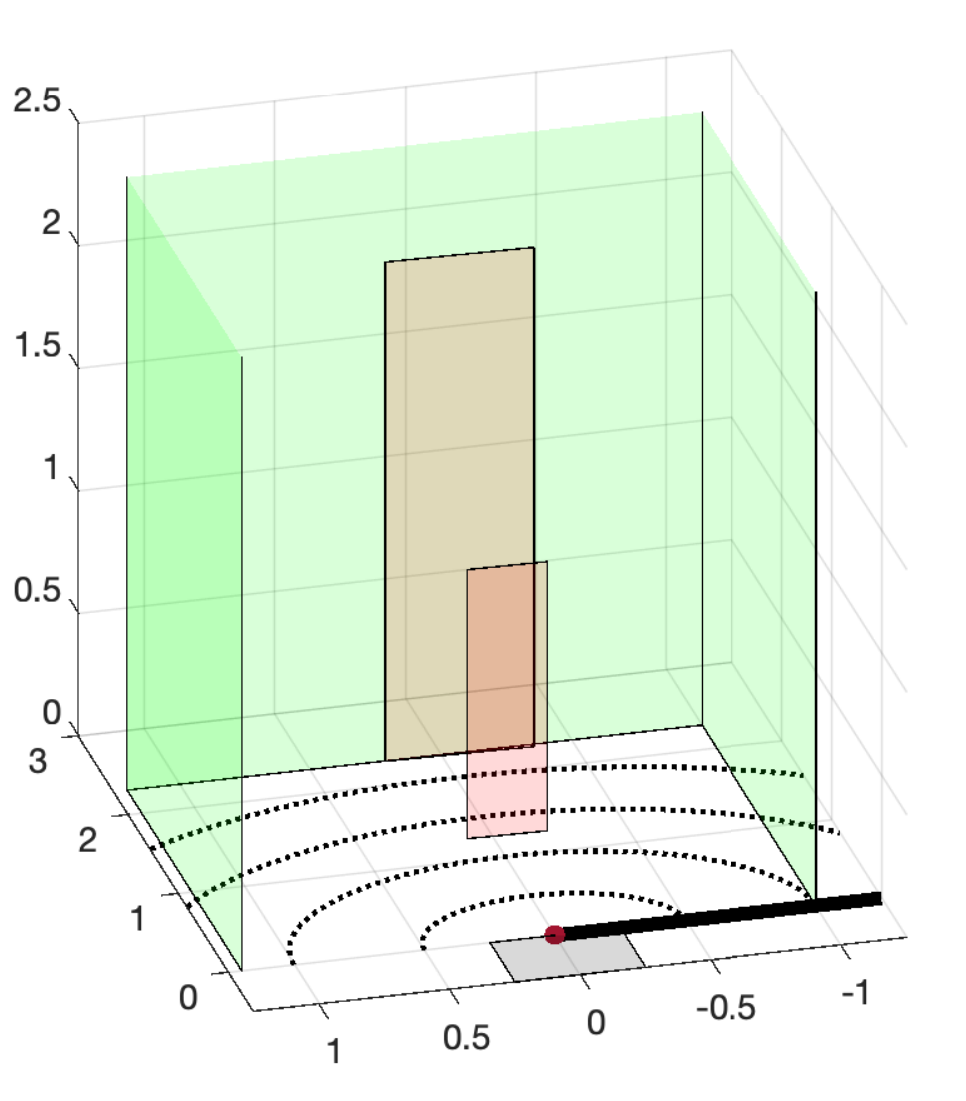}
        \caption{Reconstruction}
        \label{fig:lights_on_reconstruction}
\end{subfigure}

\caption{\label{fig:ambientLight} A demonstration of algorithm robustness to the extreme amount of ambient light introduced by turning on the overhead lighting, as shown in (a).
An LOS photo of the scene during a reference measurement is shown in (b), and a photo of the hidden scene after an object (at a range of 1.25\,m and azimuthal angle of $\pi/2$) has entered the hidden scene is shown in (c).
Spatially averaged measurements for the stationary scene (red) and one motion frame (blue) are shown in (d), with their difference shown in (e).
Despite the high background counts, the reconstruction in (f) closely matches the ground truth. The results were produced using an integration time of 3.3\,s with a reference frame integrated over 24.5\,s.}
\end{figure*}

In \Cref{fig:ambientLight}, we evaluate our algorithm with high background counts created by turning the light on in the lab. \Cref{fig:lights_on_lab_photo} shows a photograph of the entire lab with the overhead lights on,
\Cref{fig:lights_on_stationary_scene} shows just the hidden scene, and \Cref{fig:lights_on_target} shows the hidden scene once the moving object has entered.
Spatially averaged measurements for the reference and motion frames are shown in \Cref{fig:lights_on_bg_counts}, with their difference shown in \Cref{fig:lights_on_new_frame_counts}. Although the reference and motion frames and essentially indistinguishable with the naked eye, their difference, while noisy, has a discernible peak at the range of the moving target and a dip at the range of the occluded background wall;
compare with Fig.~4C of the main paper, in which the difference is much less noisy.
Single-frame reconstruction results are shown in \Cref{fig:lights_on_reconstruction}.
Despite the unmodeled effects of detector dead time and the large noise variance due to high background counts, the reconstructed object and background closely match the ground truth.

\Cref{table:countrates} provides data from representative two-minute experiments
with and without overhead lights.
The number of integrated frames is not determined entirely by the acquisition time because of the random loss of frames over the USB 2.0 interface.
Of the 1024 SPAD pixels, 39 are designated hot.
After removal of these pixels, the remaining 985 pixels have about 100 times more photon detections
with overhead lights on.
The table outlines the computations of counts per pixel per illumination pulse under each of the two conditions.
Since dead time effects are negligible, we may approximate the signal-to-ambient ratio for the results in \Cref{fig:ambientLight} as
  $\frac{ 0.0000252 }{0.00309 - 0.0000252} \approx 0.008$.


\renewcommand{\arraystretch}{1.5} 
\begin{table}
\resizebox{\textwidth}{!}{%
\begin{tabular}{l|c|c}
\hline
\hline
 & \textbf{Negligible ambient light } & \textbf{Overhead lights} \\\hline
 
Integrated frames & 
$2.118\times10^6 \,\text{frames}$ &
$2.102\times10^6 \,\text{frames}$ 
\\\hline

Detections & 
$2.10\times10^6 \,\text{counts}$ &
$2.56\times10^8 \,\text{counts}$ 
\\\hline

Pixels  & 985 & 985 \\\hline

Pulses per frame 
&  $\left(50\times 10^6 \frac{\text{pulses}}{\text{sec}}\right) \left(800\times 10^{-9} \frac{\text{sec}}{\text{frame}} \right) = 40\, \frac{\text{pulses}}{\text{frame}}$
&  $\left(50\times 10^6 \frac{\text{pulses}}{\text{sec}}\right) \left(800\times 10^{-9} \frac{\text{sec}}{\text{frame}} \right) = 40\, \frac{\text{pulses}}{\text{frame}}$
\\\hline

Detection rate & 
$\frac{ 2.10\times10^6 \,\text{counts} }
     { 2.118\times10^6 \,\text{frames} }
\left( \frac{1\,\text{frame}}{40\,\text{pulses}} \right)
\left( \frac{1}{985\,\text{pixels}} \right)
= 0.0000252 \,\frac{\text{counts}}{\text{pixel}\cdot\text{pulse}}$ &
$\frac{ 2.56\times10^8 \,\text{counts} }
     { 2.102\times10^6 \,\text{frames} }
\left( \frac{1\,\text{frame}}{40\,\text{pulses}} \right)
\left( \frac{1}{985\,\text{pixels}} \right)
= 0.00309 \,\frac{\text{counts}}{\text{pixel}\cdot\text{pulse}}$ 
\\
\hline
\hline
\end{tabular}
} 
\caption{Measurements from representative experiments with and without overhead lights,
along with derived count rates per laser pulse.
Counts from 39 hot pixels are removed.
\label{table:countrates}}
\end{table}

\clearpage
\bibliographystyle{IEEEtran}
\bibliography{nlos}

\end{document}